\documentclass[fleqn,usenatbib]{mnras}
\usepackage{newtxtext,newtxmath}
\usepackage[T1]{fontenc}
\usepackage{ae,aecompl}
\usepackage{bigints}

\usepackage{float}
\usepackage{graphicx}	
\usepackage{amsmath}	
\usepackage{bigints}
\usepackage{natbib}
\usepackage{caption}
\usepackage{ragged2e}
\usepackage{placeins}
\usepackage{bigints} 
\usepackage{physics}
\usepackage{cleveref}
\usepackage[export]{adjustbox}
\usepackage{longtable}
\usepackage{xtab}
\usepackage{bm}
\usepackage[mathscr]{eucal}
\usepackage{comment}
\usepackage{multirow}
\usepackage{tabularx}
\usepackage[table,xcdraw]{xcolor}
\usepackage{setspace}
\usepackage[inkscapearea=page]{svg}
\usepackage{longtable,rotating}
\usepackage{threeparttable}
\usepackage{booktabs}
\usepackage{tcolorbox}
\usepackage{mathtools,leftindex,tensor}
\usepackage[version=4]{mhchem}
\usepackage{makecell}
\usepackage{listings}
\usepackage{anyfontsize}
\DeclareRobustCommand{\ttfamily}{\fontencoding{T1}\fontfamily{lmtt}\selectfont}
\usepackage[normalem]{ulem}
\usepackage{CJKutf8}

\newcommand{\be}{\begin{equation}}
\newcommand{\ee}{\end{equation}}
\newcommand{\ba}{\begin{eqnarray}}
\newcommand{\ea}{\end{eqnarray}}
\newcommand{\bse}{\begin{subequations}}
\newcommand{\ese}{\end{subequations}}
\newcommand{\overbar}[1]{\mkern 1.5mu\overline{\mkern-1.5mu#1\mkern-1.5mu}\mkern 1.5mu}

\usepackage[export]{adjustbox}

\usepackage{hyperref}
\hypersetup{
    colorlinks=true,
    linkcolor=blue,
    filecolor=magenta,      
    urlcolor=cyan,
    pdftitle={Cloud crushing in expanding outflows},
    pdfpagemode=FullScreen,
}
\numberwithin{equation}{section}

\definecolor{webgreen}{rgb}{0,.5,0}
\definecolor{webbrown}{rgb}{.6,0,0}

\usepackage{etoolbox}

\title[Cloud-crushing in expanding galactic outflows]{Fading in the Flow: Suppression of cold gas growth in expanding galactic outflows}
\author[A. Dutta, P. Sharma and M. Gronke]{
Alankar Dutta,$^{1,2}$\thanks{E-mail: \href{mailto:alankard@mpa-garching.mpg.de}{alankard@mpa-garching.mpg.de}}
Prateek Sharma,$^{2}$ 
and
Max Gronke$^{1,3}$
\\
$^{1}$ Max-Planck-Institut f\"{u}r Astrophysik, Karl-Schwarzschild-Stra{\ss}e 1, 85748 Garching bei M\"{u}nchen, Germany \\
$^{2}$ Department of Physics, Indian Institute of Science, Bangalore 560012, Karnataka, India \\
$^{3}$ Astronomisches Rechen-Institut, Zentrum für Astronomie, Universität Heidelberg, Mönchhofstraße 12-14, 69120 Heidelberg, Germany
}

\date{Accepted XXX. Received YYY; in original form ZZZ}

\pubyear{2025}

\begin{document}
\label{firstpage}
\pagerange{\pageref{firstpage}--\pageref{lastpage}}
\maketitle

\defcitealias{gronke2018}{GO18}
\defcitealias{gronke2020}{GO20}
\defcitealias{Chevalier1985}{CC85}
\defcitealias{Fielding2022ApJ}{FB22}

\begin{abstract}
Multiphase outflows, revealed by multi-wavelength observations, are crucial in redistributing gas and metals within and around galaxies. These outflows are often modelled theoretically using wind tunnel simulations of a cold ($\sim 10^4$ K) cloud interacting with a uniform hot ($\sim 10^6$ K) wind. However, real outflows expand downstream, a feature overlooked in most idealised simulations. We study how an expanding wind affects the survival, morphology, and dynamics of a cloud. 
We conduct idealised hydrodynamic simulations with optically thin radiative cooling of a cloud in an expanding wind, modelled using the adiabatic Chevalier \& Clegg (1985) solution. We find that clouds remain locally isobaric with the wind, leading to a steep decline in their density contrast and eventual dissolution downstream. Compared to a plane-parallel wind, this suppresses cold gas mass growth because as clouds travel downstream, the surrounding mixed boundary layer becomes diffuse and less radiative. 
Our analytical scaling arguments show that cloud expansion and local pressure equilibrium are the key regulators of cold mass growth. Unlike traditional simulations, our model accounts for the differential expansion experienced by the long cometary tails of clouds in wind tunnels. This creates a strong head-to-tail emission gradient in the filamentary cold gas, which is more consistent with observations. We also demonstrate that the dynamics of individual clouds can substantially alter the radial properties of their host multiphase outflows.

\end{abstract}



\section{Introduction}
\label{sec:introduction}
Galactic outflows, particularly those driven by starburst activities (\citealt{Heckman1990ApJS, Lehnert1996ApJ}), are pivotal in regulating galaxy evolution by expelling gas, metals, and energy (see 
\citealt{veilleux2020, Thompson2024ARA&A} for reviews on galactic winds and \citealt{Sarkar2024A&ARv} for a review on Fermi/eROSITA bubbles 
driven perpendicular to the Milky Way disk) into the circumgalactic medium (CGM; see \citealt{tumlinson2017, Faucher2023ARA&A} for reviews on CGM) and even the intergalactic space in the case of stronger outflows, possibly powered by AGN (\citealt{Oppenheimer2020MNRAS, Davies2020MNRAS, Keller2020MNRAS}). These outflows are inherently multiphase, encompassing hot, ionised gas, warm neutral and ionised components, and cold molecular clouds (\citealt{Heckman2000ApJS, thompson2016, Roberts2020MNRAS, veilleux2020, Vijayan2020ApJ, Fluetsch2021MNRAS}). Understanding the dynamics of these cold clouds within outflows is crucial, as they not only transport mass and metals, shaping star formation rates and influencing the thermal and chemical state of the CGM and the intergalactic medium (IGM), but also alter the overall dynamics of the wind itself through momentum and energy exchange (\citealt{Hopkins2012MNRAS, Alcazar2017MNRAS, Kim2018ApJ, McQuinn2018ApJ, Zabl2020MNRAS, Belli2024Natur}).

The interaction between cold clouds and hot winds is a fundamental process not only in galactic outflows but also in many other astrophysical contexts that influence galaxy formation and evolution (\citealt{Dekel1986ApJ, Cole1994MNRAS, Erb2008ApJ, Putman2012ARA&A, Pezzulli2016MNRAS, Naab2017ARA&A}). These include fountain flows (\citealt{Shapiro1976ApJ, Bregman1980ApJ, Houck1990ApJ, Fraternali2008MNRAS, Rubin2022ApJ, Marasco2022MNRAS, Li2023MNRAS}) that potentially give rise to high-velocity clouds (\citealt{Benjamin1997ApJ, Putman1999astro.ph, Wakker2001ApJS, Putman2002AJ, Binney2009MNRAS, Heitsch2009, Richter2017A&A, Lehner2022MNRAS}), and the stripping of gas from satellite/jellyfish galaxies (\citealt{Roediger2007MNRAS, Tonnesen2010ApJ, Steinhauser2016A&A, Ramatsoku2020A&A, Franchetto2020ApJ, Poggianti2019ApJ, Campitiello2021ApJ, Deb2022MNRAS}) with the subsequent formation of cold gas (\citealt{Ghosh2024MNRAS, Roy2024MNRAS}) and stars (\citealt{Tonnesen2012MNRAS, Vulcani2018ApJ, Roberts2022ApJ}) in their tails. The fate of these cold clouds, i.e., whether they survive and grow or are destroyed, has significant implications for the star formation rate/history of their host galaxies, as well as for the mass and energy budget of galaxies and their surrounding media. The cold gas may also condense from the hot and  diffuse CGM due to thermal instability, particularly in the presence of large density fluctuations and stratification -- common in CGM-like environments (e.g., \citealt{sharma2012, McCourt2012MNRAS, Voit2018cosp, choudhury2019, Mohapatra2020MNRAS, Mohapatra2022MNRAS, Wibking2025MNRAS}). Most importantly, observations indicate the presence of multiphase gas for most sightlines through the CGM, indicating the ubiquity of cold and hot gas sitting next to each other in different parts of galactic atmospheres. At smaller scales - particularly when probing the structure and thermodynamics of multiphase gas in the interstellar medium (ISM) - numerous studies have explored the physics of cloud-wind interactions. These include interactions between clouds and supernova-driven winds within the ISM \citep{Cowie1981ApJ, Pittard2019MNRAS, Kim2017ApJ, Guo2024arXiv}, as well as with the immediate surroundings of supernova remnants \citep{Dwarkadas2007ApJ}, and in bubbles inflated by stellar winds \citep{Lancaster2021ApJ, 2022MNRAS.515.1815P, Dwarkadas2023Galax}. Remarkably, even seemingly unrelated astrophysical phenomena involving vastly different time and length scales - such as coronal rain in the solar atmosphere - exhibit striking similarities to cloud-wind interactions including multiphase gas mixing and thermal instability, also observed in galactic environments \citep{Antolin2010ApJ, antolin2020, Kolotkov2023-eg, Hillier2025AAP}. 

The importance of cold clouds interacting with the surrounding medium in shaping galaxy formation has also been recently demonstrated in cosmological simulations of galaxy formation (\citealt{nelson2020, Ramesh2024MNRAS}). Observations as well as simulations suggest that similar physics of multiphase gas mixing governs the formation of Magellanic streams surrounding the large and small Magellanic clouds (\citealt{DOnghia2016ARA&, Fox2020ApJ, Lucchini2021ApJ, Mishra2024ApJ}) as well as in the fuelling galaxies with cold gas by pristine cold filaments from the IGM (\citealt{Keres2005MNRAS, Dekel2009Natur, Voort2012MNRAS, Mandelker2016MNRAS, Mandelker2020MNRAS, Berlok2019MNRAS, Martin2019NatAs, Aung2024MNRAS}), especially around the cosmic noon ($z\sim 2-3$). 

Traditionally, the interaction between cold clouds and hot winds has been studied using the cloud-crushing setup (\citealt{klein1994, Xu1995ApJ, Schiano1995ApJ, Vietri1997ApJ, Orlando2005A&A, Nakamura2006ApJS, Scannapieco2015ApJ, Zhang2017MNRAS}), where a cold and dense cloud is subjected to a low density hot wind in an initially isobaric condition. There have been notable differences across the simulation setups in these works, for example, whether there is an initial shock sweeping through the cloud, the inclusion of thermal conduction and magnetic fields (\citealt{Shin2008ApJ, McCourt2015MNRAS, Kooij2021MNRAS, Pineda2024MNRAS}), or if there are multiple clouds (\citealt{Byung1996ApJ, Poludnenko2002ApJ, Pittard2005MNRAS, Aluzas2012MNRAS}) that have greatly improved our understanding of cloud-wind interactions.

These early `cloud-crushing' studies have revealed a common picture in which cold clouds undergo mixing and destruction due to a combination of various hydrodynamic instabilities, such as Kelvin-Helmholtz (\citealt{vonHelmholtz1868uber, Thomson01111871, Chandrasekhar1961book}), Rayleigh-Taylor (\citealt{Rayleigh, taylor1950formation, Kull1991PhR, Choudhuri1998book}), and Richtmyer-Meshkov (\citealt{Richtmyer, Meshkov1969, Yang1994PhyFl, Brouillette2002ARFM, Zhou2021PhyD, Napieralski2024JFM}), which develop as a consequence of the relative motion and subsequent momentum transfer between the cloud and the wind. A critical 
timescale in this context is the cloud-crushing time $t_{\rm cc}$, defined as the timescale over which a shock propagates through the cloud (see Eqs. 2.2 \& 2.4 from \citealt{klein1994}), leading to its compression and potential destruction of the cloud. This timescale, which also approximately equals the time for the growth of the unstable modes of the Kelvin-Helmholtz instability, is given by $t_{\rm cc} = \sqrt{\chi} R_{\rm cl}/ v_{\rm wind}$, where $\chi$ is the factor by which the cloud is over-dense compared to the wind, $R_{\rm cl}$ is the cloud size, and $v_{\rm wind}$ is the wind velocity. 

A significant breakthrough in 
the relevance of the cloud-crushing problem to the multiphase CGM came 
with the discovery of a regime in which cold clouds can grow in the wind 
in the presence of radiative cooling (\citealt{armillotta2016, armillotta2017, gronke2018, Li2020, Sparre2020, Kanjilal2021}). In order to explain the origin of high-velocity clouds (HVCs) and cold extra-planar HI gas in Milky Way-like galaxies (\citealt{Oosterloo2007AJ, Fraternali2009IAUS}), \citealt{Marinacci2010MNRAS, Marinacci2011MNRAS} carried out two-dimensional simulations of clouds (possibly uplifted from the ISM by supernovae) falling towards the Galactic disk. These are some of the earliest cloud-crushing simulations that found growth of cold gas due to radiative cooling in the wakes of clouds moving in the CGM (see review by \citealt{Fraternali2017ASSL}). 
More recent simulations have supported this finding and have consistently shown that clouds can grow in the regime of rapid radiative cooling. The relative shear between the cold cloud and the hot wind gives rise to a turbulent boundary layer with a short cooling time. This can lead to the entrainment of the diffuse gas into the dense medium. 
Such turbulent radiative mixing layers (TRMLs) have been extensively studied in recent idealised simulations covering a wide range of astrophysical conditions (\citealt{Ji2019MNRAS, Fielding_2020, Tan2020, Yang2023MNRAS, Abruzzo2024ApJ}), providing a general 
consistency with the results of radiative cloud-crushing simulations. While TRMLs driven by shear lead to mass growth at early times in radiative cloud-crushing, after entrainment, the relative velocity between the cloud and the wind decreases and yet the dense cloud growth rate peaks in such a state without shear (\citealt{gronke2020}). This late-time growth may be interpreted as the steady, subsonic cooling flow solution that exists in spherical and cylindrical flow geometries (no such steady inflow solution exists in Cartesian geometry; \citealt{Dutta2022MNRAS}).

Early works by \citealt{Marinacci2010MNRAS, Marinacci2011MNRAS}, which found growth of cold clouds,  
assume in their analytic estimate that the cooling time in the mixed phase remains very long during mixing. Later, \citealt{gronke2018, gronke2020} removed this assumption and proposed an analytic criterion required for the growth of shear-driven cold clouds -- if the cooling time of the gas in the turbulent boundary layers (formed in cloud-crushing) is shorter than the cloud-crushing/shear time, the cold gas in the cloud keeps growing, and the cloud acquires mass and momentum from the wind. This cooling criterion has since been verified by simulation results and translates to an equivalent threshold on cloud size, such that any cloud larger than this threshold size can grow due to the net cooling 
of the mixed (warm) gas. \citealt{gronke2018} proposed this threshold cloud radius $R_{\rm thres}$ as 
\begin{equation}
    \begin{aligned}
             R_{\rm thres} &= \frac{\sqrt{\frac{\gamma k_B T_{\rm cl}}{\mu m_p} }}{\gamma -1} \frac{\left(k_B T_{\rm cl}\right)^2 \mathcal{M} }{X_H^2 P \Lambda \left(\sqrt{\chi} T_{\rm cl}\right)} \chi \delta^{-1}\\
             &\approx 10\ {\rm pc} \left(\frac{T_{\rm cl}}{10^4 \rm \ K}\right)^{5/2} \left(\frac{p/k_B}{10^3 \ \rm K\ cm^{-3}}\right)^{-1}  \left(\frac{\mathcal{M}}{1.0}\right) \\
             & \ \ \ \  \left(\frac{\Lambda \left(\sqrt{\chi} T_{\rm cl}\right)}{10^{-21.29}\ \rm erg\ cm^3\ s^{-1}} \right)^{-1}
             \left(\frac{\chi}{100}\right) \delta^{-1}.
    \end{aligned}
     \label{eq:R_thres} 
\end{equation}
Here $p$ is the 
background pressure, $T_{\rm cl}$ is the cloud temperature, $\mathcal{M}$ is the Mach number of the wind (with respect to the hot gas sound speed), $\chi$ is the density contrast of the cloud with the wind, $\Lambda (T)$ is the radiative cooling function,\footnote{The radiative cooling rate per unit volume $\mathcal{L} = -n_H^2 \Lambda (T)$, where $n_H$ is the total hydrogen number density.} $X_H$ is the total hydrogen mass fraction, and $\delta$ is a dimensionless parameter. We run simulations to numerically determine the value of $\delta$ that sets the size of the cloud when it transitions from destruction to the growth regime 
(see Section \ref{subsec:vanilla-fudge}). 

However, real galactic outflows are more complex, often exhibiting adiabatic expansion as they move downstream due to the continuous energy 
and mass input from starburst regions (\citealt{Chevalier1985}). This expansion leads to spatial and temporal variations in wind properties that the cloud experiences as it moves down the wind, necessitating a more generalised approach to studying cloud-wind interactions. In this paper, we extend the traditional cloud-crushing model. Moving beyond the traditional model where the background wind was considered to be uniform (which we refer to as `vanilla' cloud-crushing), we investigate the behaviour of cold clouds in adiabatically expanding winds powered by continuous starbursts. Through a series of hydrodynamic simulations, we analyse how these realistic wind environments influence the survival, acceleration, and mixing of clouds. 
We model the radially expanding adiabatic background wind using the steady-state analytic solution of an adiabatically expanding spherical wind proposed by \citealt{Chevalier1985} (\citetalias{Chevalier1985} hereafter) for starburst-driven outflows. In a \citetalias{Chevalier1985} wind, an injection radius exists within which continuous star formation (and subsequent supernovae) injects mass and (thermal-)energy at a constant rate, which drives a steady wind. The wind transitions from a subsonic regime to a supersonic regime as it moves past the injection radius, where the Mach number of the wind is unity. In the asymptotic limit, the \citetalias{Chevalier1985} profile gives the wind pressure $p_{\rm wind} \propto d^{-2\gamma}$, density $\rho_{\rm wind} \propto d^{-2}$, and radial velocity $v_{\rm wind} \sim \rm{constant}$, where $d$ is the distance from the centre of the wind and $\gamma$ is the ratio of specific heats. 

Our paper is organized as follows. We present our numerical setup and discuss some of the novel techniques developed for our study in Section \ref{sec:num_sims}. In Section \ref{sec:results}, we present our simulation results and analyse the cloud properties as they evolve in a \citetalias{Chevalier1985} wind. We show that the growth of cold clouds is suppressed in an expanding outflow and clouds become increasingly indistinguishable from the background wind as they travel downstream. In Section \ref{sec:discuss}, we 
compare with previous simulations, and models of cloud-wind interaction and multiphase galactic outflows. We discuss models that 
evolve the clouds in an expanding wind using the analytic framework of radiatively cooling turbulent boundary layers. We 
build a physical theory of cloud evolution based on its downstream expansion, 
assuming it to be locally isobaric relation with the wind, backed by our simulation results. In this context, we also highlight the limitations of the existing models and simulations of cloud-wind interaction and how model parameters dramatically affect the properties of multiphase galactic outflows. We also compare with recent JWST observations of cold clouds in outflows (in M82). 
Finally, we summarize and conclude in Section \ref{sec:summary}. 

\section{Physical setup \& Numerical Simulations}
\label{sec:num_sims}
Our numerical simulations use a three-dimensional grid with spherical polar coordinates. Our grid is uniformly spaced along the radial as well as along the polar and azimuthal directions. The spherical polar geometry of the grid aligns with the spherical symmetry of the \citetalias{Chevalier1985} wind and reduces numerical artifacts that can otherwise perturb the steady-state \citetalias{Chevalier1985} wind profile used as the background. We have used a modified version of {\tt PLUTO v4.4p2} in our simulations. It is a conservative finite-volume Godunov (magneto) hydrodynamics code (\citealt{PLUTO}). We also include optically thin radiative cooling of gas at solar metallicity assuming photo+collisional ionisation equilibrium in the presence of a ($z=0$) Haardt-Madau extragalactic UV background (\citealt{Haardt2012}). Radiative cooling in our simulation is turned off at the floor temperature of $4 \times 10^{4}$ K, the initial temperature of the cloud. This mimics the drastic reduction in cooling efficiency below this temperature because of the heating by the extragalactic background radiation. Further, in order to prevent deviations from the steady adiabatically expanding background, we have turned off radiative cooling in the bulk of the wind (identified with passive tracer), and only the mixed gas is allowed to cool radiatively. Thermal instability in the bulk of the background wind in the presence of radiative cooling (\citealt{Wang1995ApJ, thompson2016, Schneider2018, Nguyen2024MNRAS}) is a related but different problem -- 
beyond the scope of our present study. 
We have also not considered other physical processes like magnetic fields, thermal conduction, self-gravity, etc. Since this is the first work that investigates the effect of an expanding background on cloud-crushing simulations, we restrict ourselves to the minimum set of physical effects. In this controlled simulation setup, we explore the parameter space of cloud-wind interaction in an expanding background and prepare the necessary groundwork before introducing additional physical effects in the future. An additional caveat with the choice of spherical coordinates is that the cell size increases, worsening the resolution downstream. However, on comparing cold mass across simulations at different resolutions (cf. Appendix \ref{app:convergence}), we find the mass estimate agrees within a factor of a few -- enough for our simulations to provide a robust estimate of the different scaling arguments (presented later in the paper) and reliably capture the physics of cloud-wind interaction.


\subsection{Self-similarity and parameter degeneracy}
\label{subsec:degenracy}
Cloud-crushing simulations are self-similar across multiple parameters which are degenerate, i.e., for different sets of parameters with identical initial and boundary conditions, and scaled evolution equations, the solution is identical when appropriately scaled (a general feature in many hydrodynamic problems; see, for example, \citealt{1972AnRFM...4..285B, Kavvas2016ANat, Heller04052017, VINOD2024105694}). However, differences in evolution can arise as a consequence of numerical 
errors. Determining this degeneracy across parameters is useful both in saving computational time and also in providing a robust choice of `code units'. A well-known example of such a degeneracy is the similarity of an incompressible flow with the same Reynolds numbers and the same (scaled) initial and boundary conditions.

\subsubsection{Equations solved}
\label{subsubsec:equations-solved}

The set of degenerate and independent parameters is dependent on the simulated physics. 
So, we first list the equations solved in our simulations here in this subsection. Incorporating the effect of radiative cooling, the following are the hydrodynamic equations which are solved,
\begin{subequations}
\label{eq:Hydro}
\begin{align}
\label{eq:cont}
\frac{\partial\rho}{\partial t} + \nabla . \left(\rho \textbf{v}\right) &= 0,\\ 
\label{eq:mom}
\frac{\partial}{\partial t}\left(\rho \textbf{v}\right) + \nabla . \left(\rho \textbf{v} \otimes \textbf{v} + p \mathcal{I}\right) &= 0,\\
\label{eq:enrg}
\frac{\partial}{\partial t}\left(\frac{1}{2}\rho v^2 + e\right) + \nabla . \left[\left(\frac{1}{2}\rho v^2 + e + p \right)\textbf{v}\right] &= -n_H^2 \Lambda(T).
\end{align}
\end{subequations}
Here $t$ denotes time, $\mathcal{I}$ is the identity tensor and $\rho, \textbf{v}, p,$ and $e$ denotes density, velocity, pressure and internal energy density of the gas respectively. $\Lambda(T)$ that appears in the sink term in the energy equation (Eq. \ref{eq:enrg}) is the radiative cooling rate and $n_H=\rho X_H/m_H$ is the total hydrogen number density, where $X_H$ is the hydrogen mass fraction in the universe and $m_H$ is mass of the hydrogen atom. This set of equations (Eqs. \ref{eq:Hydro}) is closed by the ideal gas equation of state $e = p/(\gamma -1)$, where $\gamma$ is the ratio of specific heats with constant pressure and that with constant volume. 

\subsubsection{De-dimensionalisation of cloud-crushing in uniform wind}
\label{subsubsec:dedimensionalise-vanilla}
\begin{figure}
\centering
\includesvg[width=\columnwidth]{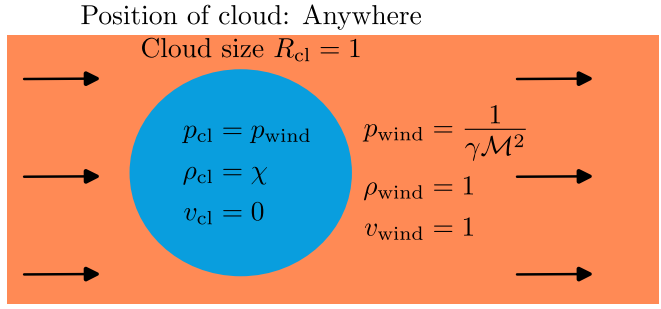}
\caption{The initialisation of a cloud-crushing simulation with the choice of code units in Eqs. \ref{eq:codeUnits}. This simulation setup has translational symmetry because the background wind is uniform. Therefore, the cloud can, in principle, be placed anywhere. However, we choose to place it a few cloud radii from the left edge of the simulation domain as the wind blows from left to right. This ensures that the initial bow shock and the cloud tail are reliably captured within the simulation domain.
}
\label{fig:iniSetup}
\end{figure}

In this section, we outline the degenerate parameters in the cloud-crushing problem. We can de-dimensionalise Eqs. \ref{eq:Hydro} in units of cloud size $R_{\rm cl}$, cloud density $\rho_{\rm cl}$, and wind velocity $v_{\rm wind}$ (which can be conveniently expressed in terms of the wind Mach number $\mathcal{M}$). Therefore, the following are the choices of units for de-dimensionalising Eqs. \ref{eq:Hydro},
\begin{subequations}
\label{eq:codeUnits}
\begin{align}
\label{eq:Lcode}
L_0 &= R_{\rm cl},\\ 
\label{eq:vcode}
v_0 = v_{\rm wind} &= \mathcal{M} \sqrt{\gamma \frac{k_B \chi T_{\rm cl}}{\mu m_p}},\\
\label{eq:rhocode}
\rho_0 = \rho_{\rm wind} &= \frac{p_{\rm wind}/k_B}{\chi T_{\rm cl}}\mu m_p, 
\end{align}
\end{subequations}
where $T_{\rm cl}$ is cloud temperature, $p_{\rm wind}$ is the wind pressure, $\chi$ is the density contrast between the cloud and the wind (such that $\chi = \rho_{\rm cl}/\rho_{\rm wind}$), $\gamma$ is the adiabatic index, and $\mu$ is the mean particle mass (in units of proton mass). The constants have their usual meaning. 
Using Eqs. \ref{eq:codeUnits}, we de-dimensionalise the hydrodynamic equations into the following (using tildes to represent de-dimensionalised quantities),
\begin{subequations}
\label{eq:HydroDD}
\begin{align}
\label{eq:contDD}
\frac{\partial \Tilde{\rho}}{\partial \Tilde{t}} + \Tilde{\nabla} . \left(\Tilde{\rho} \Tilde{\textbf{v}}\right) &= 0,\\ 
\label{eq:momDD}
\frac{\partial}{ \partial \Tilde{t}}\left(\Tilde{\rho} \Tilde{\textbf{v}}\right) + \Tilde{\nabla} . \left(\Tilde{\rho} \Tilde{\textbf{v}} \otimes \Tilde{\textbf{v}} + \Tilde{p} \mathcal{I}\right) &= 0,\\
\label{ew:enrgDD}
\frac{\partial}{\partial \Tilde{t}}\left(\frac{1}{2}\Tilde{\rho} \Tilde{v}^2 + \Tilde{e}\right) + \Tilde{\nabla} . \left[\left(\frac{1}{2}\Tilde{\rho} \Tilde{v}^2 + \Tilde{e} + \Tilde{p} \right)\Tilde{\textbf{v}}\right] &= - \zeta \Tilde{n}_H^2 \Tilde{\Lambda}(\Tilde{T}).
\end{align}
\end{subequations}
Here $\zeta = \frac{X_H \mu}{\gamma (\gamma -1)}\mathcal{M}^{-2} \chi^{-3/2} \left(\frac{t_{\rm cool, mix}}{t_{\rm cc}}\right)^{-1}\left[\frac{\Lambda \left(T_0\right)}{\Lambda \left(\sqrt{\chi} T_{\rm cl}\right)}\right]$ and $T_0 = v_0^2 \mu m_p/k_B = \mathcal{M}^2 \gamma \chi T_{\rm cl}$, where $X_H$ is the total hydrogen mass fraction, and $\mu$ is particle mass (in units of proton mass).
With $L_0$, $v_0$ and $\rho_0$ in Eqs. \ref{eq:codeUnits} as the choice of code units, the isobaric initial pressure $\Tilde{p}$ in code units is 
\begin{equation}
    \label{eq:P_code}
    \frac{p_{\rm cl}}{\rho_0 v_0^2}=\frac{p_{\rm wind}}{\rho_0 v_0^2}=\frac{1}{\gamma \mathcal{M}^2}, 
\end{equation}
where the pressures in the cloud $p_{\rm cl}$ and in the wind $p_{\rm wind}$ are initially equal because of the isobaric condition. 

%
The initialisation of cloud-crushing simulations with the choice of code units in Eqs. \ref{eq:codeUnits} is illustrated in Fig. \ref{fig:iniSetup}. The parameters that arise in the initialisation, boundary condition, and the evolution equations are the only relevant parameters in the cloud-crushing problem. The rest of the parameters, on the other hand, that do not appear in the initial or boundary conditions, as well as in the evolution equations, are degenerate. The relevant parameters are, therefore, $\chi$, $\mathcal{M}$, $t_{\rm cool, mix}/t_{\rm cc} \propto R_{\rm thres}/R_{\rm cl}$, and $T_{\rm cl}$.\footnote{The simulation parameters also depend implicitly on the cooling function $\Lambda (T)$ which goes in $\zeta$ defined earlier.} As long as these parameters are fixed, the initial and boundary conditions, and the evolution (appropriately scaled) remain unchanged. Therefore, cloud-crushing simulations are degenerate in the choice of cloud density (hence the value of initial pressure). 
For simulations without radiative cooling ($t_{\rm cool,mix}/t_{\rm cc} \rightarrow \infty$), the set of free parameters reduces to only the density contrast $\chi$ and the Mach number of the wind $\mathcal{M}$. It becomes degenerate for arbitrary choice of both the initial pressure $p_{\rm wind}$ and the initial size of the cloud $R_{\rm cl}$ (see also \citealt{Scannapieco2015ApJ}). 

\subsubsection{De-dimensionalisation of cloud-crushing in CC85 wind}
\label{subsubsec:dedimensionalise-cc85}

\begin{figure}
    \centering
    \includesvg[width=\columnwidth]{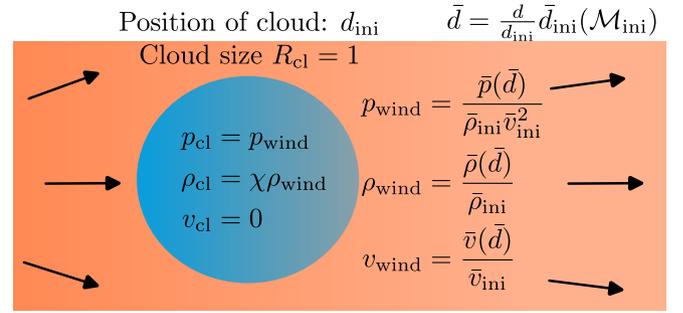}
    \caption{The initialisation of a cloud-crushing simulation with the choice of code units in Eqs. \ref{eq:codeUnitsCC85}. Unlike `vanilla' cloud-crushing, here the seed cloud is embedded in a \citetalias{Chevalier1985} wind. This introduces a background structure to the wind and breaks the translational symmetry of the initial condition of the `vanilla' cloud-crushing setup. In addition, the wind is directed radially outward instead of being plane-parallel. Refer to Section \ref{subsubsec:dedimensionalise-cc85} for an explanation of each of the terms in this cartoon.}
    \label{fig:iniSetupCC85}
\end{figure}

We extend this de-dimensionalisation technique to simulations where clouds are seeded in expanding outflows, 
specifically the steadily expanding adiabatic outflows modelled by \citealt{Chevalier1985} (\citetalias{Chevalier1985}). The main difference between these setups and `vanilla' cloud-crushing is that the background wind conditions have a spatial gradient. Additionally, instead of being plane parallel, the wind is blown radially in our simulation. 
In this case, the most convenient choice of code units for de-dimensionalisation is the wind condition at the initial position of the centre of the cloud. In the de-dimensionalised form, the hydrodynamic equations remain the same as Eqs. \ref{eq:HydroDD}.
The choice of variables for de-dimensionalisation is as follows 
\begin{subequations}
\label{eq:codeUnitsCC85}
\begin{align}
\label{eq:LcodeCC85}
L_0 &= R_{\rm cl},\\ 
\label{eq:vcodeCC85}
v_0 = v_{\rm wind,ini} &= \overbar{v}_{\rm ini} \dot{M}^{-1/2} \dot{E}^{1/2},\\
\label{eq:rhocodeCC85}
\rho_0 = \rho_{\rm wind,ini} &= \overbar{\rho}_{\rm ini}  \dot{M}^{3/2} \dot{E}^{-1/2} d_{\rm inj}^{-2}. 
\end{align}
\end{subequations}
Here, $\overbar{v}_{\rm ini}$ and $\overbar{\rho}_{\rm ini}$ are dimensionless velocity and density solutions respectively of the \citetalias{Chevalier1985} profile at the initial position of the centre of the cloud\footnote{The \citetalias{Chevalier1985} density, pressure, and velocity profiles are de-dimensionalised using $\dot{M}^{3/2} \dot{E}^{-1/2} d_{\rm inj}^{-2}$, $\dot{M}^{1/2} \dot{E}^{1/2} d_{\rm inj}^{-2}$, and $\dot{M}^{-1/2} \dot{E}^{1/2}$ respectively. They are denoted by overbars. With this normalisation, all the \citetalias{Chevalier1985} profiles exactly trace 
each other, with distance in units of $d_{\rm inj}$.} $d_{\rm ini}$. The steady 
injection rates of energy and mass are $\dot{E}$ and $\dot{M}$ respectively within an injection region with radius $d_{\rm inj}$ (sometimes also called the sonic radius as the Mach number of the wind there is unity in a \citetalias{Chevalier1985} wind). The distance from the centre of the wind in the dimensionless \citetalias{Chevalier1985} profile $\overbar{d} = d/d_{\rm inj}$ is expressed in terms of the injection radius $d_{\rm inj}$, such that $\overbar{d}_{\rm ini} = d_{\rm ini}/d_{\rm inj}$. 
In cloud-crushing simulations, the Mach number of the wind at the initial position of the centre of the cloud $\mathcal{M}_{\rm ini}$ is a free parameter. On inverting the \citetalias{Chevalier1985} profile using $\mathcal{M}_{\rm ini}$, we get the dimensionless initial position of the cloud $\overbar{d}_{\rm ini}$. It is uniquely specified by $\mathcal{M}_{\rm ini}$ as the Mach number profile of the wind $\mathcal{M}(\overbar{d})$ in the \citetalias{Chevalier1985} solution is an invertible function, i.e., $d_{\rm ini}/d_{\rm inj} = \overbar{d}_{\rm  ini} = \mathcal{M}^{-1}(\mathcal{M}_{\rm ini})$, where $\mathcal{M}^{-1}(\mathcal{M}_{\rm ini})$ is equivalently expressed as $\overbar{d}_{\rm ini}(\mathcal{M}_{\rm ini})$ in Fig. \ref{fig:iniSetupCC85}. Further, utilising the value of $\overbar{d}_{\rm ini}$ (obtained by inverting the Mach number profile), we determine the dimensionless \citetalias{Chevalier1985} wind properties $\overbar{\rho}_{\rm ini}$, $\overbar{v}_{\rm ini}$ and $\overbar{p}_{\rm ini}$ at the initial position of the cloud required for setting up the conversion between physical units and code units (see Eqs. \ref{eq:codeUnitsCC85}). The distance of an arbitrary point in the simulation domain with respect to the injection radius (needed to set the \citetalias{Chevalier1985} wind at that point) is $\overbar{d} = (d/d_{\rm inj})(d_{\rm ini}/d_{\rm inj}) = (d/d_{\rm ini}) \overbar{d}_{\rm ini}$, where $(d/d_{\rm ini})$ is the distance of the arbitrary point with respect to the initial position of the cloud in any unit (code/physical) and $\overbar{d}_{\rm ini}$ is actually a function $\overbar{d}_{\rm ini}(\mathcal{M}_{\rm ini})$ obtained from $\mathcal{M}_{\rm ini}$ as described earlier. We use this distance $\overbar{d}$ to populate the background wind condition in the simulation domain using the \citetalias{Chevalier1985} profile. The initial condition in the code units is illustrated in Fig. \ref{fig:iniSetupCC85}.

The set of independent free parameters for the cloud-crushing simulations in a \citetalias{Chevalier1985} wind remains the same as those in `vanilla' cloud-crushing, i.e., $\chi$, $\mathcal{M}_{\rm ini}$, $t_{\rm cool, mix}/t_{\rm cc}|_{\rm ini}$ and $T_{\rm cl}$ except for an additional parameter $d_{\rm ini}/R_{\rm cl}$. This parameter $d_{\rm ini}/R_{\rm cl}$ comes from the initialisation of the \citetalias{Chevalier1985} wind using the dimensionless position of every fluid element in the simulation domain $\overbar{d} = [(d/R_{\rm cl})/(d_{\rm ini}/R_{\rm cl}) ] \overbar{d}_{\rm ini}$, where $(d/R_{\rm cl})$ and $(d_{\rm ini}/R_{\rm cl})$ are positions in code units of any point in the simulation domain and the initial centre of the cloud, respectively. This wind profile, set by the free parameter $(d_{\rm ini}/R_{\rm cl})$, breaks the translation symmetry of `vanilla' cloud-crushing simulations and introduces the spatially changing background. 
Naturally, more degenerate quantities arise here -- $\dot{E}$, $\dot{M}$, $d_{\rm inj}$, and $R_{\rm cl}$, which are all absorbed in the conversion factors between the physical and code units (see Eqs. \ref{eq:codeUnitsCC85}). Specific to our parameter sets (cf. Table \ref{tab:CCinCC85_params}), the implicit relation between the free parameters of cloud-crushing simulations and the corresponding degenerate parameters are discussed in the next sub-section. 

\subsection{Initial \& boundary conditions}
\label{subsec:ini+b_conditions}
Our initial condition is a stationary, dense, spherical cloud of radius $R_{\rm cl}$ at a temperature of $T_{\rm cl} = 4 \times 10^{4}\ \rm K$ in the simulation domain. The cloud is initialised to be locally isobaric with the background wind. 
We fix the initial cloud position and wind properties (to either Set 1 or Set 2 as listed in Table \ref{tab:CCinCC85_params}) and change the cloud size by varying $t_{\rm cool, mix}/t_{\rm cc}|_{\rm ini}$ to systematically explore the growth/destruction of clouds in an expanding background. Because we are fixing the wind properties and the initial position of the cloud (hence $\mathcal{M}_{\rm ini}$), the two independent parameters in these cloud-crushing simulations, namely $d_{\rm ini}/R_{\rm cl}$ and $t_{\rm cool, mix}/t_{\rm cc}|_{\rm ini}$, no longer remain independent (cf. Section \ref{subsubsec:dedimensionalise-cc85}). We adopt the mass and energy injection rates, as well as the injection radius, from the values listed in Table \ref{tab:CCinCC85_params}. Since these parameters are degenerate, multiple combinations are possible. To enable a meaningful comparison with previous cloud-crushing simulations, we select values such that the wind temperature at the initial position of the cloud is $\chi T_{\rm cl}$. Using this constraint, we invert the dimensionless \citetalias{Chevalier1985} profile to obtain $\overbar{d}_{\rm ini} = d_{\rm ini}/d_{\rm inj}$. Based on this setup, we then compute consistent values for the Mach number at the initial centre of the cloud $\mathcal{M}_{\rm ini}$ and the ratio $d_{\rm ini}/R_{\rm cl}$, where $R_{\rm cl}$ is determined by the chosen value of $t_{\rm cool,mix}/t_{\rm cc}|_{\rm ini}$ (also listed in Table \ref{tab:CCinCC85_params}). However, it is worth noting that our results are broadly applicable and other combination of wind properties (degenerate parameters) can also produce the same set of $(\chi, \mathcal{M}_{\rm ini}, T_{\rm cl}, t_{\rm cool, mix}/t_{\rm cc}|_{\rm ini}, d_{\rm ini}/R_{\rm cl})$ which will have identical cloud-wind interaction.

Therefore, by construction, the density of the cloud is initialised to a contrast $\chi$ with the background wind at the position of the centre of the cloud. The background condition outside the cloud is determined by the steady spherically symmetric solution of a \citetalias{Chevalier1985} wind (\citealt{Chevalier1985}). Consequently, the cloud density has a slight variation from its centre to the edge approximately by a factor of $2(\gamma -1)/(d_{\rm ini}/R_{\rm cl})$, which corresponds to at most $5\%$ for the largest cloud in our parameter space (lower for smaller clouds).  
In addition, we have ensured that our computational domain always excludes the region within the injection radius (sonic point) of the \citetalias{Chevalier1985} wind. This simplifies our numerical scheme, as the fluid equations beyond the sonic point do not require the inclusion of any mass, momentum, or energy injection sources. We were able to maintain a stable and steady background wind profile in the entire simulation domain consistent with the \citetalias{Chevalier1985} solution by fixing the boundary condition to any particular \citetalias{Chevalier1985} wind profile. 

\subsubsection{Simulation parameters}
\label{subsubsec:sim_params_num}
The simulation parameters chosen in our study are listed in Table \ref{tab:CCinCC85_params}. The wind properties are chosen close to the values reported by \citealt{Chevalier1985}, appropriate for the outflow in the M82 galaxy. The cloud properties and the initial Mach number of the wind are chosen to be similar to the previous cloud-crushing simulations like \citealt{gronke2018, gronke2020, Kanjilal2021}. Although our results are broadly applicable, our choice of free (and degenerate) parameters enables meaningful comparisons with previous studies and helps isolate the effects introduced by incorporating the expanding background of the outflow in our simulations. Discussions related to convergence and resolution dependence of our simulations are presented in Appendix \ref{app:convergence}. 

\subsection{Cloud identification}
\label{subsec:cloud_def}
Previous uniform cloud-crushing simulations identify clouds as a collection of gas cells that are cold and dense. In these simulations, cold gas is found to be nearly isobaric with the hot wind and this pressure remains unchanged with time. Therefore, a denser cell is also cooler by the same factor. This is not the case in a \citetalias{Chevalier1985} wind, where the background pressure decreases with distance. Thus, as the cloud moves downstream in the wind, it remains at a temperature close to the temperature floor, but its density reduces because it is isobaric with background pressure that reduces downstream. At some point, the density contrast relative to the background can become smaller than a chosen density contrast. Considering this, we adopt two thresholds (that needs to be simultaneously satisfied) for cloud identification: \textit{(i)} cold grid cells with temperature $< 2 T_{\rm cl} = 8 \times 10^4$ K, and \textit{(ii)} locally over-dense, i.e., $\rho > 10 \rho_{\rm wind}(d)$, where $\rho_{\rm wind}$ is the analytic solution of the \citetalias{Chevalier1985} density profile at a radial distance $d$ from the wind centre of a cell in the simulation domain. The criterion \textit{(ii)} excludes cold gas that is not sufficiently dense relative to the background. Discussed elaborately in Section \ref{subsec:cloud-prop}, we find that the cold+dense `contrasting' cloud mass decreases at late times for all our simulation parameters, since a locally isobaric cloud continuously reduces its density relative to the background wind (cf. Fig. \ref{fig:cloud_profile}). 

A key caveat is that our analysis is limited to a subset of the simulation domain where the background wind temperature, based on the \citetalias{Chevalier1985} profile, remains above  $8 \times 10^4 \ \rm K$. To ensure consistency, we define a sub-domain with  $T_{\rm wind} \ge 9 \times 10^4 \ \rm K$ for all cloud identification based on this temperature threshold. Beyond this region, the wind becomes both cold and diffuse due to adiabatic expansion, making it difficult to distinguish between cold gas originating from radiative cooling in turbulent mixing layers (triggered by cloud-crushing) and that formed via adiabatic expansion of the wind. 

\subsection{Cloud tracking scheme}
\begin{figure*}
 \centering
 \includegraphics[width=\textwidth]{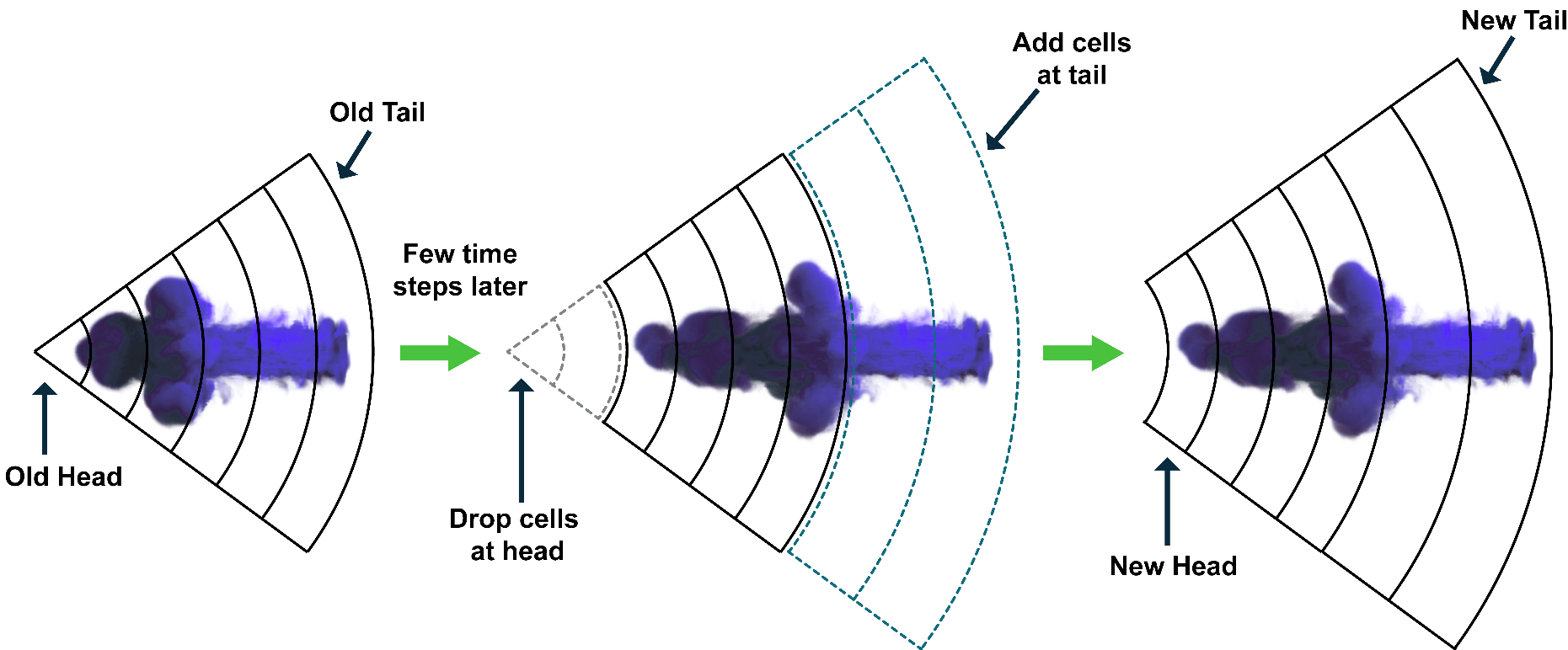}
    \caption{Illustration of our novel {\it{`cloud-tracking'}} algorithm. This algorithm is generic and can take into account a spatially varying background profile in any coordinate system. However, particular to our work, we use spherical coordinates which matches with the natural geometry of a \citetalias{Chevalier1985} wind. We track the centre of mass of the 
    cloud material, marked by a passive tracer. Whenever this moves by more than $1$ cell length and the leftmost position of the tracer is more than $16\ R_{\rm cl}$ from the leftmost edge of the domain, we `effectively' drop the cells from the left ({\it head}), introduce an identical number of cells to the right ({\it tail}) and fill them with the \citetalias{Chevalier1985} solution at their respective position. This procedure dynamically clips the simulation domain from the left (wind entry) and extends it to the right (wind exit) ensuring that a relatively small computational domain can follow a moving cloud and prevents its tail to quickly leak out of the simulation box.}
    \label{fig:track_algo}
\end{figure*}

Cloud-crushing setups generally result in an advection of the initial seed cloud along the direction of the wind. This usually causes a part of the cloud or the entirety of it to move out of the simulation domain quickly. Large simulation domains can solve this issue at the cost of computations becoming prohibitively expensive. 
In order to overcome this, cloud-crushing simulations generally employ a continuous Galilean frame transformation, which loosely follows the centre of mass of the cloud material as described in \citealt{Kanjilal2021, gronke2018, Dutta_2019} and \citealt{Shin2008ApJ}.\footnote{This procedure can also be simulated with equivalent source terms in the Euler equations and can, in principle, be extended with a time-dependent boundary condition to take into account a background having spatial variations as described and implemented by \citealt{Tan2023} in Cartesian coordinates.} 
The frame transformation procedure used in previous works changes the absolute values of the coordinate positions in the simulation domain. In Cartesian coordinates, after any frame transformation, the geometry and sizes of the grid cells, as well as the underlying fluid equations, remain unchanged, as they are independent of the absolute values of the grid positions. However, for the case of spherical coordinates (used in this work), an additional complication is that the cell size and the geometry dependent source terms in the fluid equations depend on the absolute values of the coordinate positions. Therefore, a Galilean frame transformation scheme cannot simply be implemented without a continuous change in the underlying grid. This brings additional complications to the numerical setup. On the other hand, the choice of spherical coordinates is particularly convenient to study outflows, as the grid aligns with the geometry of the background field and does not introduce numerical artifacts that cause deviation from the steady-state \citetalias{Chevalier1985} solution that we encountered on trying to initialise a \citetalias{Chevalier1985} wind in Cartesian boxes. 


Bearing these in mind, we have developed a new algorithm of \textit{`cloud tracking'} that can be applied to outflows or any other generic wind-tunnel simulations with spatially varying wind conditions in any coordinate system and does not involve continuously changing the underlying grid. In the remainder of this subsection, we describe the implementation of this cloud tracking scheme. 
A tracer field (which passively advects with the flow) tags the `initial' cloud material in our numerical setup. 
At each time step in our simulation, we calculate the leftmost (innermost/minimum) radial extent of the passive tracer by considering cells having tracer amounts above a threshold of $10^{-4}$. Our tracking scheme tries to keep this position of the minimum radial extent of the passive tracer within a fixed distance from the leftmost (innermost) radial edge of the simulation domain. This tracking procedure, therefore, switches on or off depending on the distance between the innermost radial extent of the passive tracer and the left edge of the simulation domain, which we fix to $16\ R_{\rm cl}$. 

When tracking is active, we calculate the radial distance moved by the centre of mass of the tracer material. Whenever this distance exceeds a pre-set integer number of cells in the radial direction (we term this as the `shifting threshold' and set it to 1), we shift the fluid profile by the same number of cells radially inward. This is achieved by eliminating cells at the innermost edge and introducing identical numbers of cells down the wind at the outermost edge of the simulation domain. These newly introduced cells at the right edge of the simulation domain are assigned the background \citetalias{Chevalier1985} wind condition according to their distance from the centre of the wind. We also update the coordinate positions and cell sizes of these newly introduced cells (in spherical geometry). Therefore, our tracking algorithm effectively clips the simulation domain, on-the-fly, from the left (wind entry) and extends the domain to the right (wind exit) as the cloud gradually travels downstream. Although, in principle, this algorithm can work for arbitrary cell sizes, our numerical implementation of this tracking scheme assumes uniform grid spacing both along the radial and the angular (polar and azimuthal) directions for simplicity. We start with a simulation domain of size between $400-600 R_{\rm cl}$ in the radial direction, and ensure that the total tracer material marking the initial cloud never falls below $95\%$ of the initial value over our time window of analysis. This ensures that almost the entirety of the cloud from its head to its tail remains confined within the simulation domain (see the tracer evolution following the link these videos at \href{https://youtu.be/G0jngef5CCA}{base resolution} and at \href{https://youtu.be/fPqQpE5BNcw}{one level higher}). In addition, Fig. \ref{fig:cloud-length} shows that the cloud extent along the wind saturates which ensures that the tail of the cloud is well inside the simulation domain. 

This cloud tracking scheme ensures that even with a relatively small sized simulation domain, most of the cloud material remains within the domain for the maximum amount of time. In addition, it takes care of the spatially changing background wind in spherical coordinates. Our cloud tracking algorithm is schematically illustrated in Fig. \ref{fig:track_algo}. Carried by the wind blowing from left to right, the cloud moves to the right. In response, our scheme shifts the fluid fields to the left. It is worth noting that the number of ghost cells needed for this tracking scheme to work must be more than what is required by the reconstruction stencil, plus the `shifting threshold'. This ensures that the ghost cells communicated across multiple processors (for internal boundaries) contain correct values of the fluid fields required by the reconstruction stencil after any shifting by the tracking algorithm. Once the shifting is complete, one or more ghost cells at the edge will have spurious values of the fluid fields. However, as long as these cells are not utilised in the reconstruction stencil, it is safe to allow this. It is worth noting that, in contrast to the previously used cloud tracking algorithms, we never switch the frame of reference here, and the simulation is always carried out in the initial lab frame.

\begin{table*}
\centering
\caption{Parameters of cloud-crushing simulations in CC85 wind${}^{\ddagger}$}
\label{tab:CCinCC85_params}
\resizebox{\textwidth}{!}{%
\begin{tabular}{ccccc||ccccc}
\hline
\multicolumn{5}{c||}
  {\begin{tabular}[c]{@{}c@{}}\textbf{Set 1}\\ 
  $\dot{E} = 3 \times 10^{41}\ \rm erg\ s^{-1}$, \hspace{0.3em}
  $\dot{M} = 2 \ \rm M_\odot \ yr^{-1}$, \hspace{0.3em}
  $d _{\rm inj} = 200\ \rm pc$, \hspace{0.3em}
  $d_{\rm ini} = 214\ \rm pc$, \\
  $p (d_{\rm ini}) / k_B = 2.02 \times 10^6\ \rm K\ cm^{-3}$ \end{tabular}}
  & \multicolumn{5}{c}{\begin{tabular}[c]{@{}c@{}}\textbf{Set 2}\\ $\dot{E} = 10^{42} \ \rm erg \ s^{-1}$, \hspace{0.3em}
  $\dot{M} = 5\ \rm M_\odot\ yr^{-1}$, \hspace{0.3em}
  $d_{\rm inj} = 300\ \rm pc$, \hspace{0.3em}
  $d_{\rm ini} = 371\ \rm pc$, \\
  $p (d_{\rm ini})/k_B = 1.26 \times 10^6\ \rm K \ cm^{-3}$\end{tabular}}                                                                    \\ \hline
\multicolumn{1}{c|}{$\chi$}             & \multicolumn{1}{c|}{$\mathcal{M}_{\rm ini}$} & \multicolumn{1}{c|}{$T_{\rm cl}$ [K]}       & \multicolumn{1}{c|}{$t_{\rm cool,mix}/t_{\rm cc}|_{\rm ini}$} & $d_{\rm ini}/R_{\rm cl}$ & \multicolumn{1}{c|}{$\chi$}             & \multicolumn{1}{c|}{$\mathcal{M}_{\rm ini}$} & \multicolumn{1}{c|}{$T_{\rm cl}$ [K]}       & \multicolumn{1}{c|}{$t_{\rm cool,mix}/t_{\rm cc}|_{\rm ini}$} & $d_{\rm ini}/R_{\rm cl}$ \\ \hline
\multicolumn{1}{c|}{\multirow{12}{*}{}} & \multicolumn{1}{c|}{\multirow{12}{*}{}}      & \multicolumn{1}{c|}{\multirow{12}{*}{}} & \multicolumn{1}{c|}{0.08}            & 28.268   & \multicolumn{1}{l|}{\multirow{12}{*}{}} & \multicolumn{1}{c|}{\multirow{12}{*}{}}      & \multicolumn{1}{c|}{\multirow{12}{*}{}} & \multicolumn{1}{c|}{0.1}             & 28.656   \\
\multicolumn{1}{c|}{}                   & \multicolumn{1}{c|}{}                        & \multicolumn{1}{c|}{}                   & \multicolumn{1}{c|}{0.1}             & 35.335   & \multicolumn{1}{l|}{}                   & \multicolumn{1}{c|}{}                        & \multicolumn{1}{c|}{}                   & \multicolumn{1}{c|}{0.2}             & 57.313   \\
\multicolumn{1}{c|}{}                   & \multicolumn{1}{c|}{}                        & \multicolumn{1}{c|}{}                   & \multicolumn{1}{c|}{\hspace{0.6em}0.2 ${}^{\dag}$}             & 70.671   & \multicolumn{1}{c|}{}                   & \multicolumn{1}{c|}{}                        & \multicolumn{1}{c|}{}                   & \multicolumn{1}{c|}{0.3}             & 85.696   \\
\multicolumn{1}{c|}{}                   & \multicolumn{1}{c|}{}                        & \multicolumn{1}{c|}{}                   & \multicolumn{1}{c|}{0.3}             & 106.006 & \multicolumn{1}{c|}{}                   & \multicolumn{1}{c|}{}                        & \multicolumn{1}{c|}{}                   & \multicolumn{1}{c|}{0.5}             & 143.282  \\
\multicolumn{1}{c|}{}                   & \multicolumn{1}{c|}{}                        & \multicolumn{1}{c|}{}                   & \multicolumn{1}{c|}{0.5}             & 176.677  & \multicolumn{1}{c|}{}                   & \multicolumn{1}{c|}{}                        & \multicolumn{1}{c|}{}                   & \multicolumn{1}{c|}{0.8}             & 229.252  \\
\multicolumn{1}{c|}{100}                   & \multicolumn{1}{c|}{1.496}                        & \multicolumn{1}{c|}{$4.0\times 10^4$}                   & \multicolumn{1}{c|}{0.8}             & 282.684  & \multicolumn{1}{c|}{100}                   & \multicolumn{1}{c|}{1.996}                        & \multicolumn{1}{c|}{$4.0\times 10^4$}                   & \multicolumn{1}{c|}{1.0}             & 286.565  \\
\multicolumn{1}{c|}{}                   & \multicolumn{1}{c|}{}                        & \multicolumn{1}{c|}{}                   & \multicolumn{1}{c|}{1.0}             & 353.355  & \multicolumn{1}{c|}{}                   & \multicolumn{1}{c|}{}                        & \multicolumn{1}{c|}{}                   & \multicolumn{1}{c|}{1.4}             & 401.190  \\
\multicolumn{1}{c|}{}                   & \multicolumn{1}{c|}{}                        & \multicolumn{1}{c|}{}                   & \multicolumn{1}{c|}{1.4}             & 494.697  & \multicolumn{1}{c|}{}                   & \multicolumn{1}{c|}{}                        & \multicolumn{1}{l|}{}                   & \multicolumn{1}{c|}{2.5}             & 716.411  \\
\multicolumn{1}{c|}{}                   & \multicolumn{1}{c|}{}                        & \multicolumn{1}{c|}{}                   & \multicolumn{1}{c|}{2.5}             & 883.387  & \multicolumn{1}{l|}{}                   & \multicolumn{1}{c|}{}                        & \multicolumn{1}{c|}{}                   & \multicolumn{1}{c|}{5.0}             & 1432.823 \\
\multicolumn{1}{c|}{}                   & \multicolumn{1}{c|}{}                        & \multicolumn{1}{c|}{}                   & \multicolumn{1}{c|}{8.0}             & 2826.838 & \multicolumn{1}{c|}{}                   & \multicolumn{1}{c|}{}                        & \multicolumn{1}{c|}{}                   & \multicolumn{1}{c|}{6.0}             & 1719.387 \\
\multicolumn{1}{c|}{}                   & \multicolumn{1}{c|}{}                        & \multicolumn{1}{c|}{}                   & \multicolumn{1}{c|}{}                &          & \multicolumn{1}{l|}{}                   & \multicolumn{1}{c|}{}                        & \multicolumn{1}{l|}{}                   & \multicolumn{1}{c|}{7.0}             & 2005.952 \\
\multicolumn{1}{c|}{}                   & \multicolumn{1}{c|}{}                        & \multicolumn{1}{c|}{}                   & \multicolumn{1}{c|}{}                &          & \multicolumn{1}{c|}{}                   & \multicolumn{1}{c|}{}                        & \multicolumn{1}{c|}{}                   & \multicolumn{1}{c|}{8.0}             & 2292.516 \\ \hline
\end{tabular}
}
\begin{tablenotes}
    \item ${}^{\ddagger}$ {\fontsize{7pt}{7.2pt}\selectfont The mass and energy injection rates, injection radius, and pressure at the initial position of the cloud are degenerate parameters and provided to give a point of reference to the readers. Our results are broadly applicable. As long as $(\chi, \mathcal{M}_{\rm ini}, T_{\rm cl}, t_{\rm cool, mix}/t_{\rm cc}|_{\rm ini}, d_{\rm ini}/R_{\rm cl})$ is the same, the simulations, by construction, will evolve identically.}
    \item ${}^{\dag}$ {\fontsize{7pt}{7.2pt}\selectfont The fiducial parameter set} 
    \item {\fontsize{7pt}{7.2pt}\selectfont Most of the trends discussed in our paper remain the same for the two different Mach numbers at the initial positions of the cloud in different \citetalias{Chevalier1985} winds (Sets 1 \& 2).}
    \item {\fontsize{7pt}{7.2pt}\selectfont The base resolution for all the simulations is $R_{\rm cl}/d_{\rm cell}|_{\rm ini} =8$ at the initial position of the cloud.}
    \end{tablenotes}
\end{table*}

\section{Results}
\label{sec:results}

\subsection{Radiative cooling and threshold cloud size in a uniform wind}
\label{subsec:vanilla-fudge}
Using our simulations (in spherical coordinates), we carry out `vanilla' cloud-crushing at different cloud sizes to numerically evaluate the dimensionless parameter $\delta$ in Eq. \ref{eq:R_thres}. The choice $T_{\rm mix} = \sqrt{\chi} T_{\rm cl}$ is a simple approximation (\citealt{Begelman1990}) rather than a rigorous criterion. Because of this, $\delta$ would usually depend on the detailed nature of the cooling curve and the cloud/floor temperature in a simulation. Therefore, we carry out these `vanilla' cloud-crushing simulations to get an estimate of the cloud radius above which clouds grow due to strong radiative cooling in a plane-parallel wind. These simulations also serve as a set of control setups against which cloud-crushing in an expanding wind can be benchmarked. Our simulation result on `vanilla' cloud-crushing which we use to numerically evaluate $\delta$ is illustrated in Fig. \ref{fig:vanilla_turnover} in Appendix \ref{app:size_thres}. We find that $\delta\approx 6$ is where the turnover from the destruction regime to the growth regime of
clouds occurs in uniform cloud-crushing.

\subsection{Radiative cloud-crushing in an expanding background}
\label{subsec:clcrush-exp}
\begin{figure*}
	\centering
    \includegraphics[width=\textwidth]{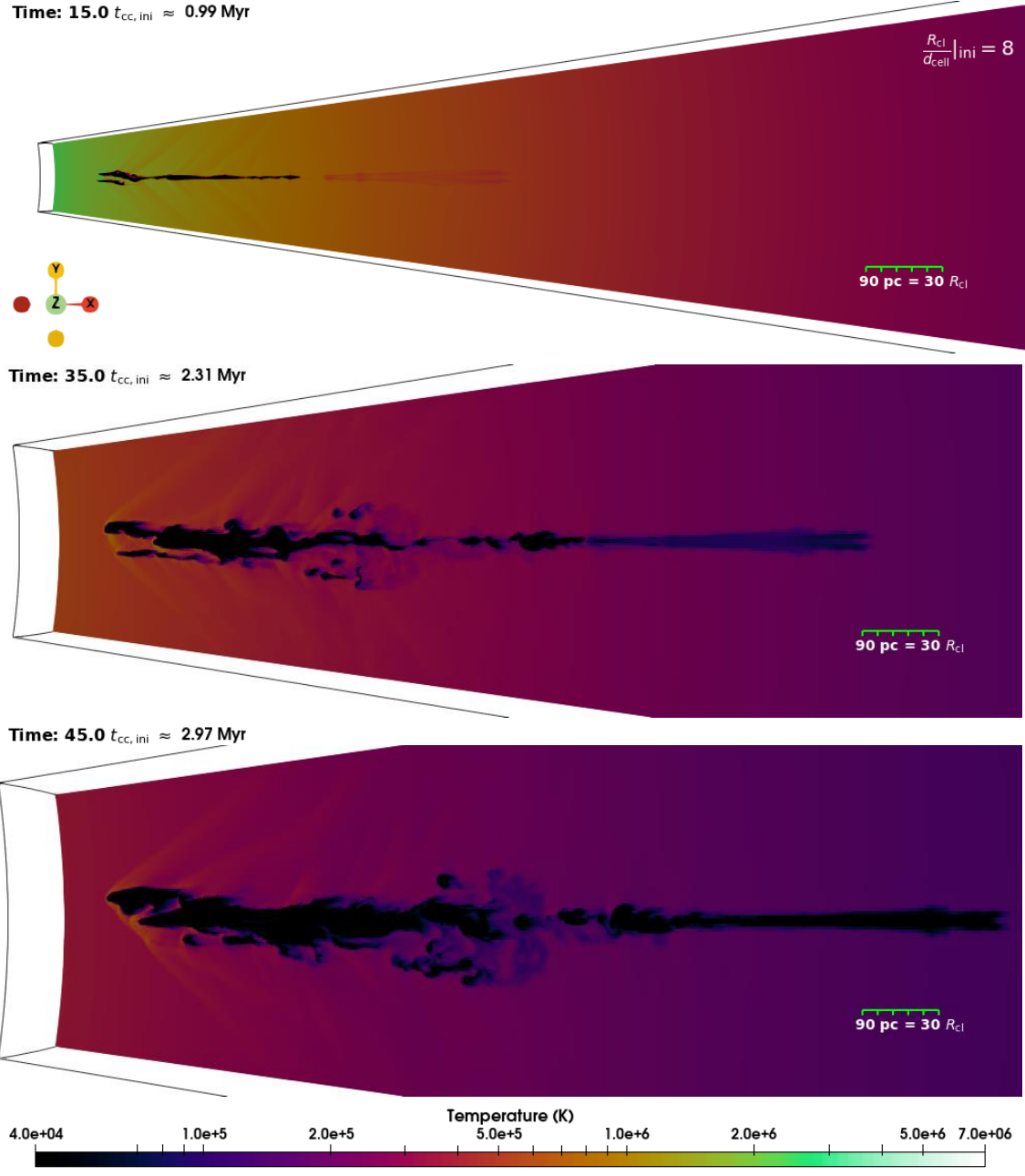}
    \caption{The temperature slices in the $z=0$ plane from our fiducial cloud-crushing simulation in a \citetalias{Chevalier1985} wind. The simulation parameters are $(\mathcal{M}, \chi, t_{\rm cool, mix}/t_{\rm cc}|_{\rm ini}) = (1.496, 100, 0.2)$; cf. Table \ref{tab:CCinCC85_params}. We show only a few selected snapshots that clearly demonstrate the formation of elongated cold structures starting from a spherical cloud (not shown here). Similar elongated structures of cold gas with cometary tails embedded in galactic winds have been recently observed in the M82 galaxy (\citealt{Bolatto2024ApJ, Fisher2024MNRAS}). Unlike in a uniform wind, the cloud encounters different wind conditions along its tail as it moves downstream. 
    The cold gas also expands orthogonal to the wind direction (see the attached scale). Follow the discussion in Section \ref{subsubsec:GO_exp-box} \& Fig. \ref{fig:cloud-spread} for a quantitative estimate on cloud expansion. Use this \href{https://youtu.be/c9lrGnPHgfI}{link here} to see a video of the full evolution (also see the evolution of the corresponding density slice following \href{https://youtu.be/SLNG99l_E6Y}{this link}). Moving beyond qualitative visualisation, in Fig. \ref{fig:cloud_profile}, we make a detailed quantitative evaluation of the evolution of clouds in a \citetalias{Chevalier1985} wind. It is worth noting that we clip the slices in our visualisation to a smaller radial extent than in the simulation for clarity. 
    }
    \label{fig:temperature_slice}
\end{figure*}

\begin{figure*}
    \centering
    \includegraphics[width=\textwidth]{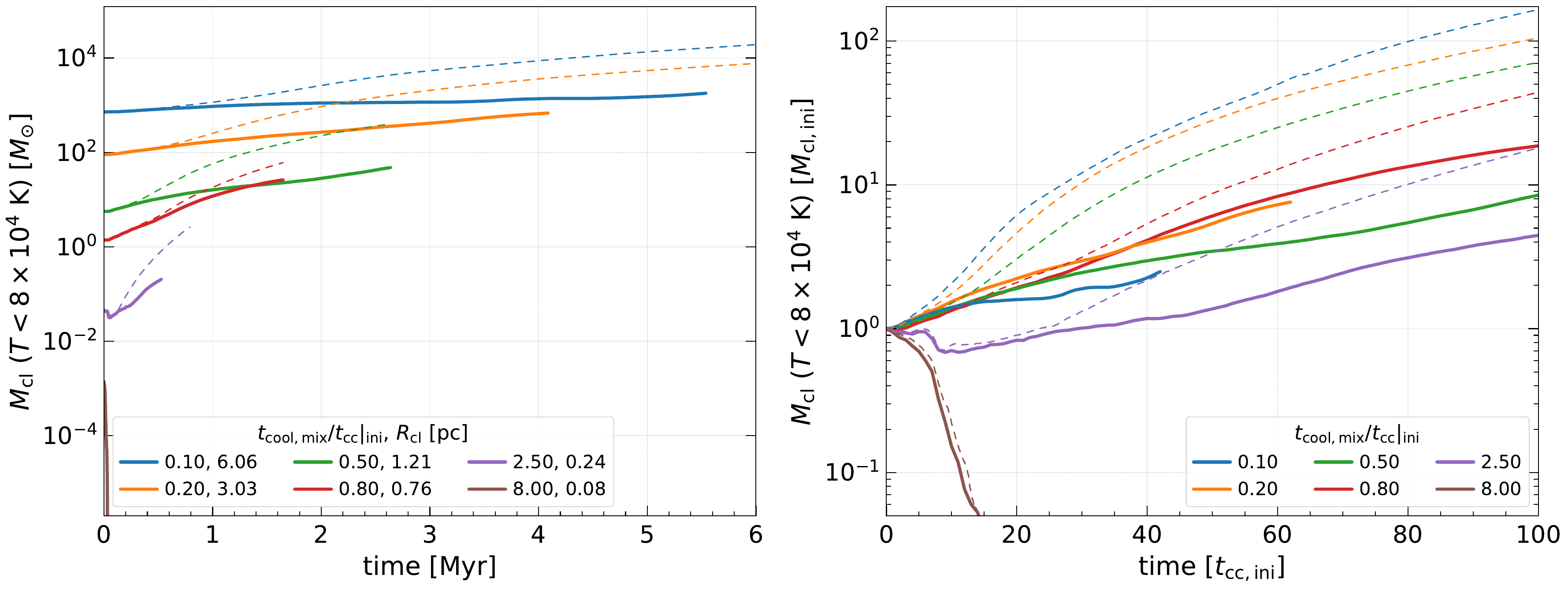}
    \caption{The temporal evolution of cold gas mass (in units of solar mass in the \textbf{left} panel and initial cloud mass in the \textbf{right} panel) in our cloud-crushing simulations for clouds of different sizes initialised in a \citetalias{Chevalier1985} wind (\textbf{solid lines}) compared with identical simulations of cloud-crushing in a uniform wind (`vanilla' cloud-crushing in \textbf{dashed lines of the same colour} as their solid counterparts). We find that the growth of cold mass is significantly suppressed in an expanding background compared to `vanilla' cloud-crushing. The background wind properties and the initial position of the cloud with respect to the wind remain the same in all these simulations. In this analysis, a cell is identified as being part of the cold cloud if they have a temperature less than $2 T_{\rm cl} = 8 \times 10^4\ {\rm K}$. The set of constant parameters chosen for these simulations is $(\mathcal{M}, \chi) = (1.496, 100)$ (cf. Table \ref{tab:CCinCC85_params}). Each line of different colour corresponds to different $t_{\rm cool, mix}/t_{\rm cc}|_{\rm ini} \propto R_{\rm thres}/R_{\rm cl}$, which translates to different cloud sizes. The time is in Myr in the \textbf{left} panel and normalised by the cloud-crushing time $t_{\rm cc, ini}=\sqrt{\chi} R_{\rm cl}/v_{\rm wind}$  evaluated at the initial position of the cloud (the \textbf{right} panel). The region of the simulation domain chosen for our analysis has wind temperatures greater than $9 \times 10^4 \ {\rm K}$, higher than the temperature below which we consider gas to be cold in our analysis. For the largest clouds ($t_{\rm cool, mix}/t_{\rm cc}|_{\rm ini} = 0.1, 0.2$), we truncate the evolution early after a significant portion of the initial tracer marking the cloud starts leaving our analysis domain. The largest clouds exhibit the greatest deviation in their cold mass content from their `vanilla' cloud-crushing counterparts. The smallest cloud from our simulation shown in this figure ($t_{\rm cool, mix}/t_{\rm cc}|_{\rm ini} = 8$) gets destroyed in both the `vanilla' cloud-crushing setup and cloud-crushing in a \citetalias{Chevalier1985} wind. It is relatively unaffected by an expanding wind and experiences a similar evolution in its cold mass content.}
    \label{fig:cc85_cold-mass}
\end{figure*}

\begin{figure}
    \centering
    \includegraphics[width=\columnwidth]{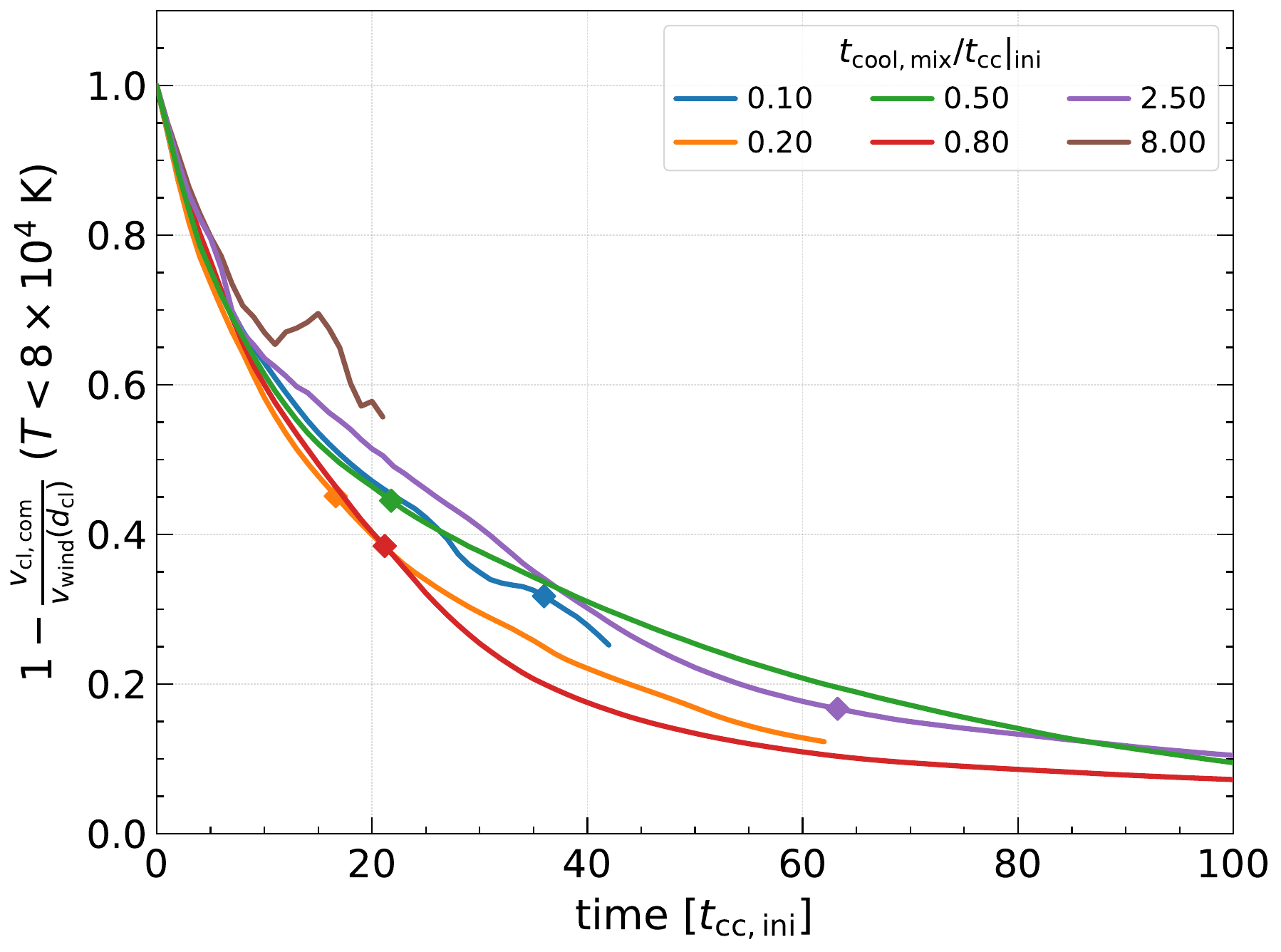}
    \caption{The temporal evolution of the centre of mass velocity of cold gas relative to the wind velocity at the position of the centre of mass of the cloud in the wind (in units of the wind velocity) in our cloud-crushing simulations for clouds of different sizes initialised in a \citetalias{Chevalier1985} wind. In this analysis, a cell is identified as being part of the cold cloud using the same definition as in Fig. \ref{fig:cc85_cold-mass} and the simulation sub-domain used in the analysis is also the same as in Fig. \ref{fig:cc85_cold-mass}. By about $30\ t_{\rm cc, ini}$, most of the clouds have have doubled in their mass content as indicated by the \textbf{diamond} marks along each of the curves and they have reached $\sim 70-80$ percent of the local wind velocity indicating that they have got sufficiently entrained in the wind. The evolution of clouds pre- and post-entrainment is significantly different as discussed later in Section \ref{subsubsec:GO_exp-box}.}
    \label{fig:cc85_cold-vel}
\end{figure}

We select clouds that are found to be growing due to radiative cooling in a uniform wind (`vanilla' cloud-crushing) and initialise them in an expanding \citetalias{Chevalier1985} wind. 
For our analysis, we consider only the sub-domain where the background wind temperature is greater than $9\times 10^4\ \rm K$ and classify all gas cells in this region with temperatures less than $2 T_{\rm cl} = 8 \times 10^4 \ \rm K$ as cold cloud material (cf. Section \ref{subsec:cloud_def}). This sub-domain is also used in the visualisation in Fig. \ref{fig:temperature_slice}, where we show the evolution of the temperature field on a slice in the $x-y$ plane passing through the centre of our simulation domain (along the $z$-axis) using a few selected snapshots. The gradient in the background wind temperature is consistent with the adiabatic expansion of a typical \citetalias{Chevalier1985} wind. Similar to \citealt{gronke2020}, for cooling dominated clouds, we find that their evolution in the wind can be broadly distinguished into two different stages -- \textit{(i)} initial `tail-formation' stage when a large relative velocity between the cloud and the wind causes the initial spherical blob of cold gas to develop a long, sheared cometary tail in the direction of the wind followed by \textit{(ii)} an `entrained' stage when the cold gas grows due to a gradual inflow through 
radiative cooling around the cloud perhaps as a local cylindrical cooling flow (\citealt{Dutta2022MNRAS}). The scales provided with the slices in Fig. \ref{fig:temperature_slice} also indicate the gradual expansion that the cloud has undergone perpendicular to the radial direction as it has moved down the expanding wind. We find that the cold gas is nearly isothermal, and most cloud cells have a temperature approximately equal to the floor temperature in our simulation, i.e., $4 \times 10^4\ \rm K$. Despite having some similarities with clouds moving in a uniform wind, an expanding wind causes the cloud properties to be qualitatively different from those in a uniform wind, which we discuss further. 

Fig. \ref{fig:cc85_cold-mass} illustrates the evolution of the cold mass of clouds with different sizes seeded in an expanding wind, and also compares it with the evolution of cold mass for the same clouds in a uniform wind (`vanilla' cloud-crushing) with identical initial conditions.  
We find that the growth of cold clouds is suppressed in an expanding wind compared to a uniform wind. An important result from our simulations is that such a suppression of cold mass for growing clouds 
occurs across clouds of all sizes. The smallest clouds that are in the destruction regime due to shear-induced mixing in the uniform wind (and inefficient radiative cooling) exhibit a similar reduction of their cold mass content over time and are relatively unaffected by the expanding background. Unlike clouds moving in a uniform wind, these clouds moving in an adiabatically expanding wind fail to remain 
dense, with contrasting features from the background as they travel down the wind (cf. Section \ref{subsec:cloud-prop} \& \ref{subsec:diff_cloud_def}). The reduction in the density of the cold cloud as it moves in the wind plays a crucial role in suppressing the growth of the cold mass in an expanding background. As the flow expands downstream, both cloud and wind densities decrease, resulting in a lower-density mixed gas that cools 
far less efficiently than it does upstream.  When the time in our simulation is normalised by the initial cloud-crushing time $t_{\rm cc, ini}$ (which is longer for larger clouds), the largest clouds travel further downstream compared to the smaller ones within the same number of cloud-crushing times. 
In other words, the change in the background wind condition for every cloud radius moved by a cloud increases with cloud size $R_{\rm cl}$. This results in the highest suppression in mass growth and departures from `vanilla' cloud-crushing for the larger clouds (smaller $t_{\rm cool, mix}/t_{\rm cc}|_{\rm ini}$) in our simulations. 

In Fig. \ref{fig:cc85_cold-vel}, we show the evolution of the centre of mass velocity of the cloud relative to the wind velocity at their instantaneous position. We find that the cloud mass increases to a few multiples of the initial cold mass for most of the growing clouds by the time they get entrained. The cloud velocity approaches about $70-80$ percent of the wind velocity by $\sim 30\ t_{\rm cc, ini}$. Once entrained, the cloud properties and its mass growth rate get significantly altered as discussed further in Section \ref{subsubsec:GO_exp-box}. It also has important implication on empirical modelling of cloud-wind (cf. Section \ref{subsubsec:DF-single_cloud}). We further investigate all these effects with quantitative estimates and speculate upon their implications for observations. 

\label{subsec:cloud-prop}
\begin{figure*}
    \centering
    \includesvg[width=\textwidth]{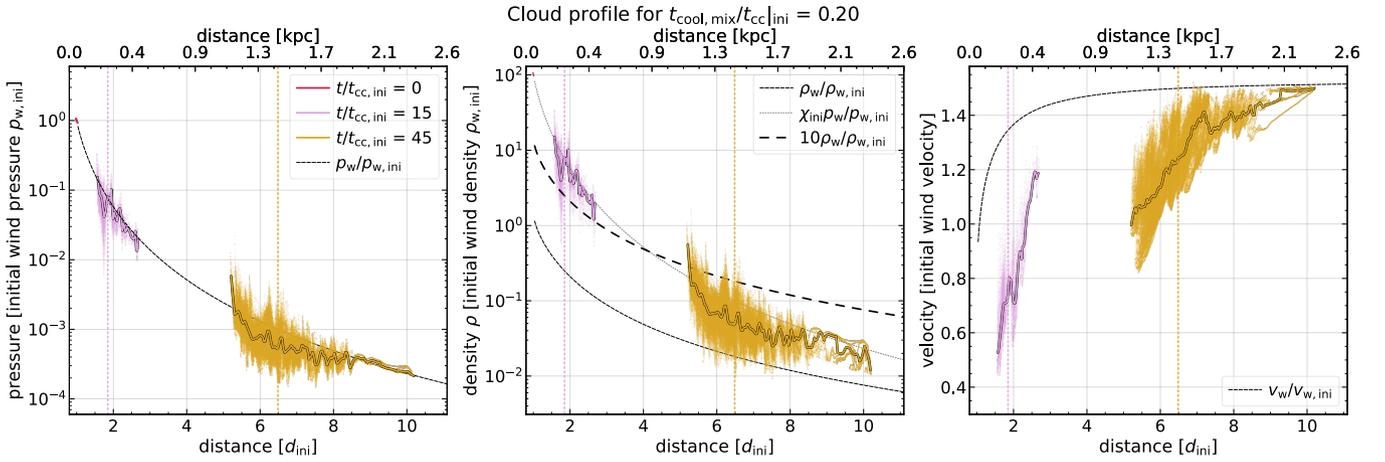}
    



    \caption{Temporal evolution of the profile of cloud ($T<2T_{\rm cl} = 8\times 10^4 \ \rm K$) pressure (\textbf{left panel}), density (\textbf{middle panel}), and velocity (\textbf{right panel}). The median cloud profiles are shown in \textbf{thin black bordered} lines, while all the cloud material is shown by dots at times 0 (\textbf{red}), 15 (\textbf{pink}), 45 (\textbf{yellow}) $t/t_{\rm cc,ini}$. The cloud profiles are normalised by the wind properties at the initial location of the cloud. Distance down the wind is expressed in multiples of the initial position of the cloud (w.r.t. the wind centre), and also in kpc (top labels). The simulation chosen for this plot corresponds to our fiducial parameters $(\mathcal{M}, \chi, t_{\rm cool, mix}/t_{\rm cc}|_{\rm ini}) = (1.496, 100, 0.2)$; cf. Table \ref{tab:CCinCC85_params}. The \textbf{black finely dashed} lines in all the panels show the background wind profile, and the \textbf{vertical} lines mark the instantaneous position of the centre of mass of the cold cloud (with the same colour as their profile). The left panel demonstrates that the cold cloud always maintains an approximate local pressure equilibrium with the wind. This implies that the cloud density is proportional to the wind pressure ($n_{\rm cl} \propto d_{\rm cl}^{-2 \gamma}$; cf. Eq. \ref{eq:cloud_density_scaling}) at any location (because the cloud temperature 
    is essentially fixed; cf. Fig. \ref{fig:cloud_pdf}) indicated by the \textbf{thin gray dotted} line in the middle panel. The cloud median density profile approximately follows this gray line. The cloud velocity profile on the right panel shows that it gradually attains the wind velocity, with its tail becoming co-moving faster than the head. In addition, the \textbf{black loosely dashed} line in the middle panel shows the profile of $10$ times the local wind density for reference as this denotes the density cut-off used in our work to identify clouds that stand out as contrasting features with respect to their background (see Section \ref{subsec:diff_cloud_def} \& Fig. \ref{fig:cloud_mass+dense} for details). Animated versions of this plot are available in the following links for the profiles of \href{https://youtu.be/r5lw4rVKWMs}{pressure}, \href{https://youtu.be/M1ixCNQ_6Qo}{density}, and \href{https://youtu.be/Ww7PMBAmDPM}{velocity}. Refer to Fig. \ref{fig:cloud_profile-high_res} for the same plot at a higher resolution.}
    \label{fig:cloud_profile}
\end{figure*}
\subsection{Properties of clouds in a typical CC85 wind}

In order to explain the suppression of cold mass growth for clouds moving in an expanding wind (see Section \ref{subsec:clcrush-exp}), we investigate the properties of the cold cloud as it moves in the \citetalias{Chevalier1985} wind. As usual, for our analysis, we consider only a sub-domain from our simulation where the wind temperature is greater than $9\times 10^4\ \rm K$, and classify all the gas cells in this region at temperatures less than $8 \times 10^4 \ \rm K$ as cold cloud material (cf. Section \ref{subsec:cloud_def}). 

In each snapshot from our simulation, starting from the minimum until the maximum radial extent of cold gas along the wind, we choose concentric radial shells in the wind of width $0.25\ R_{\rm cl}$. We calculate the median pressure (mass-weighted), density, and velocity of the cold gas ($T<8\times 10^4\ \rm K$) in each of these shells to obtain a cloud profile as a function of the downstream distance (for fiducial parameters; cf. Table \ref{tab:CCinCC85_params}). These pressure, density, and velocity profiles of the cloud are presented in Fig. \ref{fig:cloud_profile}, where we also show all the cloud cells as an underlying scatter plot to indicate the spread in the data. We use black dashed lines to show the corresponding \citetalias{Chevalier1985} wind profile for reference. Different colours are chosen to denote cloud profiles at different times in their evolution, as labelled in the accompanying legends. From these cloud profiles, we find that the cloud pressure, in the left panel of Fig. \ref{fig:cloud_profile}, closely follows the local wind pressure (except for the small portion at the head of the cloud). Therefore, the cloud evolves roughly isobarically in the wind. Note that the median pressure is slightly lower than the local wind pressure which might be responsible for the inflow of cold gas into the cloud as our fiducial parameter corresponds to a growing cloud with strong radiative cooling (see Section \ref{subsec:pdfs} \& Fig. \ref{fig:cloud_pdf} for details). 
In addition, since the cold gas itself is approximately isothermal (see Fig. \ref{fig:temperature_slice} \& Section \ref{subsec:pdfs}), we can 
approximate the cloud density as
\begin{equation}
    \label{eq:cloud_density_scaling}
    n_{\rm cl} = \frac{p_{\rm wind}/k_B}{T_{\rm cl}},
\end{equation}
which implies that $n_{\rm cl} \propto p_{\rm wind}$. Therefore, in the asymptotic limit, $n_{\rm cl} \propto (d_{\rm cl}/d_{\rm cl,ini})^{-2\gamma}$, where $d_{\rm cl}$ and $d_{\rm ini}$ are the distances of the cold gas and the initial position of the cloud respectively from the wind centre. Indeed, the cloud density profile follows our expectation as indicated by the dotted gray curves in the middle panel of Fig. \ref{fig:cloud_profile}. The most important implication is that the cloud loses its density contrast with distance. 
This is generic for growing clouds of all sizes, causing them to gradually dissolve in the background wind (cf. Section \ref{subsec:diff_cloud_def}). The cloud velocity profile in the rightmost panel in Fig. \ref{fig:cloud_profile} shows that the cloud gradually catches up with the velocity of the wind. As the wind velocity in the \citetalias{Chevalier1985} solution is not strictly a constant (but asymptotes to a constant), there is a substantial velocity difference between the head and the tail of the cloud, causing it to stretch in the direction of the wind. 

In addition to this, we find that the inflow velocity of gas due to radiative cooling feeding the cold clouds as it moves in the \citetalias{Chevalier1985} wind shows 
only a crude agreement with the theory of turbulent boundary layers introduced by \citealt{Fielding_2020, gronke2020, Tan2020} 
(see Appendix \ref{app:scaling-relations}). 
Further work is needed to test the limits of applicability 
of this theory and to come up with better models. The profiles shown in Fig. \ref{fig:cloud_profile} reveals a consistent picture even at a higher resolution (see Fig. \ref{fig:cloud_profile-high_res}) demonstrating the robustness of our scaling relations and our results on the general evolution of cold clouds in an expanding outflow. 



\subsection{Clouds as dense cold clumps embedded in a diffuse background}
\label{subsec:diff_cloud_def}
\begin{figure*}
    \includegraphics[width=\textwidth]{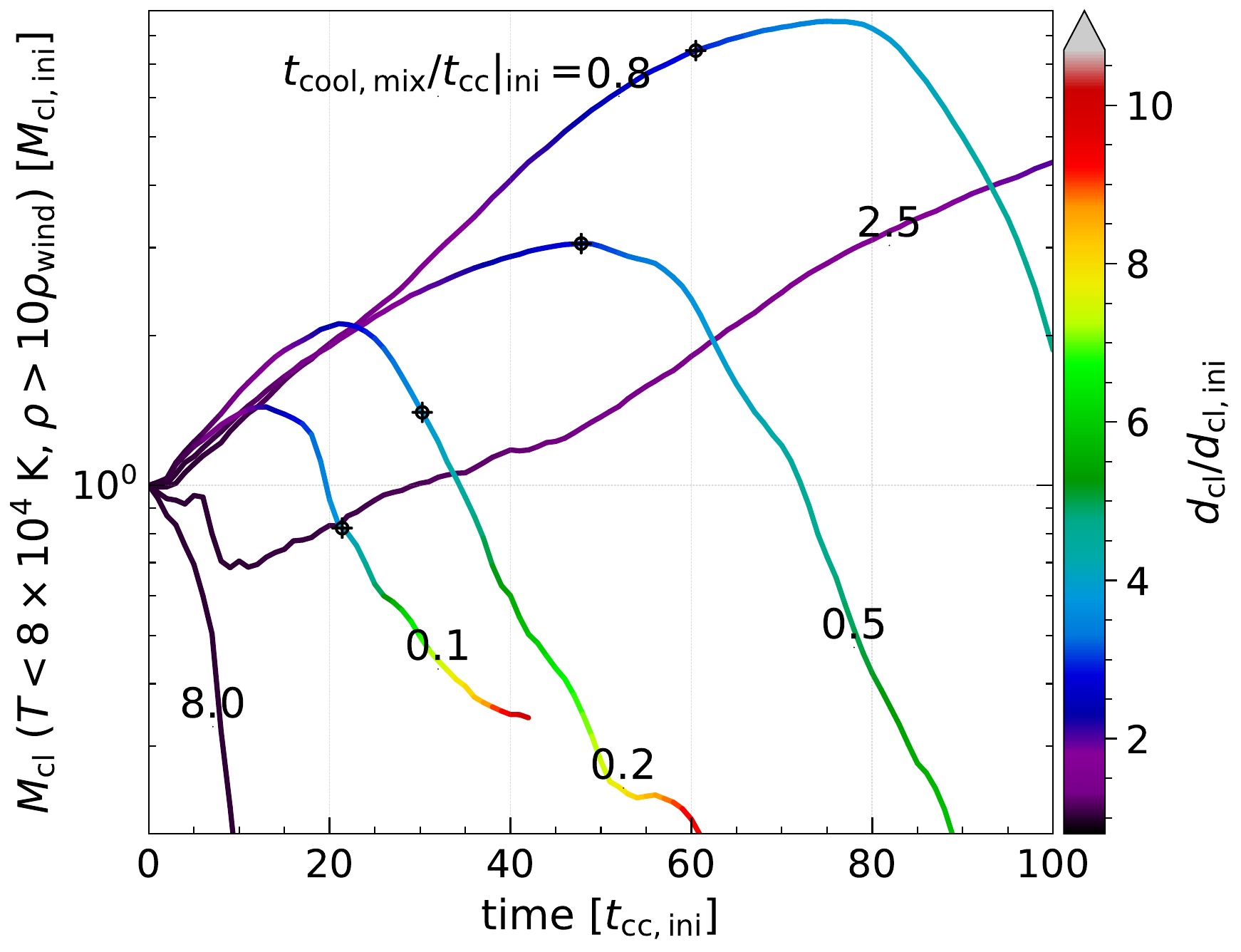}
    \caption{Temporal evolution of mass in clouds, defined as cold gas ($T<2 T_{\rm cl} = 8 \times 10^4 \ \rm K$) much denser than their local background (more than $10$ times denser than the wind at their location). The gas cells satisfying this criterion stand out as contrasting radial filaments in the wind (cf. Fig. \ref{fig:temperature_slice}). The time is normalised by the initial cloud-crushing time $t_{\rm cc,ini}$ (which is larger for larger 
    and vice versa). Each line in the plot refers to a simulation with a fixed value of $t_{\rm cool, mix}/t_{\rm cc}|_{\rm ini}$ (inversely proportional to the initial cloud size) -- \textbf{annotated} by the numbers adjacent to the curves. The lines are coloured according to the distance the cloud has travelled with respect to its initial position in the wind. Since the Mach number at the initial position $\mathcal{M} = 1.496$ of the cloud and its density contrast $\chi = 100$ are kept the same across all the simulations presented here, the initial cloud position $d_{\rm cl,ini}$ is the same.  
    The cloud density falls at a steeper rate with distance compared to the the \citetalias{Chevalier1985} wind density (cf. Section \ref{subsec:cloud-prop}). Consequently, the clouds fail to remain cold as 
    dense contrasting features in the wind as they travel downstream. Therefore, after the cloud has travelled a critical distance down the wind (approximately the same in all simulations; cf. Eq. \ref{eq:cloud_density_scaling}), the (contrasting) cloud mass transitions from an initially growing phase to a destruction phase. This transition occurs at around $d_{\rm cl}/d_{\rm cl,ini} \sim 4$, where the line colour is blue-cyan (at a time approximately predicted by Eq. \ref{eq:time_emp}; \textbf{black} crosshair) and the gray dotted and black dashed lines intersect in the middle panel of Fig. \ref{fig:cloud_profile}. Follow this \href{https://gist.github.com/dutta-alankar/ff8ccfd66c23d57f23e08721189578c3}{link} to see an alternate plot where the quantities for the x-axis and the colourbar are swapped.}
    \label{fig:cloud_mass+dense}
\end{figure*}

\begin{figure*}
    \centering
    \includegraphics[width=\textwidth]{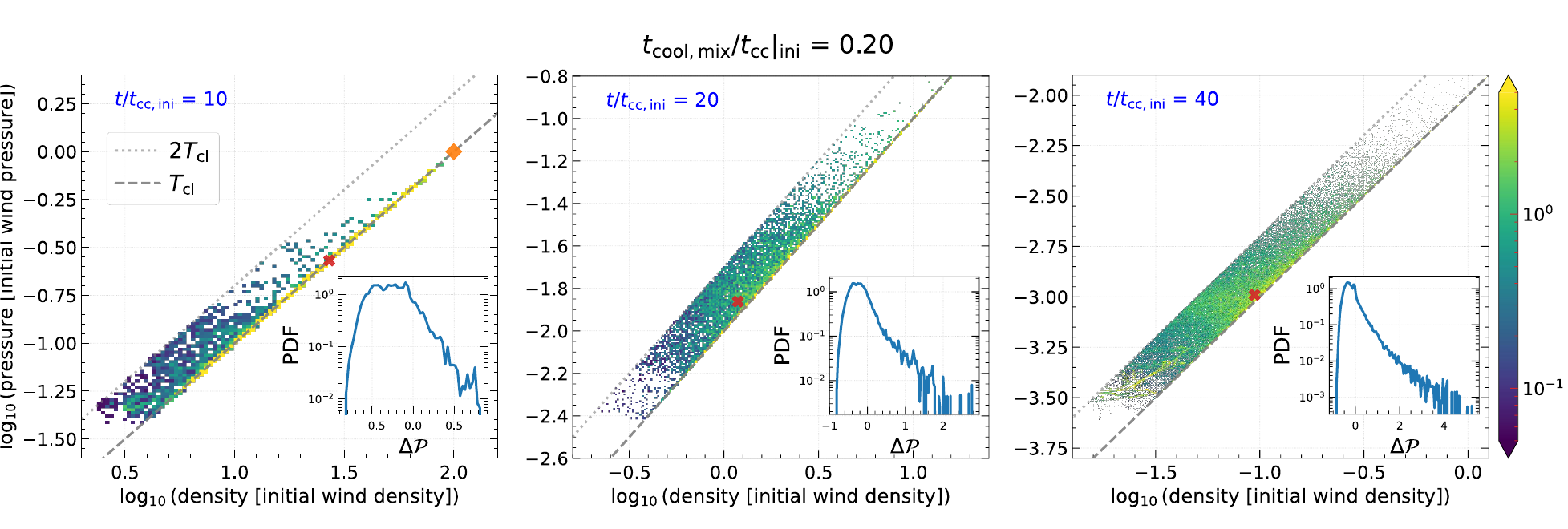}
    \caption{Temporal evolution of the phase space density-pressure distribution of the cold gas 
    ($T<2 T_{\rm cl}=8\times 10^4\ {\rm K}$) in our fiducial \citetalias{Chevalier1985} simulation. The gas density and pressure are normalised by the wind properties at the initial location of the cloud. 
    The different panels are at different times in the simulation as indicated by the \textbf{blue} text annotated in the plots. The \textbf{gray dashed} and \textbf{dotted} lines in all the panels show the isotherms of $T_{\rm cl}$ and $2 T_{\rm cl}$, respectively. The \textbf{orange} diamond in the left panel marks the position of the initial cloud material in this phase space, while the \textbf{red} cross in all the panels marks the median density and pressure of the cold gas at the instantaneous position of the centre of mass of the cloud (cf. Fig. \ref{fig:cloud_profile}). With time, the cold material becomes low in density. The red cross (centre of mass) always lies at the high probability region of the distribution and the distribution itself has a small spread about the $T=T_{\rm cl}$ isotherm due to the elongated nature of the cloud as it moves downstream. Pressure fluctuations $\Delta \mathcal{P} = (p_{\rm cl}-p_{\rm wind})/p_{\rm wind}$ in the cold gas compared to the local wind pressure are shown in the \textbf{inset} plot in each panel. In the inset figures, we plot the mass-weighted distribution of pressure fluctuation in the cold gas $1/M_0\ (dM/d\log_{10}\Delta \mathcal{P})$, revealing that the cloud is slightly under-pressurized. An animated version of this plot is available \href{https://youtu.be/ky2oNyB4G-g}{here}.}
    \label{fig:cloud_pdf}
\end{figure*}

In the context of cloud-crushing in a \citetalias{Chevalier1985} wind, the cloud encounters a 
cooler adiabatically expanding wind as it moves downstream. Consequently, the cold cloud formed by mixing and subsequent radiative cooling becomes diffuse as it moves further downstream. 
Therefore, simply identifying all the cold gas as cloud material gives an incomplete picture of the ongoing process. 
Considering this, we adopt a more restrictive definition of cloud, identifying it as cold and dense features (rather than just cold) formed via mixing and radiative cooling, relative to the featureless expanding wind (cf. Section \ref{subsec:cloud_def}).
In the remainder of this section we illustrate that in the presence of a \citetalias{Chevalier1985} wind, the clouds in the cooling regime grow in their cold mass but at a reduced rate (see Section \ref{subsec:clcrush-exp} \& Fig. \ref{fig:cc85_cold-mass}), and in addition continue to expand compared to clouds in a uniform plane-parallel wind. 
It is worth noting that the ratio of the contrast (with the background) that a cloud has between any two distances down the wind is the same for clouds of all sizes (cf. Section \ref{subsec:cloud-prop}) if the clouds start from the same initial position with the same contrast. This ensures that the qualitative picture that clouds lose contrast while moving downstream remains unaltered by the choice of threshold density contrast ($10$ here; see Section \ref{subsec:cloud_def}). 

We illustrate the evolution of the cloud mass identified using our `contrasting cloud identification' criterion (discussed in Section \ref{subsec:cloud_def}) in Fig. \ref{fig:cloud_mass+dense}. The different curves in the figure correspond to different initial cloud sizes in our simulation, labelled by $t_{\rm cool,mix}/t_{\rm cc}|_{\rm ini}$. The clouds in different runs from our simulation presented here are initially placed at the same position with respect to the wind (for same Mach number $\mathcal{M}_{\rm ini}$) with identical conditions for the background wind. The clouds are initialised with a fixed contrast of $100$ in density (or temperature) with the wind. The cloud mass evolution curves in Fig. \ref{fig:cloud_mass+dense} are coloured according to the distance travelled by the cloud (centre of mass) with respect to its initial position in the wind, as indicated by the 
colourbar. As illustrated earlier in Fig. \ref{fig:cc85_cold-mass}, although a simple identification of clouds as cold gas ($T<8\times 10^4 \ \rm K$) in a \citetalias{Chevalier1985} wind shows cloud growth (albeit at a slower rate compared to `vanilla'), the clouds fail to remain contrasting features in the wind. This trend holds for clouds of all sizes and occurs when the clouds reach a specific critical distance downstream. Since the cloud density at any instant only depends on its initial density and its distance from the initial position in the wind (see Eq. \ref{eq:cloud_density_scaling} in Section \ref{subsec:cloud-prop}), the saturation of the `contrasting' cold mass and subsequent turnover from growth to destruction/fading occur at the same distance from the injection region of the wind for all clouds presented here (as they start with the same initial contrast with the background wind). Therefore, the line colour in Fig. \ref{fig:cloud_mass+dense} (denoting the distance travelled by the cloud) at the time of 
the peak cold+dense mass is the same across simulation parameters (different cloud sizes). The larger clouds, with smaller $t_{\rm cool,mix}/t_{\rm cc}|_{\rm ini}$ and larger $t_{\rm cc, ini}$, travel farther downstream for every $t_{\rm cc, ini}$ 
and reach the turnover distance in fewer multiples of $t_{\rm cc, ini}$, as seen in Fig. \ref{fig:cloud_mass+dense} (also see the discussion near the end of Section \ref{subsec:clcrush-exp}). 

The reduction of `contrasting' cold mass constitutes a physically relevant as well as observationally detectable state in the evolution of cold clouds already growing in an expanding wind. Therefore, we term this as a `fade-away' regime in the evolution of growing clouds. It is worth noting that this is different from the usual growth/destruction regime of cold gas, which depends only on the radiative cooling of the mixed boundary layer around the cloud. 
As illustrated, clouds, which are initially growing, gradually transitions to a `fade-away' regime as they move down an expanding wind. We make an empirical estimate of the time when the mass in cold+dense clouds transition from their growing state to the `fade-away' state in a \citetalias{Chevalier1985} wind. As the cloud moves downstream, its temperature remains close to $T_{\rm cl}$ while the wind temperature $T_{\rm wind}$ scales as $d^{-2(\gamma -1)}$, where $d$ is the radial distance of the cloud from the wind centre. Since we adopt a cut-off density contrast of $10$, the transition from growth to fading is expected at a fixed position $d_f$ down the wind when the density contrast drops from an initial contrast of $\chi_{\rm ini}=100$ (in our simulation) to $\chi_f=10$ (our adopted cut-off). We propose an empirical estimate of the time elapsed $t_{\rm turn}$ when the cloud has travelled this distance downstream. The distance a cloud travels downstream starting from its initial position $d_f-d_{\rm ini}$ can be approximated as 
\begin{equation}
    \label{eq:cloud_acc}
    d_f - d_{\rm ini} = \frac{1}{2} \frac{v_{\rm wind,\infty}}{f_{\rm acc} t_{\rm drag, ini}} t_{\rm turn}^2,
\end{equation}
where we assume that the cloud accelerates uniformly to the asymptotic wind speed $v_{\rm wind,\infty}$ ($=\sqrt{2\dot{E}/\dot{M}}$; $\dot{E}$ and $\dot{M}$ are energy and mass injection rates in the wind) in some multiple $f_{\rm acc}\gtrsim1$ of the initial drag time $t_{\rm drag, ini}$. Further, using the scaling of the density contrast $\chi_f/\chi_{\rm ini} = (d_f/d_{\rm ini})^{-2(\gamma -1)}$ we obtain,
\begin{equation}
    \label{eq:time_emp}
    \frac{t_{\rm turn}}{t_{\rm cc, ini}} = \sqrt{2 f_{\rm acc} \left(\frac{d_{\rm ini}}{R_{\rm cl}}\right) \left[\left(\frac{\chi_{\rm ini}}{\chi_f}\right)^{\frac{1}{2(\gamma -1)}}-1\right]}.
\end{equation}
Appropriate to our definition of the threshold over-density contrast $\chi_f=10$, we find that $f_{\rm acc}=1.4$ approximately (within a factor of a few) explains our simulation results in estimating the time when `contrasting' cloud mass stops growing in Fig. \ref{fig:cloud_mass+dense}. 

\subsection{Pressure-density distribution of cold gas}
\label{subsec:pdfs}

We analyse the distribution of the thermodynamic state of all the cold gas, with temperature less than $2 T_{\rm cl} = 8 \times 10^4 \ \rm K$. This excludes the wind material that is initially cooler than this threshold, resulting in a smaller subdomain chosen for the analysis (cf. Section \ref{subsec:cloud_def}). We show the distribution function at three different times from one of our cloud-crushing simulations in a \citetalias{Chevalier1985} wind with $t_{\rm cool, mix}/t_{\rm cc} |_{\rm ini} = 0.2$ (the fiducial parameters; cf. Table \ref{tab:CCinCC85_params}) in Fig. \ref{fig:cloud_pdf}. As expected, we find that the pressure and density of all gas cells are bound between the two isotherms at $T_{\rm cl}$ and $2 T_{\rm cl}$ indicated by dashed lines in the figure. Initially, the cloud is isobaric with the wind, but is $100$ times denser. Therefore, all the cold gas initially occupies only one point in the phase space indicated by the orange diamond in Fig. \ref{fig:cloud_pdf}. With time, mixing and radiative cooling cause a spread in the properties of the cold gas, but the median density and temperature at the centre of mass of the cold gas, marked by the red cross in the figure, remain very close to the $T_{\rm cl}$ isotherm. As expected, the red cross is a high probability region of the distribution function and can, therefore, be used as a reliable estimate of a representative thermodynamic state for the entire cloud. The inset plots in Fig. \ref{fig:cloud_pdf} show the distribution of the mass-weighted deviation of the pressure in the cold gas compared to the local wind pressure. These plots reveal that the bulk of the cold gas in the cloud attains approximate pressure equilibrium with the local background. 
Note that there exists a small pressure gradient between the cloud and the wind because the cloud is marginally under-pressurized. This could be either a signature of a local subsonic cooling flow that can potentially siphon cold gas into these clouds (\citealt{Dutta2022MNRAS}), especially since such steady solutions exist in cylindrical geometry and turbulent transport (\citealt{Tan2024MNRAS}) or partly a consequence of insufficient resolution to resolve the local cooling length (\citealt{Fielding_2020}). Such under-pressurized clouds persist even at higher resolution (cf. Fig. \ref{fig:cloud_profile-high_res}), making it less likely to be a simple resolution artifact. 

\section{Discussion}
\label{sec:discuss}

We discuss the relevance and implications of our findings in light of the existing literature in the field. We compare our simulation results and point out the discrepancies between our results and existing models/simulations. We speculate upon possible reasons and simplifying assumptions that are responsible for these differences and what they collectively imply for the multiphase cloud-wind mixing in galactic outflows. We also draw parallels from our simulations with observations of galactic outflows using synthetic emission maps from our simulation. 
At the end of this section, we discuss some of the caveats in our simulations and highlight some open questions and challenges in existing simulations and models that need further work in the future. 

\subsection{Comparison with previous works}
\label{subsec:prev-works}

There have been numerous `cloud-crushing' studies (\citealt{bruggen2016, armillotta2017, gronke2018, Li2020, Sparre2020, Kanjilal2021, Abruzzo2023arXiv}) and a close comparison with the individual ones is beyond the scope of this section. Here, we will focus on \textit{(i)} a general comparison with previous cloud-crushing studies that employ a homogeneous, non-varying background, and \textit{(ii)} a closer comparison with studies that use a changing background in time and/or space similar to the one used here. 

\subsubsection{Homogeneous cloud-crushing simulations}
\label{subsubsec:GO_exp-box}

\begin{figure}
    \centering
    \includegraphics[width=\columnwidth]{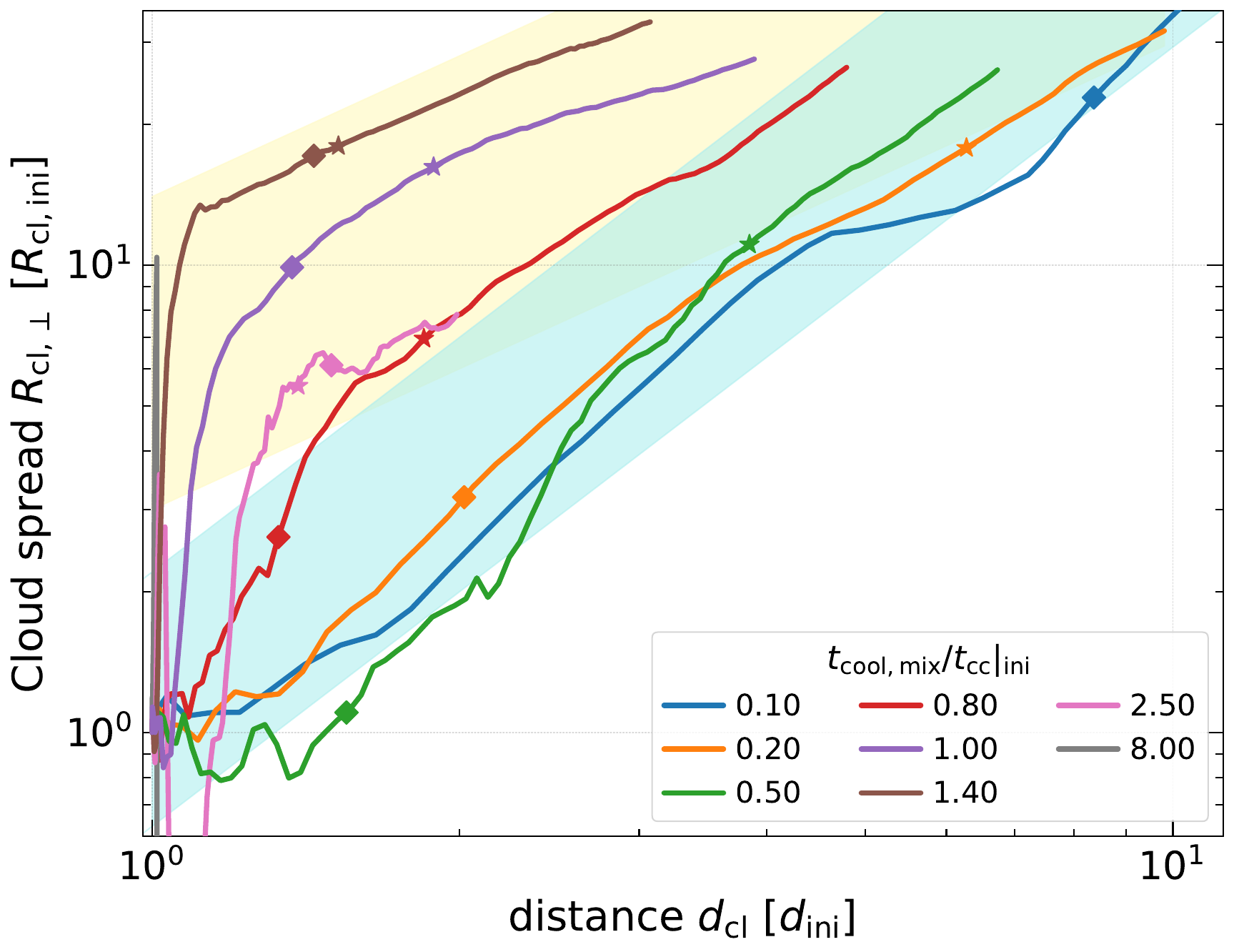}
    \caption{Evolution of the lateral extent of a cloud $R_{\rm cl, \perp}$ (in units of the initial cloud radius) as a function of the distance travelled by the centre of mass of the cloud (normalised by the initial cloud distance from the centre of the \citetalias{Chevalier1985} wind). The lateral extent of the cloud has been calculated as the difference between the $95$ and $5$ percentile values of the perpendicular distance of the cloud material from the axis passing through the centre of the cloud, parallel to the wind. The cloud increases in its lateral extent as it moves downstream. Along each of the curves, the \textbf{diamonds} denote the instant when the clouds attain twice their initial mass and the \textbf{stars} mark the instant when the centre of mass of the clouds are travelling at $80$ percent of the local wind speed -- the clouds have entered an `entrained' stage. Once the clouds are entrained, we expect the geometry of the \citetalias{Chevalier1985} wind to make the clouds spread proportional to the distance they travel, as indicated by the \textbf{yellow} band. Most of the clouds in their entrained state seem to follow this trend ($\propto d_{\rm cl}$), except for the largest clouds (small $t_{\rm cool,mx}/t_{\rm cc}|_{\rm ini}$) which are not entrained in a significant part of their journey as well as their lateral expansion is smaller compared to their initial size. Lateral expansion in these largest clouds is, therefore, not determined by the \citepalias{Chevalier1985} wind geometry but by their isobaric expansion with the wind following a much steeper relation $R_{\rm cl, \perp}\propto d_{\rm cl}^\gamma$, indicated by the \textbf{cyan} band and discussed in detail in the text.} 
    \label{fig:cloud-spread}
\end{figure}
\begin{figure}
    \centering
    \includegraphics[width=\columnwidth]{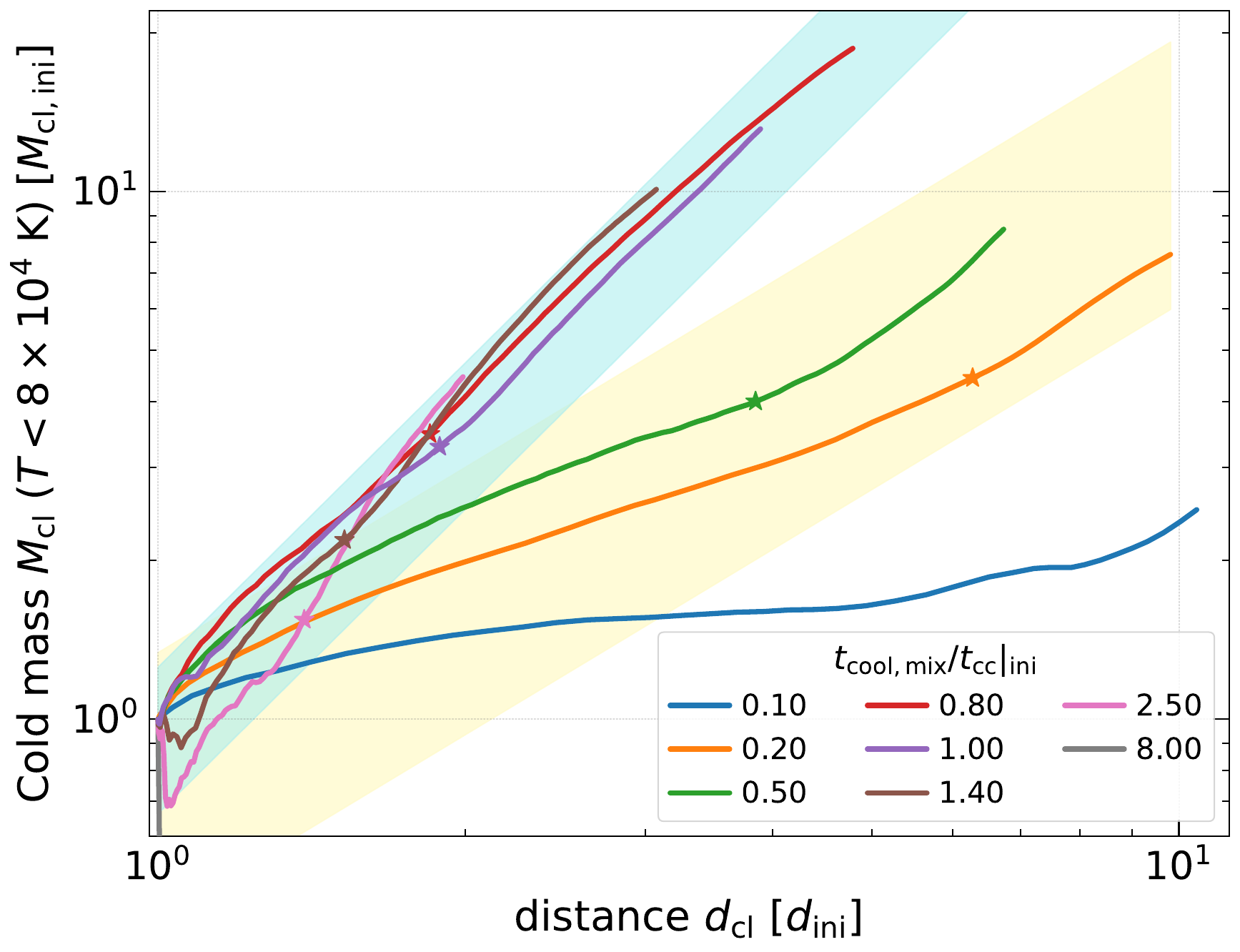}
    \caption{Evolution       of the cloud mass (in units of the initial cloud mass) as a function of the distance travelled by the centre of mass of the cloud (normalised by the initial cloud distance from the centre of the \citetalias{Chevalier1985} wind). As discussed in Section \ref{subsubsec:GO_exp-box}, the increase in the lateral spread of the cloud (cf. Fig. \ref{fig:cloud-spread}) results in a steeper increase in its mass $m_{\rm cl}$ with distance $d_{\rm cl}$ ($m_{\rm cl}\propto d_{\rm cl}^{23/12}\approx d_{\rm cl}^2$; \textbf{cyan} band) compared to the previous estimate by \citealt{gronke2020} in a \citetalias{Chevalier1985} wind ($m_{\rm cl}\propto d_{\rm cl}^{7/6}\approx d_{\rm cl}$; \textbf{yellow} band; \citetalias{gronke2020}). Along each of the curves, the \textbf{stars} have the same meaning as in Fig. \ref{fig:cloud-spread}. The clouds which have travelled a long distance after entrainment approximately follows $m_{\rm cl}\propto d_{\rm cl}^2$, in line with our analytic estimate (cyan band). For the largest clouds, the lateral expansion is lower compared to their initial size (closer to the $t_{\rm sc, cl}\sim \rm const$ assumption of \citetalias{gronke2020}; yellow band). This also leads to a different scaling relation between the cloud surface area and distance travelled by the cloud in a \citetalias{Chevalier1985} wind compared to \citetalias{gronke2020} (cf. Fig. \ref{fig:cloud_scaling}). Note that both of these mass scaling relations with distance are flatter than what is expected in a uniform wind ($m_{\rm cl} \propto d_{\rm cl}^3$) -- explaining the reduced cold mass formation in an expanding wind.} 
    \label{fig:cloud-mass-distance}
\end{figure}

While a homogeneous and non-varying background is not particularly realistic in astrophysical settings, the prior studies employing this setup focused on several different physical mechanisms at play during a cloud-wind interaction. One notable difference from simulations of cloud-crushing in a homogeneous background is that, due to the expansion of the domain, the thickness of the tail increases as the cloud moves downstream (the distance of its centre of mass $d_{\rm cl}$ increases). In a \citetalias{Chevalier1985} wind, we note that once the cloud is co-moving with the background wind, the orthogonal spread of the cloud $R_{\rm cl,\perp}$ increases with distance $\propto d_{\rm cl}$. This is illustrated in Fig. \ref{fig:cloud-spread} which shows the lateral extent of the clouds, calculated as the difference between the $95$ and $5$ percentiles of the perpendicular distance (with respect to the wind direction) of all the cold gas ($T<8\times10^4 \ \rm K$; cf. Section \ref{subsec:cloud_def}), as a function of the position of the centre of mass of the cloud. In contrast, $R_{\rm cl,\perp}$ is a constant in a uniform wind. The relation between the transverse cloud spread and the distance it travels introduces significant differences in the rate at which mass grows in entrained clouds compared to clouds in a uniform wind. We discuss using scaling relations from previous simulations of radiative turbulent mixing layers, how the orthogonal expansion of clouds can explain the suppression of cold mass growth in a \citetalias{Chevalier1985} wind compared to `vanilla' cloud-crushing (cf. Section \ref{subsec:clcrush-exp} \& Fig. \ref{fig:cc85_cold-mass}). 

In the entrained state (i.e., after substantial mass and momentum exchange between the cold cloud and the background wind), the rate of mass growth of a cloud $\dot{m}_{\rm cl}\sim A_{\rm cl} \rho_{\rm wind} v_{\rm mix}$ depends on the inflow velocity into the radiative mixing layers $v_{\rm mix}$, the surface area of the cloud $A_{\rm cl}$, and the density of the wind $\rho_{\rm wind}$. Once the cloud is entrained, its velocity is almost constant and close to the asymptotic wind velocity $v_{\rm wind, \infty}$. We first discuss the analytic scaling relations of cloud properties and turbulent boundary layers presented in \citealt{gronke2020} (\citetalias{gronke2020} hereafter) in light of our simulation results, and then compare our findings with the simulations carried out by \citetalias{gronke2020} approximating an expanding outflow. 
\citetalias{gronke2020} found that the inflow velocity in the mixing layer follows $v_{\rm mix} \sim c_{\rm s,cl} (t_{\rm sc, cl}/t_{\rm cool, cl})^{1/4}$, 
roughly consistent with later mixing layer simulations (\citealt{Fielding_2020, Tan2020}). \citetalias{gronke2020} assumed that the sound crossing time across the shortest dimension of the cloud remains constant.
However, in a \citetalias{Chevalier1985} wind, we expect the clouds, once entrained, to follow the wind geometry and have their lateral extent to increase with distance ($R_{\rm cl, \perp}\propto d_{\rm cl}$; cf. Fig. \ref{fig:cloud-spread}),\footnote{This scaling of the lateral extent of a cloud, proportional to the distance travelled by it down the wind, is also the basis of the anisotropic expanding box setup used by \citetalias{gronke2020} and discussed in Section \ref{subsubsec:GO_exp-box} \& Appendix \ref{app:exp_box}.} implying $t_{\rm sc, cl} \sim R_{\rm cl, \perp}/c_{\rm s, cl} \sim d_{\rm cl}/c_{\rm s,cl}$.
Once the cloud is entrained, similar to \citetalias{gronke2020}, we find that its surface area $A_{\rm cl} \propto V_{\rm cl}^{2/3} \sim (m_{\rm cl}/\rho_{\rm cl})^{2/3}\sim m_{\rm cl}^{2/3} d_{\rm cl}^{4 \gamma /3}$ ($V_{\rm cl}$ is the cloud volume; see upper panel of Fig. \ref{fig:cloud_scaling} in Appendix \ref{app:scaling-relations}) due to the pressure drop $p_{\rm cl} \propto d_{\rm cl}^{-2\gamma}$ in a \citetalias{Chevalier1985} wind. 

Hence, the mass of a cold cloud in the entrained state $m_{\rm cl} \propto d_{\rm cl}^{5\gamma/2-9/4}\sim d_{\rm cl}^{23/12}\approx d_{\rm cl}^2; \gamma=5/3$. This is a significantly steeper relation with cloud distance than what is expected from \citetalias{gronke2020} that uses a constant sound crossing time -- $m_{\rm cl} \propto d_{\rm cl}^{5\gamma/2-3}\sim d_{\rm cl}^{7/6}\approx d_{\rm cl}; \gamma=5/3$). Consequently, post entrainment ($A_{\rm cl} \propto V_{\rm cl}^{2/3}$), because of geometric the expansion orthogonal to the wind, small clouds follow $m_{\rm cl} \propto d_{\rm cl}^{5\gamma/2-9/4}$, which results in $A_{\rm cl}\propto d_{\rm cl}^{3(\gamma-1/2)}$. In contrast, for the largest clouds, undergoing lesser orthogonal expansion compared to their initial size, the sound crossing time across them remain constant (such that $m_{\rm cl} \propto d_{\rm cl}^{5\gamma/2-3}$), and their surface area $A_{\rm cl}\propto d_{\rm cl}^{3\gamma -2}$ (see lower panel of Fig. \ref{fig:cloud_scaling} in Appendix \ref{app:scaling-relations}). In Fig. \ref{fig:cloud-mass-distance}, we demonstrate that with the inclusion of cloud expansion in the sound crossing time, we are able to explain the steeper scaling (compared to the analytic arguments in \citetalias{gronke2020}) of cold mass with distance travelled by clouds that are entrained for a long time.\footnote{This inclusion of cloud expansion, however, weakly alters the scaling of cloud surface area with the distance travelled as the cloud traverses downstream (see Appendix \ref{app:scaling-relations}; lower panel of Fig. \ref{fig:cloud_scaling}).} We can also explain the steeper lateral extent with distance travelled for the largest clouds (cf. Fig. \ref{fig:cloud-spread}) by noting that for these large clouds the expansion of the \citetalias{Chevalier1985} background compared to their size is of lesser importance. Instead, as clouds maintain pressure equilibrium with the background, $R_{\rm cl, \perp} \propto \rho_{\rm cl}^{-1/2} (m_{\rm cl}/R_{\rm cl, \parallel})^{1/2}\sim d_{\rm cl}^{\gamma}$ -- a steeper scaling with distance for the large clouds (cyan band in Fig. \ref{fig:cloud-spread}). This scaling follows from the fact that for the largest clouds $m_{\rm cl}\propto d_{\rm cl}$ (cf. Fig. \ref{fig:cloud-mass-distance}) and they are not sufficiently entrained in the wind to follow the wind expansion. In addition, due to being not entrained, their length along the wind $R_{\rm cl, \parallel}\propto d_{\rm cl}$ (as expected from `pure stretching') -- keeps growing and does not saturate to a fixed value unlike the smaller clouds (see Fig.~\ref{fig:cloud-length} for the $R_{\rm cl,\parallel}(d_{\rm cl})$ evolution). 

A key distinction of clouds in a \citetalias{Chevalier1985} wind from homogeneous cloud-crushing studies lies in the spatial gradients of background wind properties that the cloud experiences -- both along its downstream trajectory and across its internal extent from head to tail. The latter leads to interesting and potentially observable emission gradients along the cloud body (discussed further in Section \ref{subsec:emission_cloud}), while the former may significantly influence the net cold mass growth. As demonstrated by \citetalias{gronke2020}, although the background density drops with distance as $\propto d_{\rm cl}^{-2}$, this decline is remarkably offset by the concurrent increase in the cold gas surface area and cooling time ($t_{\rm cool, cl}\propto d_{\rm cl}^{10/3}$) due to changing pressure conditions.
However, our analytic treatment -- supported by simulation results -- shows that the orthogonal expansion of the cloud causes the cold mass to scale as $m_{\rm cl} \propto d_{\rm cl}^2$, rather than the $\propto d_{\rm cl}$ scaling proposed by \citetalias{gronke2020}. 
In the uniform wind scenario, both $v_{\rm mix}$ and $\rho _{\rm wind}$ are constant, resulting in an even steeper scaling of cloud mass with distance, $m_{\rm cl} \propto d_{\rm cl}^3$, compared to the shallower relation in an expanding wind. 

\begin{figure}
    \centering
    \includesvg[width=\columnwidth]{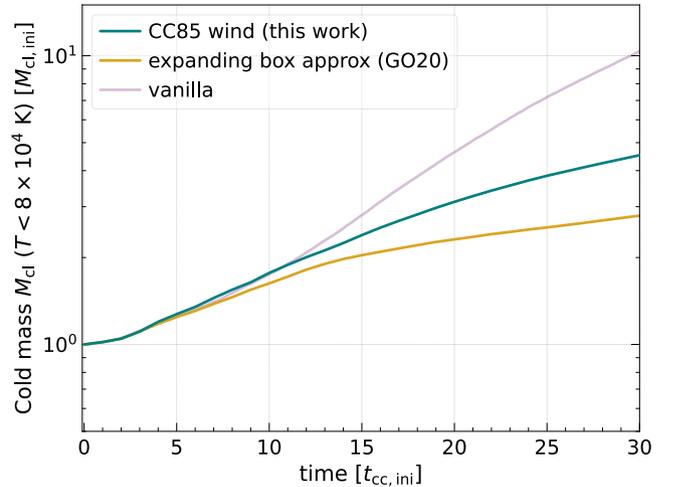}
    \caption{Comparison of the cold mass evolution of gas from our cloud-crushing simulation in \citetalias{Chevalier1985} wind with the cloud-crushing simulation employing an approximate expanding box model from \citealt{gronke2020} (\citetalias{gronke2020}; see Section \ref{subsubsec:GO_exp-box} \& Appendix \ref{app:exp_box} for details). For this comparison, we re-implemented the expanding box model used by \citetalias{gronke2020} in \texttt{PLUTO} and re-simulated the cloud-crushing problem using our fiducial parameters, i.e., $t_{\rm cool, mix}/t_{\rm cc}|_{\rm ini} = 0.2$ and an initial Mach number $\mathcal{M}=1.496$ in the wind (cf. Table \ref{tab:CCinCC85_params}). In both the simulations, the cloud radius is initially resolved by $16$ grid cells. Follow \href{https://youtu.be/HoqEV4Y1adc}{this link} to see the evolution of this cloud in the expanding box simulation.} 
    \label{fig:ourvsGO20_sim}
\end{figure}

Moving beyond analytic scaling estimates, we now directly compare simulation results from \citetalias{gronke2020} with ours. Specifically, \citetalias{gronke2020} studied the evolution of cold clouds in expanding outflows by employing a model of an expanding box in their simulations that approximates the azimuthal and polar expansion that the clouds undergo as they move downstream due to the geometry of the wind (also, see, similar setups by \citealp{Heitsch2009, Tan2023} for in-falling clouds). 
This model employs an anisotropic expansion in the simulation domain to model the \citetalias{Chevalier1985} wind which expands in the polar and azimuthal directions but not in the radial direction (cf. \citealt{Dutta_2019}; also \citealt{Scannapieco2017ApJ} who erroneously assumed the wind to be isotropically expanding). This model assumes that the size of the cloud (and the simulation domain containing and following it) is always small enough compared to the length scale on which the wind property changes, and the wind can be assumed to be plane parallel. Therefore, under these assumptions, the entire cloud always experiences a uniform expansion perpendicular to the wind due to its movement downstream. Consequently, a local simulation domain that follows the cloud gets expanded in all directions perpendicular to the direction of the wind, determined by an expansion factor that depends on the instantaneous position of the cloud with respect to its initial position.

In contrast, our cloud-crushing simulations in an expanding \citetalias{Chevalier1985} wind do not make such simplifying assumptions and can account for the differential expansion of a cloud along its tail. This differential expansion of a cloud along the wind and the curvature of the wind (i.e., variation in wind properties along the cloud) become increasingly important for large clouds deeper 
in the cooling regime. We find that this differential expansion can lead to a lower expansion in the cloud head compared to its tail (cf. Fig. \ref{fig:compareGO20-slice-plot} in Appendix \ref{app:exp_box}), which, in turn, can result in an overall increase in the tail length of the cometary cloud structures (after several $t_{\rm cc}$) compared to the uniform expansion considered by \citetalias{gronke2020}. Moreover, the background wind properties -- such as density, temperature, and pressure -- vary significantly with the distance $d_{\rm cl}$ travelled by the cloud. Consequently, even modest discrepancies in estimating how this distance changes with time (arising from differences in how mixing/cooling between the cold cloud and hot wind progresses, which affects mass and momentum exchange) can lead to substantial differences in the predicted cold mass evolution, especially at late times. 

Another difference from \citetalias{gronke2020} is that we initialise our clouds at a distance where the wind Mach number is between 1-2, not too far from the sonic point. Therefore, the wind is not in the asymptotic scaling ($\rho \propto d^{-2}$, $p \propto d^{-2\gamma}$) regime, an assumption made in \citetalias{gronke2020}.
All of these factors potentially contribute to differences in the cold mass growth between the two simulation setups, which increase as the simulations progress further in time. This is illustrated in Fig. \ref{fig:ourvsGO20_sim}, where we compare the evolution of the cold mass 
in the two simulation setups for our fiducial parameters ($t_{\rm cool, mix}/t_{\rm cc}|_{\rm ini}=0.2$, and $\mathcal{M} = 1.496$; cf. Table \ref{tab:CCinCC85_params}; see Appendix \ref{app:exp_box} for our implementation of \citetalias{gronke2020} setup). 
Since the cold mass estimate is slightly lower in the anisotropic expanding box of \citetalias{gronke2020} compared to our simulation domain, it might explain why \citetalias{gronke2020} reported a flatter mass-distance relation ($m_{\rm cl}\propto d_{\rm cl}$). We expect the differences in the cold mass estimate between these two types of simulation setups to increase further in time. 

\subsubsection{Semi-analytic multiphase wind models}
\label{subsubsec:DF-single_cloud}

\begin{figure}
    \centering
    \includegraphics[width=\columnwidth]{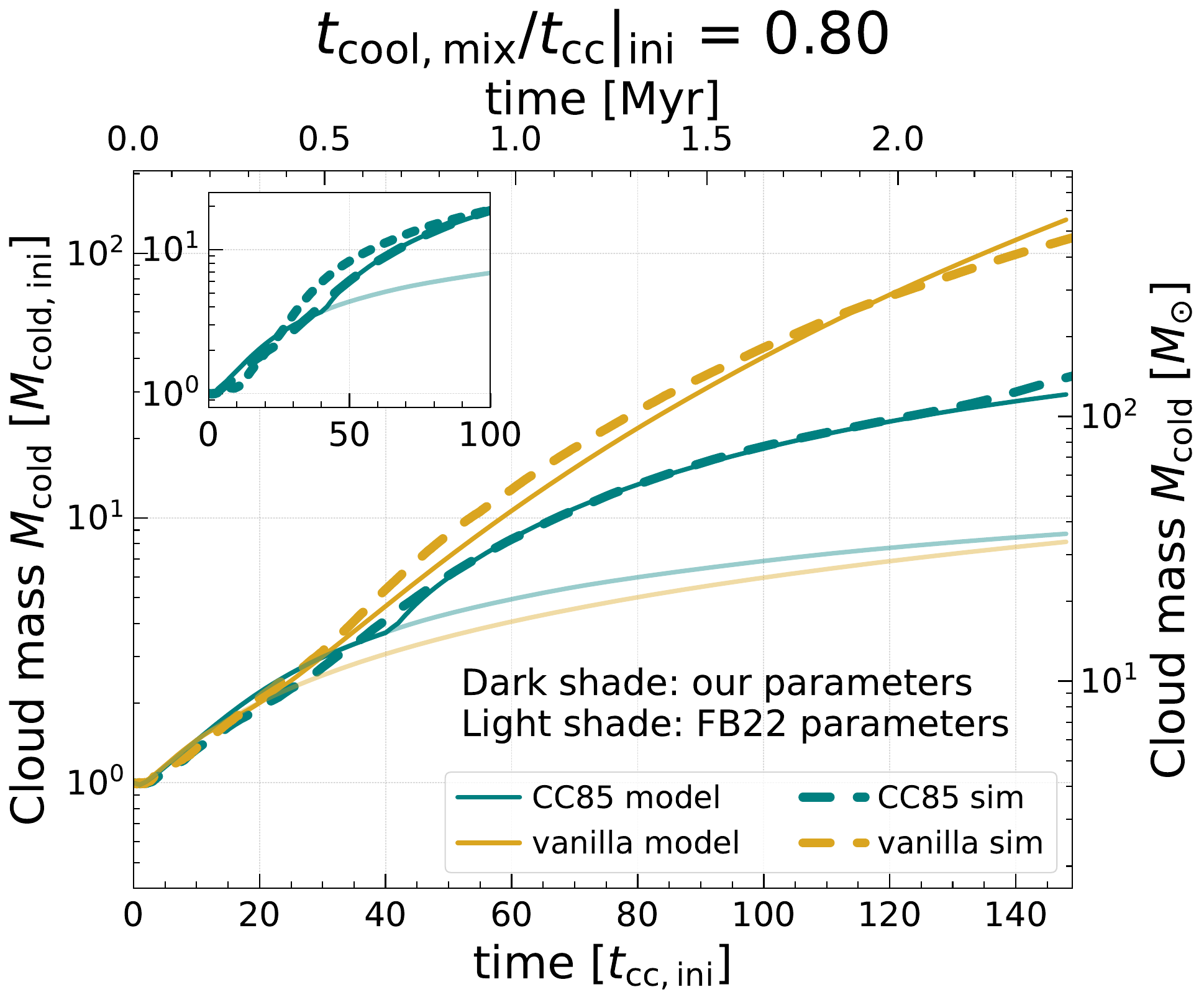}
    \caption{Comparison of the cloud mass evolution from our cloud-crushing simulation with that predicted from the \citetalias{Fielding2022ApJ} single-cloud model for `vanilla' cloud-crushing (in uniform wind) and for cloud-crushing in an expanding \citetalias{Chevalier1985} wind (see Section \ref{subsubsec:DF-single_cloud} for details). The \textbf{dashed} curves are the simulation results for `vanilla' cloud-crushing (in \textbf{yellow}) and cloud-crushing in a \citetalias{Chevalier1985} background (in \textbf{teal}). As a specific example which would clearly demonstrate departure from model prediction, we choose the simulation result with $t_{\rm cool, mix}/t_{\rm cc}|_{\rm ini} = 0.8$ and $\mathcal{M}=1.496$ (cf. Table \ref{tab:CCinCC85_params}; not the fiducial parameter set). The corresponding model predictions are shown with \textbf{solid} lines in the same colour as their dashed counterparts. The \textbf{thick, solid light}-coloured curves use the original \citetalias{Fielding2022ApJ} single-cloud model and does not match well with the simulation results. On changing to our `modified' model that floors the inflow velocity according to Eq. \ref{eq:model_modify} (described in Section \ref{subsubsec:DF-single_cloud} \& Appendix \ref{app:FB22_model}), we are able to achieve a better match with the simulation results -- the \textbf{solid thin dark} coloured curves. For simplicity, the free parameters of the model are chosen by visual inspection and 
    we simply demonstrate the need to construct a robust single-cloud model -- a 
    necessary step to explain the evolution of clouds in realistic outflows (cf. Fig. \ref{fig:model_wind_fielding_compare}). The \textbf{inset} plot shows an additional blue tightly dotted curve up to $100\ t_{\rm cc,ini}$, which corresponds to the same cloud-crushing simulation in a \citetalias{Chevalier1985} wind as the other simulation (loosely dashed blue) but run at a higher resolution of $R_{\rm cl}/d_{\rm cell}|_{\rm ini}=16$ instead of $8$. Changes in the cloud mass evolution due to increase in simulation resolution are too minimal to account for the discrepancy between the \citetalias{Fielding2022ApJ} prediction and our simulation result.} 
    \label{fig:model_fielding_compare}
\end{figure}

\begin{figure*}
    \centering
    \includegraphics[width=\textwidth]{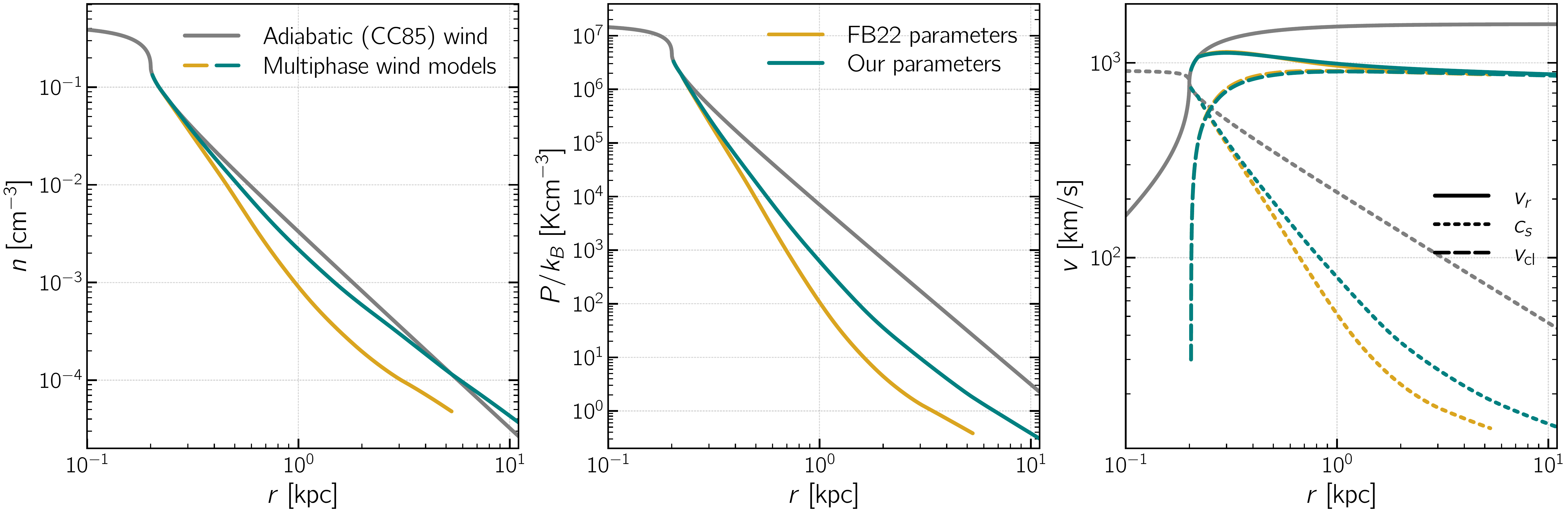}
    \caption{Comparison of the \citetalias{Fielding2022ApJ} multiphase wind profile (density: \textbf{left}, pressure: \textbf{middle}, velocity: \textbf{right} panel) due to the modification of the underlying single-cloud evolution introduced in our work (in \textbf{teal}) from the original \citetalias{Fielding2022ApJ} single-cloud model (in \textbf{yellow}). The right panel shows different velocities, namely the wind speed $v_r$, sound speed $c_s$, and the cloud speed $v_{\rm cl}$ as indicated in the legend attached to the figure. The wind parameters chosen for both the yellow and teal curves are the same and mentioned in the main text. 
    Only the single-cloud model (cf. Fig. \ref{fig:model_fielding_compare} \& Eq. \ref{eq:single_cloud_vanilla}) that controls the mass and energy loading due to the multiphase mixing of clouds seeded in an outflow is different between the yellow and the teal curves. This highlights that the evolution of clouds in an outflow provides a back-reaction to the wind and plays a crucial role in setting the large scale properties of their host outflow. These clouds become increasingly important as the efficiency of mass and energy loading in the wind increases. 
    For reference, the \textbf{gray} curves show the classic single phase adiabatic \citetalias{Chevalier1985} wind profile. Follow the \href{https://gist.github.com/dutta-alankar/72f21b02a8c9e34e64d94a9c65992fc2}{link here} for a full image of all the relevant quantities similar to Figs. 5 \& 6 from \citealt{Fielding2022ApJ}. 
    }
    \label{fig:model_wind_fielding_compare}
\end{figure*}

\citealt{Fielding2022ApJ} (\citetalias{Fielding2022ApJ} hereafter) introduced an empirical semi-analytic model that describes the evolution of a cold cloud that starts with some initial velocity (or from rest) in a hot outflow (the `single-cloud' model; cf. Section 3 in their paper). 
The \citetalias{Fielding2022ApJ} single-cloud model provides a source/sink of energy, mass, and momentum, in a steady wind driven by supernovae in order to capture the back-reaction of clouds on their host outflow. Along with the assumption that clouds are continuously introduced at a fixed rate close to the centre of the wind, \citealt{Fielding2022ApJ} utilised their single-cloud evolution model in developing a general framework (wind equation; cf. Fig. 3 in their paper) to predict the structure of a typical supernovae-driven multiphase galactic wind. 

Although in \citetalias{Fielding2022ApJ} the single-cloud model was utilised to take into account the evolution of a collection of clouds in a spatially changing background wind (that act as source terms in the wind equation), the detailed evolution of an individual cloud in a non-uniform wind, and its comparison with simulations remain largely unexplored. 
Since our work focuses on simulations of how individual clouds evolve in both a uniform and an expanding \citetalias{Chevalier1985} wind, we are able to compare the \citetalias{Fielding2022ApJ} model prediction of cloud-wind interaction with our simulation results. Furthermore, the motivation is that constraining the single-cloud model using simulations in diverse wind conditions would enable the construction of realistic wind profiles. This would consequently ensure an accurate modelling of the back-reaction of the clouds in the multiphase wind equation. Prior to a comparison with \citetalias{Fielding2022ApJ}, it is worth noting that our simulations focus on the detailed behaviour of a cloud, considering a neighbourhood close to the cloud. The back reaction of the cloud on the volume-filling gas that comprises the wind is beyond the scope of this work. In this context, we are mostly interested in simulating the evolution of an individual cloud and do not consider the collective impact of multiple clouds that can potentially alter the large-scale properties of the wind, as discussed further. 

On comparing the original \citetalias{Fielding2022ApJ} single-cloud model with our simulation results, we find an unsatisfactory model prediction for the evolution of cold mass growth (see the discussion on Fig. \ref{fig:model_fielding_compare} later in this section for details).\footnote{For simplicity and also to compare with our simulations, we do not consider different metallicities for the cloud and the wind.} 
Therefore, we modify the \citetalias{Fielding2022ApJ} single-cloud model to better reproduce the evolution of the cold mass from our simulations. 
In the remainder of this section (and Appendix \ref{app:FB22_model}), we briefly discuss the \citetalias{Fielding2022ApJ} single-cloud model. Specifically, we highlight the modifications that we introduce to the original model, and motivate these modifications in order to obtain a significantly improved match with our simulation results. We also discuss tentative shortcomings of the current model and point at directions for further improvement. 

At its core, the \citetalias{Fielding2022ApJ} single-cloud model is a set of coupled ordinary differential equations describing the evolution of the position, velocity, and mass of a cloud in a hot wind which can be numerically solved with some initial condition (described elaborately in Appendix \ref{app:FB22_model}). One of the most important components in this model is radiative cooling of the turbulent mixing layer which forms as the cloud moves and mixes in the wind. This essentially captures the growth/destruction of cold clouds in moving in a hot wind. 
Fig. \ref{fig:model_fielding_compare} shows a comparison of our numerical simulations -- both in an expanding \citetalias{Chevalier1985} (teal curves) as well as uniform plane-parallel wind (yellow curves) -- with the prediction of the \citetalias{Fielding2022ApJ} single-cloud model (shown in thick, solid lines with a lighter shade in Fig. \ref{fig:model_fielding_compare}). One can see that the numerical results do not match the model sufficiently, with $\gtrsim1$ dex more cold gas mass in the simulations compared to the model. In addition, the \citetalias{Fielding2022ApJ} model is also unable to reproduce the suppression of cold mass growth of clouds in an expanding wind when compared to a uniform plane-parallel wind -- a definitive trend in our simulation. This is concerning, since the \citetalias{Fielding2022ApJ} is used as a basis for multiphase sub-grid model of galactic winds with the aim of being eventually incorporated into cosmological simulations  (\citealt{Smith2024MNRASa, Smith2024MNRASb}) as well as to interpret observations \citep{Lopez2025arXiv, Xu2023ApJ}. This mismatch in the cold gas mass also propagates to the overall wind properties -- as we discuss below. 

As a cloud moves down the wind, it can be either in the slow ($\xi\leq1$) or in the fast ($\xi>1$) cooling regime (cf. Eq. \ref{eq:single_cloud_vanilla}), depending on a competition between radiative cooling and turbulent mixing. Simulations have found that the growth rate of the clouds are different in these two regimes (\citealt{Ji2019MNRAS, Tan2020}) and is interpreted as fast cooling causing a change in the fractal nature of the turbulent boundary layer around cold clouds (\citealt{Sreenivasan1989RSPSA, Fielding_2020}). An important point to note is that the difference between the original model prediction and the simulation results increase with time as clouds become increasingly co-moving with the wind. 
This might be because the \citetalias{Fielding2022ApJ} single-cloud model only considers shear-driven hydrodynamic instabilities (\citealt{Mandelbrot1975JFM, Federrath2009ApJ, Fielding_2020, Tan2024MNRAS}) to be causing the turbulent mixing. Over time, as shear diminishes, the model predicts lower production of efficiently cooling mixed gas, resulting in a reduced growth of cold gas in a cloud. However, simulations have routinely shown that cold clouds always maintain a turbulent boundary layer and keep growing in mass even when they are co-moving with the background wind. Once entrained, boundary layers around cold clouds can be maintained by pulsations (\citealt{Waters2019ApJ, Gronke2023MNRAS}), or local cooling flows (\citealt{Tan2021MNRAS, Dutta2022MNRAS}). 

To support this speculation, we show in Fig. \ref{fig:model_fielding_compare}, using solid dark coloured curves, a new model prediction which produces a substantially improved match with the simulation data. Motivated from the semi-analytic work by \citealt{Nikolis2024MNRAS} (see also \citealp{Tan2023}), we propose that the inflow velocity due to cooling, when shear diminishes depends, does not diminish to zero but instead depends on the sound speed in the cold phase. This modification to the model makes it reproduce the general trend of mass growth in the simulations, in particular the suppression of cold mass in clouds in an expanding \citetalias{Chevalier1985} wind compared to a uniform one. It is worth noting that this is an empirical prescription and further work is needed to build an adequate understanding of the underlying physics that maintain a turbulent layer around cold clouds after entrainment. 

The exact modification to the original \citetalias{Fielding2022ApJ} single-cloud model is as follows (see Eq. \ref{eq:single_cloud_vanilla} for details). Instead of defining the inflow velocity $v_{\rm in}$ that causes mass growth in the cloud as $v_{\rm in}=f_{\rm turb} |v_{\rm rel}| \xi^{\alpha}$, where $v_{\rm rel}=v_{\rm wind}-v_{\rm cl}$, we re-define inflow velocity 
\begin{equation}
    \label{eq:model_modify}
    v_{\rm in}={\rm max}\left[f_{\rm turb} v_{\rm rel} \xi^{\alpha}, f_{\rm sat} c_{\rm s, cl} \left(\frac{l'}{c_{\rm s, cl} t_{\rm cool, cl}}\right)^{1/4}\right].
\end{equation}
We found that in the case of a uniform wind (`vanilla'), $f_{\rm sat}=0.075$ and $l'$ set to the initial cloud size best reproduces the simulation results (which is in agreement with previous simulations; see \citealp{gronke2020, Fielding_2020, Tan2020} and references therein for details on the $1/4$ scaling). However, in case of a \citetalias{Chevalier1985} wind, $f_{\rm sat}=0.19$ and $l'=R_{\rm cl} = (3M_{\rm cl}/4\pi\rho_{\rm cl})^{1/3}$ gives a good match with simulations. Since $l'$ represents the characteristic length scale for turbulence in the boundary layers, an expression for $l'$ that scales with cloud size is intuitively expected. Nevertheless, the fact that we need different values for the free parameters for different wind conditions reveals the limitation of the current model to accurately capture the cloud geometry and its evolution in diverse cloud-wind conditions. 

One possibly limiting assumption in the model is that a cloud instantaneous stretches along the wind as it cools and it is determined by a constant factor $f_{\rm cool}$ in the \citetalias{Fielding2022ApJ} model (see Appendix \ref{app:FB22_model} for details). Clouds can however undergo differential expansion and stretch over time in directions both along and orthogonal to the wind (\citealt{Huang2020MNRAS,Nikolis2024MNRAS}), as also seen in our simulations. The model can, therefore, be improved by position/time dependent free parameters to better capture the cloud geometry and its evolution downstream. The free parameters listed here are, therefore, not robust across diverse wind or cloud properties (and would need to be adjusted depending on the wind+cloud parameters) in accurately predicting the mass evolution nor a representative best fit value. Instead, the values for these free parameters used here is intended as a demonstration in support of the idea that there is a need to floor the inflow velocity (that causes the mass growth in cold clouds) and not to let it diminish over time. Note that the agreement between the original \citetalias{Fielding2022ApJ} model and our modified prescription is significantly better in the cloud velocity than in mass. In fact, the cloud velocity in our `modified' model matches the original \citetalias{Fielding2022ApJ} model on a percent level. We detail this further in Appendix \ref{app:FB22_model}. A detailed exploration of the model’s parameter space as well as developing a more robust physical accurate model is beyond the scope of this work and is left for future studies. Our primary goal is to emphasize the need for improving existing models by calibrating them against cloud-crushing simulations in evolving background wind conditions. In the remainder of this section, we discuss that changes in the single-cloud model can potentially result in non-negligible differences in the multiphase nature of their host outflows.

\citealt{Fielding2022ApJ} also developed a framework for evolving a population of cool clouds moving in a hot wind. In this framework, the single-cloud model (described in the earlier part of this section) serves as the basis for modelling the source (or sink) terms that determine the radial profile of the wind (see the equations illustrated in Fig. 3 \& Eq. 17 from \citetalias{Fielding2022ApJ}). These source/sink terms directly affect the energy and mass loading of the outflows. Therefore, the way individual clouds interact with the background wind gives rise to a back reaction on their host outflow and determines the multiphase nature of the wind. The \citetalias{Fielding2022ApJ} prescription assumes a collection of cold clouds that are continuously injected in the outflow. These clouds introduce energy and mass in both the cold and hot phases of the wind, depending on how 
they cool, mix and evaporate. To this end, we demonstrate that changes in the single-cloud evolution model can directly change the properties of the wind by solving the wind equation from \citetalias{Fielding2022ApJ} (Fig. 3 \& Eq. 17 in their paper) with different models of the single-cloud evolution. This is illustrated in Fig. \ref{fig:model_wind_fielding_compare}, where we compare the \citetalias{Fielding2022ApJ} single-cloud model, set as source/sink terms in the wind equation, with our single-cloud model. Specifically, we solve the \citetalias{Fielding2022ApJ} wind equation with $5\ \rm M_\odot yr^{-1}$ star formation and energy and mass loading efficiencies of $0.5$ and $0.2$ respectively. The coupling efficiency of the cloud with the wind $\eta_{\rm M, cold}$ is set to $0.15$. We set the initial cloud mass to $100 \ \rm M_\odot$ ($\xi_{\rm ini}=2.63$) introduced at about $220\ \rm pc$ with a temperature of $8\times 10^3\ \rm K$ and an initial radial velocity of $30 \ \rm km\ s^{-1}$ in a steady \citetalias{Chevalier1985} wind which has a sonic radius of $200 \ \rm pc$ opening over a full $4\pi$ steradians in solid angle. Additionally, the gravity is set by a circular velocity of $150 \ \rm km\ s^{-1}$. The wind and the cloud metallicities are set to $2\ Z_\odot$ and $Z_\odot$, respectively. In the Fig. \ref{fig:model_wind_fielding_compare}, we compare the wind profiles on changing from the original \citetalias{Fielding2022ApJ} single-cloud model (in yellow) to our `modified' single-cloud model (in teal). The two wind profiles differ substantially from one another ($\sim 1 \ \rm dex$ in pressure and $\sim 0.5\ \rm dex$ in density and velocity) over almost their entire downstream distance ($\gtrsim 1 \rm \ kpc$).

Since cloud evolution directly affects the mass and energy loading of an outflow and its multiphase nature, winds with higher efficiency of mass loading are increasingly affected by the behaviour of an increased number of clouds that are seeded in them. It is worth noting that our discussion in this section is far from being a comprehensive modelling of multiphase galactic outflows. We simply demonstrate that wind properties can significantly influence the evolution of embedded clouds, which in turn modify the large-scale structure of the host outflow -- setting up a self-adjusting interplay that further shapes subsequent cloud evolution. Overall, our comparison highlights the need to accurately test models of individual cloud-wind evolution in order to get an 
accurate model of multiphase galactic outflows. Such accurate models can greatly improve sub-grid prescriptions of cold gas in cosmological simulations or simulations with limited resolutions (\citealt{Weinberger2023MNRAS, Smith2024MNRASa, Butsky2024MNRAS}). 

\subsection{Emission from clouds in a CC85 wind}
\label{subsec:emission_cloud}
\begin{figure*}
    \centering
	\includegraphics[width=0.88\textwidth]{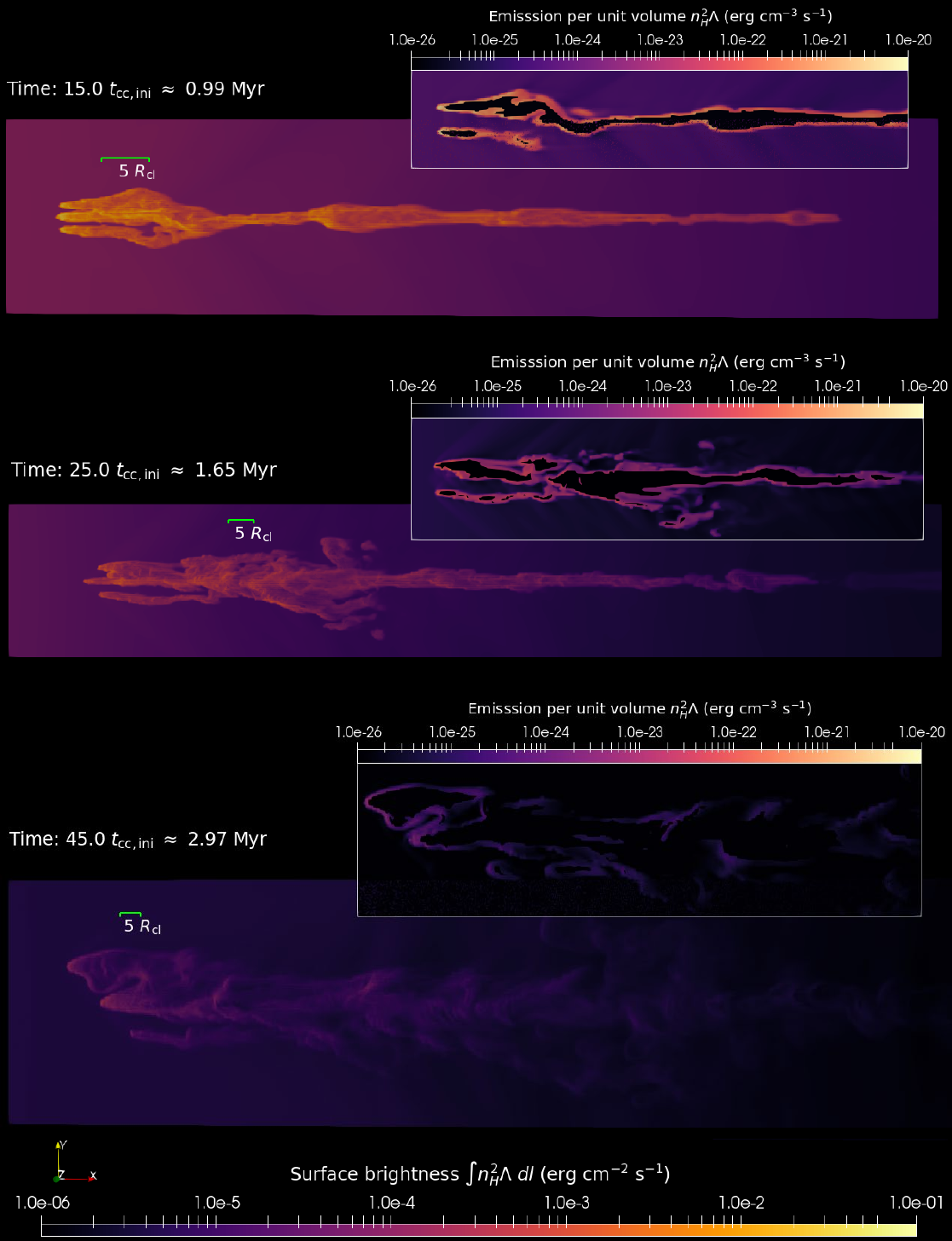}
    \caption{The temporal evolution of the total surface brightness from our cloud-crushing simulation in a \citetalias{Chevalier1985} wind using fiducial parameters $(\mathcal{M}, \chi, t_{\rm cool, mix}/t_{\rm cc}|_{\rm ini}) = (1.496, 100, 0.2)$; cf. Table \ref{tab:CCinCC85_params}. Here, we show only a few selected snapshots that clearly demonstrate the trend that the emissivity from the cloud gradually fades with time. This is due to the weakening of radiative cooling downstream leading to a reduction in the amount of dense, cold mass growing due to shear-driven mixing in the turbulent boundary layers between the cloud and the wind. The main body of the cloud is very close to the floor temperature of our simulation, at which the emission is set to zero. This is clearly demonstrated in the inset figures showing emissivity per unit volume of the gas on a slice passing through the centre of our simulation domain with normal along the $z$-axis. Emission is dominated by the turbulent boundary layers. Similar layers are integrated along the $z$-axis to obtain the projected surface brightness plots in this figure. Note that the zoom levels are different across snapshots, indicated by the scales attached along with them.}
    \label{fig:emission_slice}
\end{figure*}

The turbulent mixing layers and cooling flows surrounding the cold cloud radiate energy, powered by the enthalpy flux from the hot to the cold medium. In a steady state when a cloud is entrained, the bolometric luminosity is proportional to the mass flux. In a TRML or a local cooling flow, the total radiative cooling rate equals the enthalpy inflow rate (both proportional to the mass flux), which is regulated by the density and temperature distribution within the cooling layer. Thus, because of the lower mass growth rates we found for clouds embedded in a \citetalias{Chevalier1985} wind compared to a uniform one (cf. Fig. \ref{fig:cc85_cold-mass} \& Section \ref{subsubsec:GO_exp-box}), we expect the potentially observable surface brightness levels to be accordingly lower.


Another equivalent way to interpret this is that the surface brightness of the gas depends on its emissivity, which is proportional to the square of the gas density.\footnote{This is emissivity integrated across all wavelengths and might not be strictly valid for any individual emission lines like HI.} Since the density of the cold gas formed due to cloud-crushing in an adiabatically expanding background keeps falling, its surface brightness is expected to be significantly diminished as the clouds move down the wind. In addition, the enthalpy flux and hence the emission can slightly differ from (the non-entrained, dense) head to (entrained, diffuse) tail, in both the homogeneous cloud-crushing and the \citetalias{Chevalier1985} wind 
because of the variation in density and $\dot{m}_{\rm cl}$ (cf. \citealp{Tan2020}). This is certainly not the case once the cold cloud is fully entrained. Even in this state, our expanding wind setup still exhibits a strong emission gradient not present for clouds in a homogeneous background. 

Fig. \ref{fig:emission_slice} shows the surface brightness of cold gas from our fiducial cloud-crushing simulation (cf. Table \ref{tab:CCinCC85_params}) at different times. To calculate the surface brightness map (projected in the $z$-direction orthogonal to the wind), we calculate $n_H^2\Lambda (T) \Delta V$ for every cell in the simulation domain and add them along the $z$-direction, where $\Delta V$ is cell volume. The radiative loss for gas at the floor temperature is set to zero, i.e., $\Lambda \left(T_{\rm cl}\right) = 0$. 
As the cloud moves in the wind, it expands, and its surface brightness diminishes, consistent with the above discussion. 
The insets with each of the surface brightness plots in Fig. \ref{fig:emission_slice} show a slice of emission rate per unit volume 
through the centre of our simulation domain. These clearly demonstrate that emission is dominated by the radiatively cooling 
boundary layers between the cloud and the wind. Contributions from several such slices of emission have been integrated along the line of sight ($z$-axis, here) to generate the surface brightness maps. The bulk of the cloud is at the cooling floor of our simulation ($T=4\times 10^4 \ \rm K$), emission from which is set to zero. Fig. \ref{fig:emission_slice} shows a strong gradient in surface brightness in the elongated tails of the cloud, 
with the surface brightness of the long cometary cloud tails dimming along the wind. 

We point out that the surface brightness gradient in Fig. \ref{fig:emission_slice} is for an individual cloud. For a more realistic comparison, a semi-analytical wind model including several clouds (cf. Section \ref{subsubsec:DF-single_cloud}) has to be convolved with the global surface brightness profile. Such wind models need to be multiphase by implicitly taking into account mass and energy loading by several such clouds. 
Comparing predicted and observed surface brightness maps is an interesting future avenue to constrain the models and our understanding of galactic winds. In particular, a statistical measure of emission gradients can reveal the properties of the hot medium otherwise inaccessible at higher redshift.

The M82 galaxy is a classic example of a starburst-driven galaxy harbouring strong outflows (\citealt{Lynds1963ApJ}), which we use as a benchmark for comparing simulation results. Clouds simulated in an expanding wind have a morphological similarity with the elongated cold gas `chimneys' observed by \citealt{Bolatto2024ApJ, Fisher2024MNRAS} in the M82 galaxy. The infrared emission from the cold gas and dust observed using JWST in M82 by \citealt{Bolatto2024ApJ, Fisher2024MNRAS} is formed in a starburst-driven outflow environment, similar to what we have simulated. Long, elongated plumes of cold gas also show a similar gradient in their emission along the direction of the wind (see Figs. 4 \& 6 from \citealt{Fisher2024MNRAS}). As discussed earlier, such a strong gradient in the surface brightness of the cold clouds is not possible in `vanilla' cloud-crushing because the clouds in homogeneous cloud-crushing setups lack differential expansion parallel to the wind due to the absence of a spatial gradient in the wind. It is worth noting that we are unable to reproduce the loop-like features towards the centre of the wind, which are very likely magnetically dominated structures -- beyond the scope of our pure hydro simulations. 
A detailed, quantitative comparison with the measured emission gradients can help us constrain the mass transfer rate between the phases, and in particular, study the hot, accelerated phase, which is otherwise inaccessible but crucial in order to constrain feedback models.


\subsection{Caveats and Future Directions}
\label{subsec:cc85_caveat}





The \citealt{Chevalier1985} (\citetalias{Chevalier1985}) model of an adiabatically expanding outflow 
makes dramatic simplifications. Observations show that the thermodynamic state of the outflow can be significantly different from a steady 
\citetalias{Chevalier1985} wind (\citealt{Xu2023ApJ}). For instance, \citealt{Lopez2020ApJ} observe much shallower density and temperature profiles (also in free electron density profile observed by Deanne Fisher; private communication) for the X-ray bright hot gas beyond $1 \ \rm kpc$ in M82 compared to the \citetalias{Chevalier1985} model. Such deviations can be generic and wind profiles can potentially be non-steady in various starburst-driven outflows. Our work 
highlights that the expansion of the wind bears non-trivial implications for the survival/growth of cold clouds in outflows.\footnote{For instance, in an isothermally expanding wind, we expect a significantly reduced cloud mass growth than in an adiabatically expanding wind. From our analytic considerations, we expect $m_{\rm cl} \propto d_{\rm cl}^{1/4}$ in an isothermal wind (not validated against any simulations carried out in this paper).} This is not surprising, and depending on the background profile, clouds behave significantly differently in how they mix, shred, or grow (cf. also \citealt{Tan2023} for clouds in-falling in a stratified medium).

An extensive study of the behaviour of cold clouds in diverse environments of astrophysical relevance under a unified framework has yet to be explored. Note that for the majority of our runs, the cooling time of the hot medium is longer than the runtime. However, in some runs (with the largest clouds) the runtime is comparable with the wind cooling time, and we would expect thermal instabilities to occur. Given that we ignore additional heating sources (like the UV background radiation and strong stellar radiation from starburst regions irradiating the low density wind) as well as the back-reaction of the cold clouds on the wind (\citealt{Schneider2018, Fielding2022ApJ}), a self-consistent investigation of the hot gas cooling is beyond the scope of this work. Furthermore, the assumptions that go into the \citetalias{Chevalier1985} model, such as spherical symmetry and continuous energy and mass injection, 
are extremely simplistic. Supernovae can be clustered or spread out in the galactic disk, and the injection of mass and energy by them can be in the form of bursty episodes (\citealt{schneider2020, Schneider2024ApJ, Mohapatra2025ApJ}), for which the outflow does not attain a smooth \citetalias{Chevalier1985} profile (\citealt{Sharma2014}). All of these effects can add additional complications, including driving a significant amount of turbulence in the wind (\citealt{Ghosh2025arxiv}) and transport of metals (\citealt{Sharda2024MNRAS}) that can potentially alter how the individual clouds or their population evolve in outflows. Therefore, different models of supernovae feedback (\citealt{Warren2024ApJ}) and turbulence give rise to a great deal of spatial and temporal variations in the background conditions (\citealt{mohapatra2019, Mohapatra2020MNRAS}) that different cold clouds encounter. These must be carefully taken into account before robust comparison can be made with observations and simulations (both large-scale outflow simulations of isolated galaxies; \citealt{schneider2017a, Schneider2018, schneider2020, Kim2018ApJ, vijayan2018, Vijayan2020ApJ, Steinwandel2024ApJ, Wang2024arXiv, Li2025ApJ, Vijayan2025MNRAS} or cosmological simulations of galaxy formation; \citealt{Dave2020MNRAS, Ramesh2024AAP}) of cold clouds and their host galactic winds. Although not directly related to the idealised simulations discussed here, similar caveats must be considered while testing the applicability of different sub-grid models of cold gas in cosmological simulations (\citealt{Weinberger2023MNRAS, Smith2024MNRASa, Smith2024MNRASb, Bennett2024arXiv, Butsky2024MNRAS, Das2024arXiv}) under various kinematic and thermodynamic conditions.  

We make qualitative parallels of our cloud-crushing simulation results with recent JWST observations of infrared emission line maps (from cold clouds and dust) in the M82 outflow by \citealt{Fisher2024MNRAS}. However, a detailed comparison would require cloud-crushing simulations to include the molecular phase ($\lesssim 10^4\ \rm K$) which brings in additional physics like self-gravity, self-shielding, dust sputtering and creation (\citealt{Kannan2021MNRAS,Girichidis2021MNRAS.505.1083G,farber2022MNRAS, Chen2024MNRAS}), which we have not taken into account.

In addition, the effects of thermal conduction and magnetic fields can be significant in determining the small-scale structure of cold clouds. Several recent works have demonstrated that magnetic fields suppress mixing-induced hydrodynamic instabilities, reducing the rate at which the cold mass grows in cloud-crushing simulations (\citealt{McCourt2015MNRAS, Kooij2021MNRAS}) -- a picture also supported by recent idealised simulations of radiative turbulent mixing layers with magnetic fields by \citealt{DasHK2024MNRAS} and wind-blown bubble simulations by \citealt{Lancaster2024ApJ} where suppression of hydrodynamic mixing gets manifested as a reduction in the fractal structure of turbulent boundary layers. Moreover, magnetic fields give rise to additional drag on the cloud due to draping by magnetic fields (\citealt{Dursi2008ApJ}) which is very sensitive to not just the magnetic field strength but more importantly to the orientation of the magnetic fields (\citealt{Pineda2024MNRAS, Kaul2025MNRAS}) with respect to the direction of motion of the clouds or the direction of spatial gradients in the background medium. Strength and orientation of magnetic fields might also be responsible for the generation of cold filamentary structures observed in galactic outflows (like M82) that bear striking similarity with solar prominences. In systems similar to cloud-crushing, magnetic fields have also been found to modulate the properties of ram-pressure stripped tails in jellyfish galaxies (\citealt{Muller2021Galax}). Therefore, magnetic fields and their associated effects can give rise to slow movement of long-lived clouds, leading to higher cold-mass seeding at any position in the bulk of the outflow, albeit at a lower rate. Furthermore, a distribution of multiple clouds (common in the CGM; cf. \citealt{Liang2020MNRAS, Bisht2024MNRAS, Hummels2024ApJ, Tan2024MNRAS}) can interact with each other (\citealt{Nguyen2024MNRAS}) and can also change the bulk properties of the wind by inducing precipitation (owing to thermal instability; \citealt{thompson2016, Schneider2018, Dannen2020ApJ, lochhaas_2021MNRAS}). Such effects can give rise to multiphase outflows with significantly different kinematic and thermodynamic properties, 
not considered in our work. Careful consideration of such effects and their relative importance 
and diverse outflow environments should be considered in the future. This would enable simulations to make realistic predictions of cloud properties in different regimes and bridge the discrepancies between predictions from small-scale simulations and results of large-scale simulations of individual galaxies or cosmological simulations. 

Other HST observations and analysis of cool clouds (bright in H$\alpha$) in the M82 outflow reveal that their properties (like density contrast, size, pressure and cooling times) are comparable to the clouds in cloud-crushing simulations (\citealt{Mutchler2007PASP, Thompson2024ARA&A}). However, a significant amount of work needs to be done to explain the detailed morphology of these observed structures. In particular, a large fraction of these observed clouds seem to be immersed in an inherently turbulent medium (\citealt{Lv2024ApJ}) and form arc-like structures (\citealt{Lopez2025arXiv}) that lack the distinct comet-like head-tail morphology predicted by cloud-crushing simulations. Moreover, even for predicting the emissivity or column densities of certain ions tracing the cold gas (like $\rm NV$) from simulations, one might need to consider not just extra-galactic sources, but additional sources of photo-ionising UV background radiation from stars, especially when the star formation rate is very high, like in M82 (\citealt{Villarruel2024arXiv}). Additional photo-heating can also be dynamically important in determining the cooling rates in the turbulent radiative mixing layers (TRMLs) as well as the multiphase nature of the galactic winds that host them (\citealt{Sarkar2022ApJ}). 

Despite all these caveats, the expansion of galactic outflows is expected to be generic and we expect our simulation results to be a qualitative indicator of cloud properties in multiphase outflow environments. 

\section{Summary \& Conclusion}
\label{sec:summary}
The formation and survival of cold clouds in galactic outflows are crucial for explaining the ubiquity of cold gas in the CGM. Observations suggest that cold clouds are accelerated farther from the galactic disk by powerful bursts of supernovae. The gas in an outflow is routinely found to have an expansion in the wind away from the centre of the galactic disk. In this work, we simulate the evolution of cold clouds seeded in an expanding wind, which is modelled by a steady-state \citealt{Chevalier1985} (\citetalias{Chevalier1985}) profile. The adiabatically expanding outflow shreds, stretches, and mixes the cold clouds seeded in the wind. In general, we find that
\begin{enumerate}
    \item Compared to the indefinite growth of cold mass seen in 'vanilla' cloud-crushing simulations (\citealt{bruggen2016, armillotta2017, gronke2018, Li2020, Sparre2020, Kanjilal2021, Abruzzo2023arXiv}), clouds with the same initial condition but exposed to a steady adiabatically expanding (\citetalias{Chevalier1985}) wind, instead of a uniform wind, undergo significantly lower cold mass growth (can be lower by almost an order of magnitude; see Fig. \ref{fig:cc85_cold-mass}). 

    \item As a cloud travels down the \citetalias{Chevalier1985} wind, it expands, becomes roughly isobaric relative to the local wind, and its properties depend on its instantaneous position in the wind (see Fig. \ref{fig:cloud_profile}). The cloud maintains `local' pressure equilibrium with the wind. This causes it to continuously lose its density contrast with the wind proportional to the distance travelled $d_{\rm cl}^{-2 \gamma}$ by it. Eventually, the cloud fades in the background wind and become indistinguishable from its surroundings (see Fig. \ref{fig:cloud_mass+dense}) -- a picture also supported by the evolution of the phase space distribution of cloud pressure and density (see Fig. \ref{fig:cloud_pdf}). 
    

    \item We find that the expansion associated with the \citetalias{Chevalier1985} wind results in clouds embedded in this wind to spread in directions orthogonal to the wind. Once entrained, the cloud expansion is proportional to the distance travelled by the cloud $d_{\rm cl}$ (see Fig. \ref{fig:cloud-spread}), and the cold gas mass $m_{\rm cl}$ grows proportional to the square of the distance they travel, i.e., $m_{\rm cl} \propto d_{\rm cl}^2$ (for entrained clouds; see Fig. \ref{fig:cloud-mass-distance}). This also explains the suppression of cold mass growth compared to the uniform wind `vanilla' cloud-crushing, where $m_{\rm cl} \propto d_{\rm cl}^3$ -- a significantly steeper cold mass growth with distance. 
    \item We compare our simulation results with the prediction of the phenomenological model of single-cloud evolution in galactic outflows proposed by \citealt{Fielding2022ApJ} (\citetalias{Fielding2022ApJ}; Section 3 in their paper). Upon comparing cloud mass growth with time with our simulations using the parameters chosen for their model (Section 3.2 in their paper), we 
     are unable to reproduce the suppression of cold mass growth in a \citetalias{Chevalier1985} wind compared to their `vanilla' cloud-crushing counterparts (uniform wind) for all our simulation parameters. The \citetalias{Fielding2022ApJ} model is very sensitive to the fractal dimension of the radiatively cooling turbulent boundary layer, which can be tweaked to match our simulation result (see Fig. \ref{fig:model_fielding_compare}). This highlights that the treatment of an individual cloud evolution in \citetalias{Fielding2022ApJ} (Section 3.1 in their paper) is generic and can potentially be improved to match simulation results in the future. However, this is not robust across different simulation parameters and further work is needed towards developing a sound physical interpretation of any specific choice of single-cloud evolution models. 
    \citetalias{Fielding2022ApJ} also developed a general framework for the profile of a multiphase outflow (equations in Fig. 3 \& Eq. 17 in \citetalias{Fielding2022ApJ}) with different energy and mass source terms in the wind. These source terms can come from any model of single-cloud evolution in galactic outflows. In particular, we introduce source terms to account for the cloud-wind interaction from the phenomenological cloud-crushing model that we discussed earlier. Using the general multiphase wind framework from \citetalias{Fielding2022ApJ}, we find that changes in the behaviour of individual clouds in the wind lead to substantial changes in the global properties of their host outflow (see Fig. \ref{fig:model_wind_fielding_compare}). This has important implications for enhancing the applicability of the \citetalias{Fielding2022ApJ} model as a sub-grid prescription across the diverse range of conditions typically encountered in cosmological simulations -- one of the primary motivations for developing such models. 
    
    \item We estimate the surface brightness of the clouds in our expanding cloud-crushing simulations (see Fig. \ref{fig:emission_slice}). We find that, just like in the `vanilla' cloud-crushing, most of the emission is confined to the radiatively cooling mixing layers. However, the mixing layers gradually become lower in density as the cloud encounters a lower-density wind in an adiabatically expanding outflow. Therefore, the surface brightness of the cloud continues to drop and they gradually become indistinguishable from the background. In addition, the surface brightness of the cold cloud shows a gradient that gradually dims downstream of the wind. This feature bears a striking resemblance to the recent JWST observation of the cold-cloud emission map in the outflow of M82 (\citealt{Fisher2024MNRAS}).
\end{enumerate}

We hope that this study will be a stepping stone in our effort to bridge the gap between idealised simulations and the complex, multiphase nature of real galactic winds. A great deal of work is still needed to generalise the behaviour of cold clouds in diverse 
environments and constrain their collective effect on large-scale properties of their host outflows so that simulations and models can faithfully reproduce current and upcoming observations. 

\section*{Acknowledgements}
AD acknowledges support from the Max-Planck-Gesellschaft. AD acknowledges his colleague Ayan Ghosh for helping with the cartoon illustration used in this paper. AD also expresses his gratitude to his colleague Ritali Ghosh for help with visualisations and also for several hours of useful discussions. AD also acknowledges Arif Babul, Kartick Sarkar, Evan Scannapieco, Matthew Abruzzo, Volker Springel, Evan Schneider, Christoph Pfrommer, Tiago Costa, Filippo Fraternali, Greg Bryan, Rainer Weinberger, Ruediger Pakmor, Matthew Smith, Sunmyon Chon, Drummond Fielding, Simon White, and our anonymous referee for their useful comments and discussions. The authors acknowledge support from the Supercomputer Education and Research centre (SERC) at the Indian Institute of Science (IISc), Bangalore, where some of the production runs of our simulations were carried out. AD would like to express his gratitude to the super-responsive and helpful people at \href{https://discourse.paraview.org/}{Paraview forum} for all the technical support needed for some of the visualisations and simulation data analysis. Research of AD as a graduate student, when this work initiated, was supported by the Prime Minister's Research Fellowship (PMRF)  from the Ministry of Education (MoE), Govt. of India. AD acknowledges the support from the Max Planck Institute for Astrophysics, Garching, for accepting him as a visiting student and providing access to their Freya \& Orion supercomputing clusters when this work was initiated. Several analysis scripts in our work make use of {\tt Numpy} (\citealt{Harris}), {\tt Matplotlib} (\citealt{Hunter}), and {\tt Scipy} (\citealt{2020SciPy-NMeth}) whose communities we thank for continued development and support. 
MG thanks the Max Planck Society for support through the Max Planck Research Group, and the European Union for support through ERC-2024-STG 101165038 (ReMMU).

\section{Data Availability}
All the relevant data associated with this article will be shared on a reasonable request to the authors. Most of the relevant analysis, plotting and visualisation codes are hosted on \texttt{GitHub} for public access. Follow \url{https://github.com/dutta-alankar/CCinCC85/} for the cloud-crushing simulation setup in a spherical \citetalias{Chevalier1985} wind and also the \texttt{exp\_box} branch in \url{https://github.com/dutta-alankar/cloud-crushing_PLUTO/tree/exp_box}, where a re-implementation of the \citetalias{gronke2020} expanding box setup has been hosted. Several videos accompanying this work are hosted in the following playlist: \url{https://www.youtube.com/playlist?list=PL35YR13Ha_K1P721H_ub_13fmsfp6EAA2} and follow \url{https://github.com/dutta-alankar/FB22-model-comparison.git} for other relevant codes related to \citetalias{Fielding2022ApJ} model comparison. 

\bibliographystyle{mnras}
\bibliography{references} 

\appendix
\section{Threshold size for cloud mass growth in a uniform wind}
\label{app:size_thres}
\begin{figure}
   \centering
	\includesvg[width=\columnwidth]{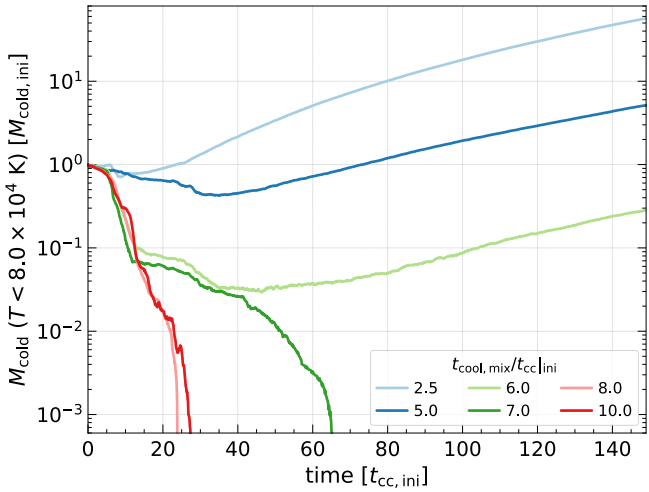}
    \caption{The temporal evolution of cold and dense gas mass (in units of initial mass) in our `vanilla` cloud-crushing simulations (in spherical coordinates) for clouds of different sizes. The background wind is uniform and plane-parallel. In the simulations, cold and dense gas is identified as gas cells with a 
    temperature lower than $2 T_{\rm cl} = 8 \times 10^4 \rm \ K$. The set of constant parameters chosen for these simulations is $(\mathcal{M}, \chi) = (1.496, 100)$. The parameter $t_{\rm cool, mix}/t_{\rm cc} \propto R_{\rm thres}/R_{\rm cl}$ is set to different values by changing the cloud size. Time is normalised by the cloud-crushing time $t_{\rm cc}=\sqrt{\chi} R_{\rm cl}/v_{\rm wind}$. This plot shows that the transition from destruction to the growth regime takes place at $t_{\rm cool, mix}/t_{\rm cc} \lesssim 6$ giving the dimensionless parameter $\delta \approx 6$ in Eq. \ref{eq:R_thres}.}
    \label{fig:vanilla_turnover}
\end{figure}

We carry out cloud-crushing simulation in a uniform plane parallel wind in spherical coordinates (`vanilla' cloud-crushing) for different cloud sizes to numerically evaluate the dimensionless parameter $\delta$ in Eq. \ref{eq:R_thres}. Fig. \ref{fig:vanilla_turnover} shows that the transition from growing clouds to destroyed clouds (and vice versa) occurs at $t_{\rm cool,mix}/t_{\rm cc}|_{\rm ini} \approx R_{\rm thres}/R_{\rm cl} \approx 6$, where the symbols have their usual meaning. We have used this numerically determined threshold to choose different cloud sizes in all the simulations presented in this paper. Such a calibration is necessary because of slight differences in the cooling function and other parameters in various existing radiative cloud-crushing simulations.


\section{Cloud surface \& scaling relations}
\label{app:scaling-relations}

\begin{figure}
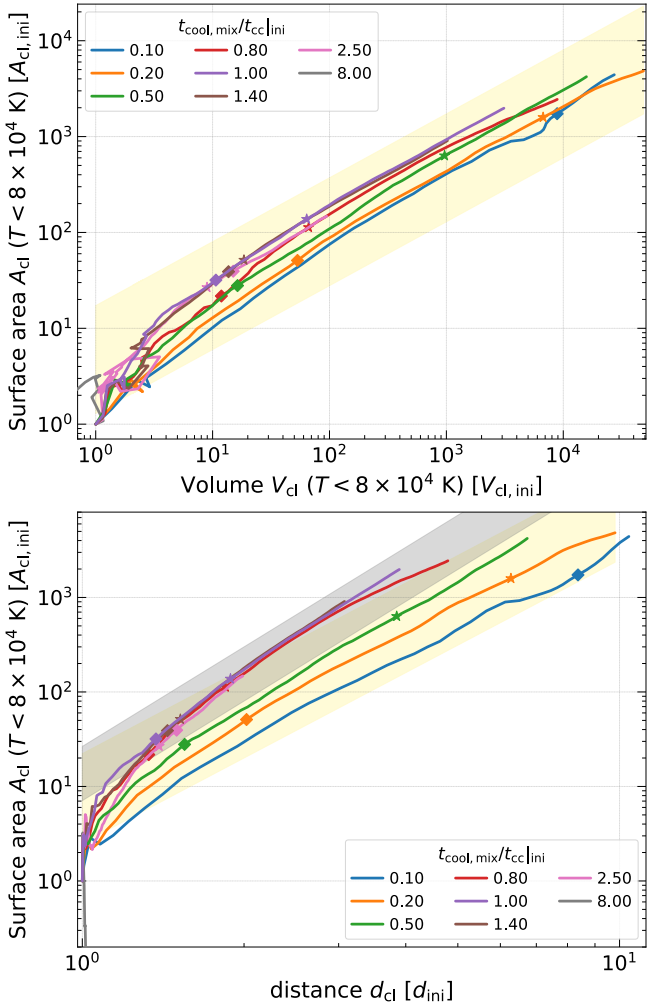

    \centering
    \includesvg[width=\columnwidth]{figures/results-analysis/new-analysis/area-cold-trunc_volume.svg}
    \includesvg[width=\columnwidth]{figures/results-analysis/new-analysis/area-cold-trunc_distance.svg}
    \caption{Evolution of the surface area (in units of the initial cloud surface area) of the radiatively cooling turbulent boundary layer, identified as the bounding surface of all gas below a temperature of $2 T_{\rm cl} = 8\times 10^4\ \rm K$, as a function of cloud volume (\textbf{upper} panel) the distance (\textbf{lower} panel) travelled by the centre of mass of the cloud (normalised by the initial cloud distance from the centre of the \citetalias{Chevalier1985} wind). Along each of the curves, the diamonds and stars have the same meaning as in Fig. \ref{fig:cloud-spread}. The surface area of the clouds, after entrainment (beyond the star marks), follows $A_{\rm cl}\propto V_{\rm cl}^{2/3}$ as indicated by the \textbf{yellow} band in the upper panel. However, prior to entrainment, the area of the boundary layer (formed by shear induced instabilities) increases much steeply with volume, as indicated by the \textbf{cyan} band in the upper panel which marks $A_{\rm cl}\propto V_{\rm cl}^{5/6}$, due to radiative cooling which gives a fractal nature to the cloud surface (\citealt{Fielding_2020}). Although the differences in the two slopes in the area-volume relation is small, there is a slight flattening of the slope beyond the star marks in the upper panel. As discussed in Section \ref{subsubsec:GO_exp-box}, 
    one gets $A_{\rm cl}\propto d_{\rm cl}^{3 \gamma -2}\sim d_{\rm cl}^3; \gamma=5/3$, 
    which is approximately followed by the largest clouds (\textbf{yellow} band in the lower panel) that do not experience as much orthogonal expansion compared to their initial size as the smaller clouds (cf. Fig. \ref{fig:cloud-spread}). While for small clouds, this relative expansion is important and $m_{\rm cl}\propto d_{\rm cl}^{23/12}\approx d_{\rm cl}^2$ (cf. Fig. \ref{fig:cloud-mass-distance}), which implies $A_{\rm cl}\propto d_{\rm cl}^{3(\gamma - 1/2)}\sim d_{\rm cl}^{3.5}; \gamma=5/3$ as indicated by the \textbf{cyan} band in the lower panel, albeit the difference in the slopes between the yellow and the cyan bands is small (compare with Fig. \ref{fig:cloud-mass-distance} where the difference in the slopes is large).}
    \label{fig:cloud_scaling}
\end{figure}

\begin{figure}
    \centering
    \includesvg[width=\columnwidth]{figures/results-analysis/new-analysis/trml/length-cold-trunc_distance.svg}
    \caption{Evolution of the extent of a cloud $R_{\rm cl,\parallel}$ in the direction of the wind (in units of the initial cloud radius) as a function of the distance travelled by the centre of mass of the cloud (normalised by the initial cloud distance from the centre of the \citetalias{Chevalier1985} wind). The length of the cloud has been calculated as the difference between the $60$ and $5$ percentile values of the position of the cloud material along the axis passing through the centre of the cloud parallel to the wind. Along each of the curves, the diamonds and stars have the same meaning as in Fig. \ref{fig:cloud-spread}. For clouds that are entrained (beyond the stars), the length of the cloud in the wind direction approximately saturates to a constant. On the other hand, the largest clouds which do not get entrained in their journey down the wind keep growing in their extent in the wind direction proportional to the distance they travel as indicated by the \textbf{yellow} band. This evolution can be compared with the orthogonal extent of the cloud (cf. Fig. \ref{fig:cloud-spread}) which shows a different scaling for the two types of clouds categorized according to their size.}
    \label{fig:cloud-length}
\end{figure}

We 
investigate how the surface area scales with the volume of the cold gas in a \citetalias{Chevalier1985} wind. To evaluate the cloud surface area, we first identify the all the cells in the simulation domain with temperature less than $8\times 10^4\ \rm K$ within a region where the \citetalias{Chevalier1985} wind temperature is greater than $9\times 10^4\ \rm K$ using the \texttt{VTK} `Threshold' filter that comes as a part of the \texttt{Paraview} data analysis/visualisation software. We then isolate the outer surface of all these `cold' cells by selecting only the cell faces that belong to one cell and are not shared by any other cells using the `Extract Surface' \texttt{VTK} filter also available in \texttt{Paraview}. Additionally to calculate the inflow velocity, we interpolate the fields to the face centres of the faces tessellating this outer surface and calculate the surface area of every face as well as their respective surface normals.\footnote{Sanity checks on the initial spherical cloud using these \texttt{VTK} filters give consistent results for the cloud volume and surface area.} 

Once the cloud is sufficiently entrained, from geometric considerations, we propose that the cloud surface area $A_{\rm cl}$ to follow $A_{\rm cl} \propto V_{\rm cl}^{2/3}$, where $V_{\rm cl}$ is the volume of the cold cloud. This scaling relation is highlighted in a yellow band in the upper panel of Fig. \ref{fig:cloud_scaling}, which shows that as clouds start to grow 
with strong radiative cooling, their surface area follows the simple geometric (non-fractal) scaling relation. However, initially, the cloud area increases much rapidly with volume. For the largest clouds that do not get sufficiently entrained the surface area is dominated by shear-driven turbulent mixing which radiatively cools and has a fractal nature (\citealt{Fielding_2020}), where $A_{\rm cl}\propto V_{\rm cl}^{5/6}$ (fractal dimension of $2.5$) -- a steeper power law dependence between surface area and cloud volume. Although the differences in the slope are small and there is a lot of scatter in the data, a gradual flattening of the scaling relation post entrainment between surface area and cloud volume is evident. Large clouds that do not get entrained also behave differently in terms of their size along the wind which keeps growing with the distance they travel compared to the smaller clouds, whose extent along the wind saturates (cf. Fig. \ref{fig:cloud-length}). Further investigation is needed to determine the influence of simulation resolution on the fractal surface geometry of clouds and its potential consequences for the emergent scaling relation. 

\begin{figure}
    \centering
    \includegraphics[width=\columnwidth]{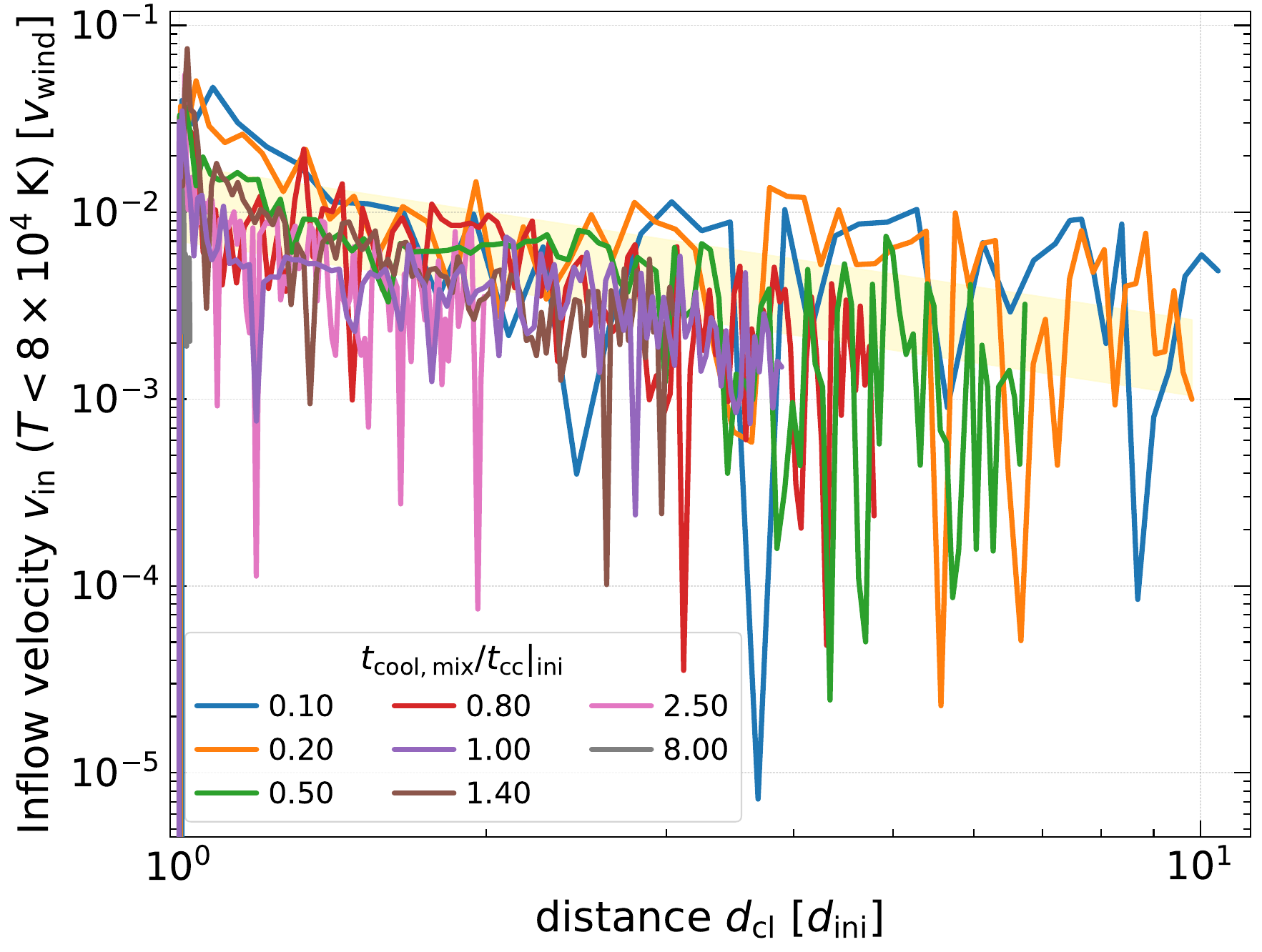}
    \caption{Evolution of the inflow velocity (in units of the wind speed at the initial cloud position) across the surface of the radiatively cooling turbulent boundary layer, identified in the same way as in Fig. \ref{fig:cloud_scaling}, as a function of the distance travelled by its centre of mass (normalised by its initial distance from the centre of the \citetalias{Chevalier1985} wind). The inflow velocity $v_{\rm in}$ has been calculated as the average mass-weighted velocity along the local normals used to tessellate the iso-surface of temperature at $8 \times 10^4 \ \rm K$ bounding the cloud. Albeit with a lot of scatter, $v_{\rm in}$ scales with the position of the centre-of-mass $d$ of the clouds as $v_{\rm in} \propto d_{\rm cl}^{-\gamma /2} \sim d_{\rm cl}^{-5/6}; \gamma=5/3$, as indicated by the yellow band. 
    }
    \label{fig:cloud_vin}
\end{figure}

We also calculate the (mass) average inflow velocity $v_{\rm in}$ across the surface of the cold gas at different instants as a function of the centre-of-mass of the cloud $d_{\rm cl}$ with respect to the initial position. To calculate this inflow velocity, we identify the iso-surface bounding all gas below $2 T_{\rm cl} = 8\times 10^4 \ \rm K$ and calculate the local normal for every tile in the tessellation of the iso-surface. The inflow velocity $v_{\rm in}$ is the mass averaged velocity along the local normal for all the tiles bounding the iso-surface (see \citealt{Abruzzo2024ApJ} for details). This inflow velocity $v_{\rm in}$, although employing different definitions, was previously shown by \citealt{Fielding_2020, gronke2020, Tan2020} to follow the scaling relation $v_{\rm in} \propto t_{\rm cool,cl}^{-1/4}$ in the strong cooling regime, where $t_{\rm cool,cl}$ is the cooling time of the cold gas -- the fastest cooling phase. 
Using Eq. \ref{eq:cloud_density_scaling} and the arguments presented in Section \ref{subsec:cloud-prop} (that clouds are at $T_{\rm cl}$ and locally isobaric), we get $t_{\rm cool,cl} \propto d_{\rm cl}^{2 \gamma}$. Therefore, the inflow velocity $v_{\rm in}$ follows $v_{\rm in} \propto d_{\rm cl}^{-\gamma /2}$, as indicated by the yellow band in Fig. \ref{fig:cloud_vin}. There is a significant scatter in the simulation data and this trend is only crudely followed by the clouds. 
The dependence of $R_{\rm cl, \perp}$ ($\sim d$ (see Section \ref{subsubsec:GO_exp-box}) introduces a weak additional dependence of $v_{\rm in}$ ($\sim R_{\rm cl, \perp}^{1/4} t_{\rm cool,cl}^{-1/4} \sim d^{1/4-\gamma/2}\approx d^{-\gamma/2}; \gamma=5/3$) on the distance travelled by the cloud, which is even shallower but 
difficult to tell apart due to the noisy nature of $v_{\rm in}$. 

\section{Single-cloud model from FB22}
\label{app:FB22_model}

\begin{figure}
    \centering
    \includesvg[width=\columnwidth]{figures/model-fielding/distance-time_FB22vsOur-cloud.svg}
    \caption{Comparison of the distance travelled by the centre of mass of the cloud from our cloud-crushing simulation with that predicted from the \citetalias{Fielding2022ApJ} single-cloud model for `vanilla' cloud-crushing (in uniform wind) and for cloud-crushing in a \citetalias{Chevalier1985} wind (see Section \ref{subsubsec:DF-single_cloud} for details). The \textbf{dashed} lines are the simulation results for `vanilla' cloud-crushing (in \textbf{yellow}) and cloud-crushing in a \citetalias{Chevalier1985} background (in \textbf{teal}). As a specific example which would clearly demonstrate departure from model prediction, we choose the simulation result with $t_{\rm cool, mix}/t_{\rm cc}|_{\rm ini} = 0.8$ and $\mathcal{M}=1.496$ (cf. Table \ref{tab:CCinCC85_params}; not the fiducial parameter set). The corresponding model predictions are shown with \textbf{solid} lines in the same colour as their dashed counterparts. Although Fig. \ref{fig:model_fielding_compare} shows that the mass of clouds has a large variation between the original parameter set from \citetalias{Fielding2022ApJ} and our set of parameters (described in Section \ref{subsubsec:DF-single_cloud}), we find that the distance the cloud travels with time differs from our parameters (\textbf{solid dark} lines) and the original one (\textbf{thick, solid light}-coloured curves) by less than $1\%$.} 
    \label{fig:model_fielding_compare_distance}
\end{figure}

\citealt{Fielding2022ApJ} (\citetalias{Fielding2022ApJ} hereafter) introduced the following set of ordinary differential equations\footnote{We ignore gravity and assume the same metallicity for both the cloud and the wind and present these equations in an equivalent but modified form.} to model how an individual cloud evolves in a hot wind.

\begin{align}
\label{eq:single_cloud_vanilla}
\renewcommand{\arraystretch}{2.0}
\rm \frac{d}{dt} \begin{bmatrix} d_{\rm cl} \\ v_{\rm cl} \\ M_{\rm cl} \end{bmatrix} = \begin{bmatrix} v_{\rm cl} \\  v_{\rm rel} \left(\frac{\dot{M}_{\rm cl, grow}}{M_{\rm cl}}\right) + \frac{3}{8} C_{\rm drag} \frac{v_{\rm rel}^2}{\chi R_{\rm cl}} \\ \dot{M}_{\rm cl, 0} \left(\frac{v_{\rm in}}{f_{\rm turb} |v_{\rm rel}|} - \frac{f_{\rm mix}}{f_{\rm cool}}\right) \end{bmatrix},
\end{align}
where $v_{\rm in}$ is the inflow velocity across the turbulent boundary layer, where the ansatz $v_{\rm in} = f_{\rm turb} |v_{\rm rel}| \xi^\alpha$ is used in the original \citetalias{Fielding2022ApJ} single-cloud model. We modify $v_{\rm in}$ and propose a `modified' single-cloud model as represented in Eq. \ref{eq:model_modify}. This sets a floor value to $v_{\rm in}$ such that the model prediction can better match the trend in mass evolution of cold clouds from the simulation data (cf. Section \ref{subsubsec:DF-single_cloud}). 

In this set of equations, $d_{\rm cl}$ is the distance travelled by a cloud, and $v_{\rm cl}$ and $M_{\rm cl}$ are its velocity and mass respectively. The relative velocity between the cloud and the wind is denoted by $v_{\rm rel} = v_{\rm wind} - v_{\rm cl}$, where $v_{\rm wind}$ is the wind velocity. The radius of the cloud $R_{\rm cl}$ is a time dependent quantity and is defined as $R_{\rm cl} = (3 M_{\rm cl}/(4\pi \chi \rho_{\rm wind}))^{1/3}$, where $\rho_{\rm wind}$ and $\chi$ are wind density and the density contrast between the cloud and the background wind, respectively. The cloud is assumed to be always in an instantaneous pressure equilibrium with the background wind. $\dot{M}_{\rm cl, grow}$ is the rate at which cloud mass grows due to radiative cooling and is defined as $\dot{M}_{\rm cl, grow} = \dot{M}_{\rm cl, 0} \xi^\alpha$, where $\dot{M}_{\rm cl, 0} = 3 f_{\rm turb} f_{\rm cool} M_{\rm cl} |v_{\rm rel}|/\left(\sqrt{\chi} R_{\rm cl}\right)$, as prescribed in \citealt{Fielding2022ApJ}. $C_{\rm drag}$ is the drag coefficient due to the relative velocity between the cloud and the wind. This relative velocity also generates turbulent instabilities (like Kelvin-Helmholtz and Rayleigh-Taylor), resulting in the development of turbulent velocity 
with $f_{\rm turb}$ being the proportionality constant (similar notation was previously used by \citealt{Mandelkar2019MNRAS, Mandelker2020MNRAS}, 
such that the mass loss rate from the cloud to the wind $\dot{M}_{\rm cl, loss} = 3 f_{\rm mix} f_{\rm turb} M_{\rm cl} |v_{\rm rel}|/(\sqrt{\chi} R_{\rm cl})$. The area available for cooling $A_{\rm cool}$ ($\propto R_{\rm cl} v_{\rm rel} t_{\rm cc}$) increases as the cloud stretches and mixes in the wind and forms a wake in its path defined as $A_{\rm cool} = f_{\rm cool} 4 \pi R_{\rm cl}^2 \sqrt{\chi}$, where the cloud-crushing time $t_{\rm cc} = \sqrt{\chi} R_{\rm cl}/|v_{\rm rel}|$ is the characteristic timescale for the growth of hydrodynamic instabilities that mix a cloud and $f_{\rm cool}$ represents a factor that uniformly stretches the `effective' area of the entire cloud that is available for cooling. Here $\xi = R_{\rm cl}/ (f_{\rm turb} |v_{\rm rel}| t_{\rm cool})$, where we consider $t_{\rm cool}$ as the isobaric cooling time of the mixed gas, which is assumed to be at the temperature of $\sqrt{\chi}$ times the cloud temperature $T_{\rm cl}$.\footnote{$t_{\rm cool} = \gamma/(\gamma -1) \left(p_{\rm wind}/((\sqrt{\chi} \rho_{\rm wind}X_H/m_p)^2 \Lambda[\sqrt{\chi T_{\rm cl}}] )\right)$, where $p_{\rm wind}$ is the pressure in the wind (as well as the cloud), $X_H$ is the hydrogen density fraction, $m_p$ is proton mass, $\Lambda$ is the radiative cooling function, and $\gamma$ is the adiabatic index.} This factor $\xi$ bears similarity to the Damk\"ohler number from turbulent combustion literature, introduced in the context of cloud-crushing by \citealt{Tan2020}. Clouds where radiative cooling is dominant correspond to large values of $\xi$ (i.e., small $t_{\rm cool}$ or large $R_{\rm cl}$).

In Fig. \ref{fig:model_fielding_compare_distance}, we show a comparison of the model prediction of the distance travelled by the cloud using \citetalias{Fielding2022ApJ} single-cloud model (with the original parameters as well as our modified one) with simulation results (similar to Fig. \ref{fig:model_fielding_compare}). Interestingly, we find that unlike evolution of cloud mass, the distance travelled by the cloud predicted by the model is almost unaffected (variation $\lesssim 1\%$) on changing from the original \citetalias{Fielding2022ApJ} parameters to our. Both models predict that a cloud would travel a greater distance downstream in an expanding \citetalias{Chevalier1985} wind compared to a uniform wind -- consistent with the trend found in the simulations. It is worth noting that we fit our model prescription to only match the mass evolution, while the distance evolution comes as a natural consequence of the chosen model. In addition, cloud mass $m_{\rm cl}\propto d_{\rm cl}^3$ in a uniform wind compared to $m_{\rm cl}\propto d_{\rm cl}^2$ (or $m_{\rm cl}\propto d_{\rm cl}$; cf. Fig. \ref{fig:cloud-mass-distance}) in a \citetalias{Chevalier1985} wind causing clouds in uniform wind to grow more in mass even though they travel lesser distance downstream compared to clouds in an expanding wind.

\section{Clouds in an anisotropically expanding box}
\label{app:exp_box}

\begin{figure*}
    \centering
    \includegraphics[width=\textwidth]{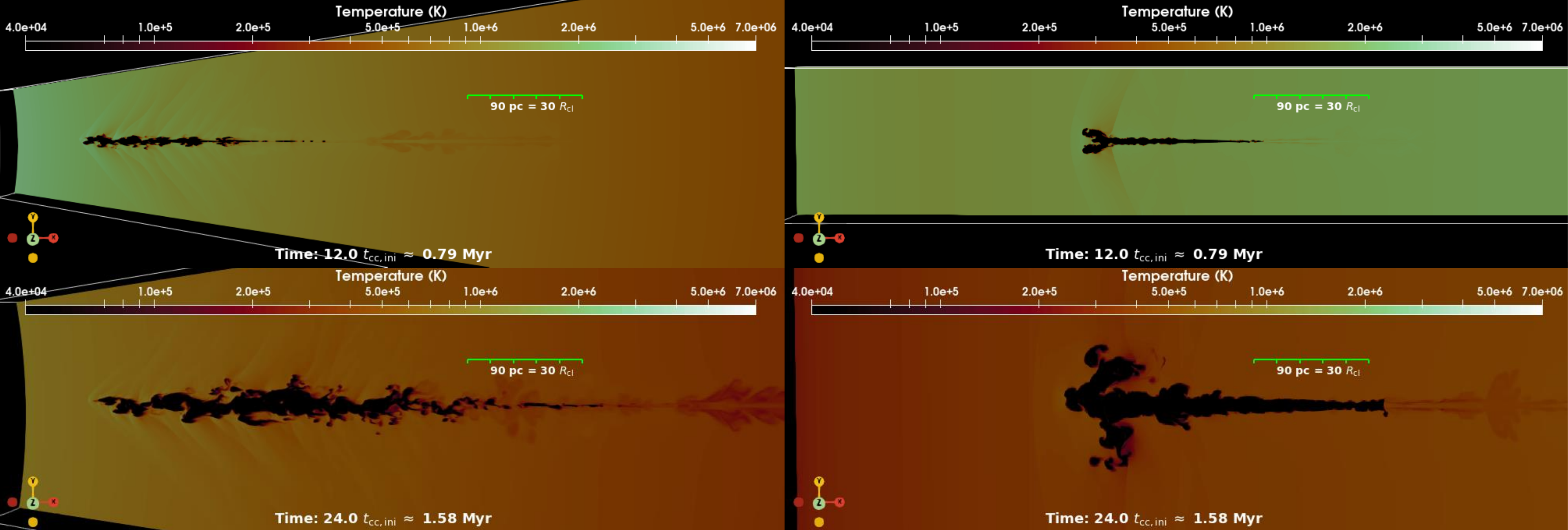}
    \caption{Slice plots of the temperature field comparing cloud-crushing simulations in a \citetalias{Chevalier1985} wind from our simulation (\textbf{left} panel) with the anisotropically expanding box model from \citealt{gronke2020} (\textbf{right} panel) re-simulated here using \texttt{PLUTO} MHD code (\citealt{PLUTO}). The anisotropically expanding box approximates the cloud dynamics in a \citetalias{Chevalier1985} wind using a uniform expansion of the entire simulation domain in all directions orthogonal to the direction of the wind as the cloud moves downstream (no expansion in the wind direction). Although the cold gas formed are comparable (see Section \ref{subsubsec:GO_exp-box}), the cloud morphologies are significantly different between the two simulations. In our simulations, a cloud experiences spatially different wind conditions at any particular instant, in contrast to a homogeneous wind in \citealt{gronke2020}. Except for the region around the head, which is more spread out laterally in \citetalias{gronke2020} setup, the cloud forms cometary tails, which are longer in our simulations, with comparable expansion orthogonal to the wind in both simulations. Follow the \href{https://youtu.be/HOBN_ft66as}{link here} to see a video of the full evolution.}
    \label{fig:compareGO20-slice-plot}
\end{figure*}
\citealt{gronke2020} (\citetalias{gronke2020} hereafter) employed a model of a homogeneous anisotropically expanding computational domain to approximate the evolution of a cloud in a \citetalias{Chevalier1985} wind. The 
cloud is assumed to be much smaller than its from the wind centre. 
In addition, the entire cloud is assumed to experience a uniform wind condition, which only changes over time as the cloud moves down the wind. 
We re-simulated the cloud-crushing problem with the anisotropic expansion model used by \citetalias{gronke2020} in \texttt{PLUTO} for our fiducial parameters and ran it up to $30\ t_{\rm cc, ini}$. The extent of the simulation domain is $250 R_{\rm cl}$ in the direction of the wind and $22 R_{\rm cl}$ in directions orthogonal to it. Along with that, we utilise the frame boost method from \citealt{Kanjilal2021} to track the cloud. 


Comparing the evolution of clouds from our simulation and from \citetalias{gronke2020}, we find that the simplifying assumptions made by \citetalias{gronke2020} lead to significant differences in the morphology of the cloud formed in the wind (see Fig. \ref{fig:ourvsGO20_sim}). We use identical initial conditions and resolution ($R_{\rm cl}/d_{\rm cell}|_{\rm ini} = 16$) in both the simulation setups. 
As illustrated in Fig. \ref{fig:compareGO20-slice-plot}, we find that clouds in a \citetalias{Chevalier1985} wind in our simulations form much longer cometary tails. Although the mass formed (cf. Section \ref{subsubsec:GO_exp-box} \& Fig. \ref{fig:ourvsGO20_sim}) and the expansion of the tail of the cloud orthogonal to the wind between the simulation setups are comparable, there exists a difference in the estimate of the distance moved by the cloud and possibly in the rate of mass growth of clouds in the cooling regime. 

\section{Convergence}
\label{app:convergence}

\begin{figure*}
    \centering
    \includegraphics[width=\textwidth]{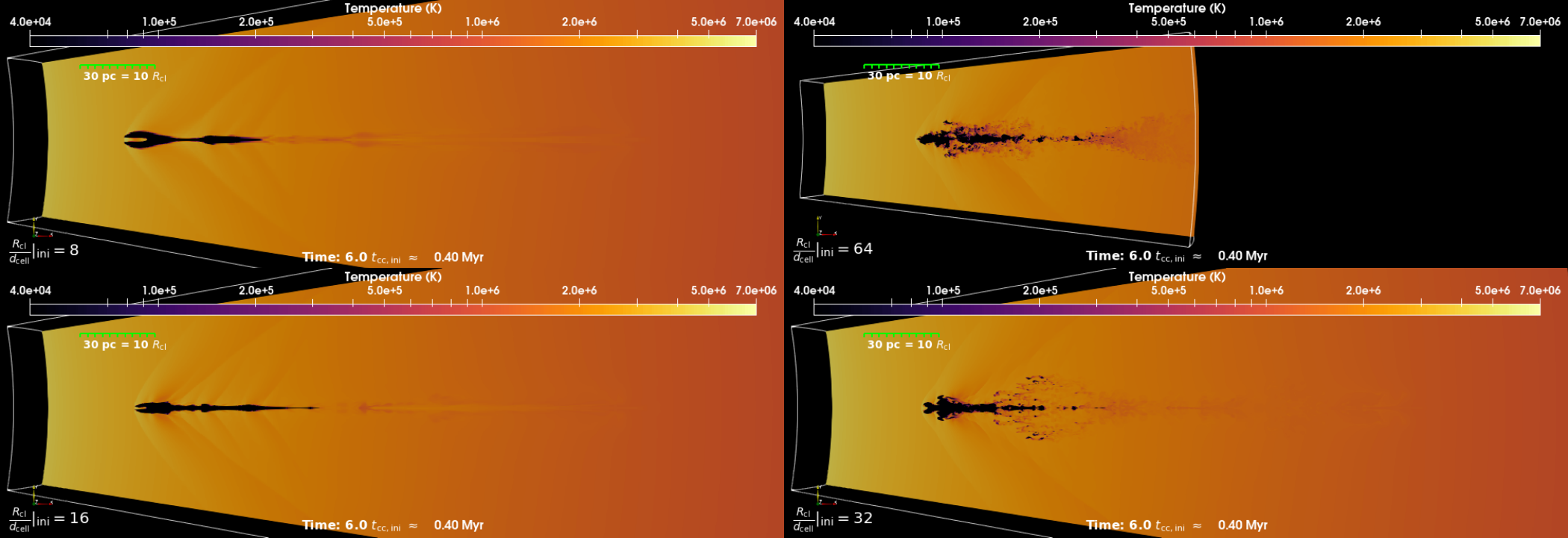}
    \caption{Slice plots of the temperature field comparing cloud-crushing simulations in a \citetalias{Chevalier1985} wind at four different resolutions -- $R_{\rm cl}/d_{\rm cell}|_{\rm ini} = (8, 16, 32, 64)$ at time $t/t_{\rm cc, ini} = 6$. As expected, on increasing resolution, mixing at smaller scales is better resolved. For these simulations, a domain with an extent of about $250\ R_{\rm cl}$ in the radial direction and $22\ R_{\rm cl}$ at the initial position of the cloud in the orthogonal direction had been considered, with the exception of the simulation with the highest resolution of $R_{\rm cl}/d_{\rm cell}|_{\rm ini} = 64$, where the radial extent of the box has been reduced to $100\ R_{\rm cl}$ due to computational limitations. Follow the \href{https://youtu.be/ehb4QDGVPA8}{link here} to see a video of the full evolution.}
    \label{fig:convergence-slice-plot}
\end{figure*}

 \begin{figure*}
    \centering
    \includesvg[width=\textwidth]{figures/results-analysis/new-analysis/convergence-all-3p-res.svg}
    \caption{Comparison of the evolution of cold cloud mass $m_{\rm cl}(t)$ (\textbf{upper left panel}), its growth rate $\dot{m}_{\rm cl}(t)$ (\textbf{lower left panel}) and distance travelled by the centre of mass of the cold gas $d_{\rm cl}(t)$ (\textbf{right panel}) in \citetalias{Chevalier1985} wind from our simulation at different resolutions using the fiducial parameter  $t_{\rm cool, mix}/t_{\rm cc}|_{\rm ini} = 0.2$ (cf. Table \ref{tab:CCinCC85_params}). The resolution marked in the legend denotes the number of cells by which the cloud radius is resolved at the initial location of the cloud; a higher value corresponds to a higher resolution. The highest resolution curve with $R_{\rm cl}/d_{\rm cell}|_{\rm ini} = 64$ has not been considered beyond $6\ t_{\rm cc,ini}$ as a significant amount of cold gas leaves the computational domain due to a smaller domain size compared to the simulations at lower resolutions. 
    The highest resolution curve $R_{\rm cl}/d_{\rm cell}|_{\rm ini} = 64$ is confined within the $R_{\rm cl}/d_{\rm cell}|_{\rm ini} = 16$ and $R_{\rm cl}/d_{\rm cell}|_{\rm ini} = 32$ curves, indicating that the cold mass growth is converged. 
    The mass estimated at different resolutions agree with one another within a factor of a few, and their difference diminishes as the simulation resolution increases (compare, for example, the difference between the $R_{\rm cl}/d_{\rm cell}|_{\rm ini} = 32$ and $R_{\rm cl}/d_{\rm cell}|_{\rm ini} = 64$) indicating convergence in cloud mass. In fact, the mass growth rates $\dot{m}_{\rm cl}$ are even better converged across different resolutions. Increased mixing at high resolution causes increased momentum transfer from the hot to the cold phase causing the cloud to move faster in the wind at higher resolution.
    }
    \label{fig:convergence-plot}
\end{figure*}

\begin{figure*}
    \centering
    \includesvg[width=\textwidth]{figures/results-analysis/new-analysis/snap-profile-c100,m1.496,T4e4,t0.20,r70.671-res16.svg}
    \caption{Evolution of the cloud profile at a resolution $R_{\rm cl}/d_{\rm cell}|_{\rm ini}=16$; one level higher than our fiducial resolution with the same fiducial parameters (cf. Table \ref{tab:CCinCC85_params}) as in Fig. \ref{fig:cloud_profile}. The clouds maintain local pressure equilibrium with the background wind and falls steeply in their density contrast with respect to the background. In terms of these general features, the cloud properties remain unchanged across resolution demonstrating the robustness of our simulation results and the analytic scaling relations presented in Section \ref{sec:results}. Although the cloud pressure profiles track the local wind pressure, it is always slightly under-pressurized with respect to the ambient medium which is sustained even after increasing resolution -- tentative evidence of a local cooling flow around these clouds (\citealt{Dutta2022MNRAS}), rather than an artifact of low resolution.}
    \label{fig:cloud_profile-high_res}
\end{figure*}

We demonstrate results from our convergence test on the evolution of the cold cloud mass in our simulation for the fiducial parameter, $t_{\rm cool, mix}/t_{\rm cc}|_{\rm ini} = 0.2$ (cf. Table \ref{tab:CCinCC85_params}). For all simulations in our convergence test, a domain with an extent of about $250\ R_{\rm cl}$ in the radial direction and $22\ R_{\rm cl}$ at the initial position of the cloud in the orthogonal directions had been considered. The only exception is the simulation with the highest resolution of $R_{\rm cl}/d_{\rm cell}|_{\rm ini} = 64$, where the simulation domain is reduced to $100\ R_{\rm cl}$ in the radial direction due to computational limitations. We find that with higher resolution (increasing the value of $R_{\rm cl}/d_{\rm cell}|_{\rm ini}$ at the initial position of the cloud), the cold mass in the cloud is mostly higher -- except for the $R_{\rm cl}/d_{\rm cell}|_{\rm ini}=32$ and $64$ simulations where $m_{\rm cl}(t)$ seems to agree.
Also with the other simulations, the results across different resolutions agree within a factor of a few in $m_{\rm cl}(t)$ and $\dot{m}_{\rm cl}(t)$ (see left panels in Fig.~\ref{fig:convergence-plot}) in the time window of our analysis. At high resolutions, the evolution of the cold mass tends to converge, although the clouds differ considerably in their morphology. The qualitative picture of cloud growth due to mixing and strong radiative cooling is well captured across all resolutions.

The convergence of cold gas mass growth is not surprising -- and has been studied extensively in prior cloud-crushing \citep[e.g.][]{gronke2020} and high resolution shearing layer simulations \citep[e.g.][]{Tan2020}. Although we do not explicitly model thermal conduction, therefore, the thermal diffusion across a cold gas front is purely numerical in our simulations. Nevertheless, this is sub-dominant compared to turbulent diffusion. In other words, the `bottleneck' in the mass transfer is due to the turbulent mixing and not due to the (rapid) cooling. This implies that resolving the mixing at the largest scales ($\sim R_{\rm cl}$) is sufficient to converge in the mass growth rate $\dot{m}_{\rm cl}$. This has been explicitly checked with the above mentioned high-resolution simulation which partially also included thermal conduction \citep{Tan2020}. 
The small increase in cold mass $m_{\rm cl}$ with increasing resolution is unsurprising as well, because we are able to capture the formation and evolution of smaller cold clumps when the resolution is higher. Thus, the overall cold gas surface area formed by radiative cooling, can be higher in these simulations -- which is a major difference to the turbulent mixing layer simulations where the (effective) surface area is fixed. To which degree this additional fragmentation leads to a (non-)convergence in mass transfer depends on the exact geometry and other simulation parameters. Although it is an interesting avenue for future investigations, it does not affect our main conclusions here. 

Interestingly, this slightly larger $\dot{m}_{\rm cl}$ for the simulations with $R_{\rm cl}/d_{\rm cell}|_{\rm ini}=8-32$ also leads to a higher transfer of momentum
 as seen in the position of the centre of mass of the cloud in the right panel of Fig. \ref{fig:convergence-plot}. Therefore, the cloud travels faster and gets quickly entrained at higher resolution. A similar trend has also been previously found in  our earlier works on cloud-crushing (cf. Fig. B1 in \citealt{Kanjilal2021}). However, \citetalias{Chevalier1985} wind introduces an additional complication due to the faster travel of the cloud material -- the wind condition depends strongly on the instantaneous position of the cloud (for example, $T_{\rm wind}\sim d^{-2(\gamma-1)}$), which might lead to differences in cloud properties across simulations at different resolutions as time progresses. 

 Fig. \ref{fig:cloud_profile-high_res} shows the temporal evolution of the cloud ($T<2T_{\rm cl}=8\times 10^4\ \rm K$) profile from our fiducial simulation set (cf. Table \ref{tab:CCinCC85_params}). It is similar to Fig. \ref{fig:cloud_profile}, but at a resolution of $R_{\rm cl}/d_{\rm cell}|_{\rm ini}=16$; one level higher than the base fiducial resolution. The qualitative picture that clouds follow the local wind pressure as the they move downstream, and that they are slightly under-pressurized causing them to keep on growing (albeit gradually losing their density contrast, a.k.a `fading') robustly carries over across simulations at different resolution. This demonstrates that our scaling arguments and the general description of cloud properties are not sensitive to the resolution chosen for the simulation. 

\bsp	
\label{lastpage}
\end{document}